\documentclass[aps,showpacs,amsfonts,amsmath, amssymb,twocolumn,floatfix,superscriptaddress]{revtex4}
\usepackage[final]{graphicx}
\usepackage{color}

\newcommand{\vct}[1]{\mathbf{#1}}

\renewcommand\Re{\operatorname{Re}}
\renewcommand\Im{\operatorname{Im}}
\newcommand\Tr{{\rm Tr}}

\newcommand{\be}{\begin{equation}}
\newcommand{\ee}{\end{equation}}
\allowdisplaybreaks

\DeclareSymbolFont{bbgreek}{U}{bbold}{m}{n}
\DeclareMathSymbol{\bbmu}{\mathbb}{bbgreek}{'26}
\DeclareMathSymbol{\bbeps}{\mathbb}{bbgreek}{'17}
\begin{document}

\title{Trace formulae for non-equilibrium Casimir interactions, heat radiation and heat transfer for arbitrary objects}

\date{\today}

\author{Matthias Kr\"uger}
\affiliation{Massachusetts Institute of Technology, Department of
  Physics, Cambridge, Massachusetts 02139, USA}

\author{Giuseppe Bimonte}
\affiliation{Dipartimento di Scienze Fisiche, Universit{\`a} di Napoli Federico II, Complesso Universitario MSA, Via Cintia, I-80126 Napoli, Italy and INFN Sezione di Napoli, I-80126 Napoli, Italy}

\author{Thorsten Emig}
\affiliation{Laboratoire de Physique Th\'eorique et Mod\`eles
  Statistiques, CNRS UMR 8626, B\^at.~100, Universit\'e Paris-Sud, 91405
  Orsay cedex, France}

\author{Mehran Kardar}
\affiliation{Massachusetts Institute of Technology, Department of
 Physics, Cambridge, Massachusetts 02139, USA}

\begin{abstract} 
We present a detailed derivation of heat radiation, heat transfer and (Casimir) interactions for $N$ arbitrary objects in the framework of fluctuational electrodynamics in thermal non-equilibrium.
The results can be expressed as basis-independent trace formulae in terms of the scattering operators of the individual objects. 
We prove that heat radiation of a single object is positive, 
and that heat transfer (for two arbitrary passive objects) is from the hotter to a colder body.
The heat transferred is also symmetric, exactly reversed if the two temperatures are exchanged.
Introducing partial wave-expansions, we transform the results for radiation, transfer and forces into traces of matrices that can be evaluated in any basis, analogous to the equilibrium Casimir force. The method is illustrated by (re)deriving the heat radiation of a plate, a sphere and a cylinder.  We analyze the radiation of a sphere for different materials, emphasizing that a simplification often employed for metallic nano-spheres is typically invalid. 
We derive asymptotic formulae for heat transfer and non-equilibrium interactions for the cases of a sphere in front a plate and for two spheres, extending previous results. As an example, we show that a hot nano-sphere can levitate above a plate with the repulsive non-equilibrium force overcoming gravity -- an effect that is not due to radiation pressure.  
\end{abstract}

\bibliographystyle{plain}

\maketitle

\section{Introduction}

Quantum thermal fluctuations of electromagnetic waves lie at the heart of statistical physics, 
accounting for seminal discoveries such as Planck's law for thermal radiation of black bodies, introduced almost a century ago~\cite{Planck}. 
The equilibrium Casimir force~\cite{Casimir48} between parallel metallic plates can be equivalently
attributed to the fluctuations of the  electromagnetic field, or to the charge and current fluctuations
in the plates~\cite{Lifshitz56}. 
At separations much smaller than the thermal wavelength, which is roughly 8 $\mu$m at room temperature, these forces are generally dominated by quantum zero point fluctuations, whereas at larger separations  thermal fluctuations also need to be considered~\cite{Lifshitz56,Milonni}.  
In thermal equilibrium the tools of statistical physics can be exploited to ascribe
a (Helmholtz) free energy to a collection of objects; derivative of the free energy with respect to separation (or orientation) of the objects yields the force (or torque). 
The electromagnetic free energy  can itself be compactly expressed in terms of the scattering operators of the objects (see, e.g., Refs.~\cite{Emig07,Neto08,Rahi09}). 

A notable property of the equilibrium formulae is that integrals over frequency can be evaluated
along the imaginary axis~\cite{Lifshitz56}, where expressions for the response functions are
much smoother, and forces are not very much influenced by the precise position of material resonances
(in contrast to thermal non-equilibrium). 
Another feature of equilibrium forces is that stable situations are impossible under rather general conditions, e.g., the free energy as a function of the assembly of objects in vacuum has no minima~\cite{Rahi10}. (Unstable repulsion can still be obtained in certain cases~\cite{LevinMc10}.)

The improved precision of measurements of force and heat transfer at sub-micron scales have provided renewed incentive in the past decade to examine fluctuational electrodynamics (FE) for objects at different temperatures~\cite{Bordag, Antezza08}  (but also other forms of non-equilibrium, e.g., objects in motion~\cite{dalvit10}). The current status of theories for thermal non-equilibrium have in general two commonalities: additional assumptions about the system have to be made (e.g. the assumption of local equilibrium within each object~\cite{Rytov3}) to enable any prediction, and the basic tools of statistical physics, e.g., the free energy, cannot be employed complicating analysis. 
Assuming local equilibria, the current fluctuations in each body are treated separately at the objects' temperature, e.g., using fluctuational electrodynamics introduced by Rytov over 60 years ago~\cite{Rytov3}. Recently, out of equilibrium Casimir forces have been computed in a number of cases including parallel plates~\cite{Antezza08}, deformed plates~\cite{Bimonte09}, as well as a plate and an atom in different setups~\cite{Henkel02,Antezza05,Ellingsen10}. There also exists a large body of work on forces between atoms or molecules in non-equilibrium~\cite{Kweon93,Power94,Cohen03, Sherkunov09, Behunin10, Haakh11, Sanchez12}. Formalisms for treating FE for arbitrary objects at different temperatures have been recently presented~\cite{Messina,Kruger11,Kruger11b, Messina11b}.  In particular, for compact objects, radiation from the environment contributes to the force and has to be incorporated. 
Quite generally  thermal non-equilibrium can be repulsive~\cite{Henkel02,Antezza08,Messina,Kruger11b,Bimonte11,Messina11b}, and allow for stable zero force points~\cite{Kruger11b,Bimonte11}.
Repulsion can even occur at separations far below the thermal wavelength if the resonances of the materials are suitably detuned~\cite{Bimonte11}. 
It is also possible for a hot and cold sphere to exert equal mutual forces in the same direction,
leading to a self-propelled state~\cite{Kruger11b}. 
Two parallel nanotubes where shown to be good candidates for experimental detection of non-equilibrium forces, as these can be made relatively strong~\cite{Golyk12}. Recently, the high temperature limit was investigated in thermal non-equilibrium \cite{Dean12}.

Heat radiation and transfer are of particular interest when the size and/or separation of the objects is comparable to, or smaller than, the thermal wavelength, because then they differ strongly from the predictions of the Stefan-Boltzmann law.
For example, as confirmed experimentally~\cite{Sheng09, Rousseau09,Ottens11}, there is
a considerably larger near-field heat transfer due to tunneling of evanescent waves.  
Theoretical computations of heat transfer were only recently extended from two parallel plates~\cite{Polder71} or dipoles~\cite{Volokitin01} to compact objects of finite size, more precisely to two spheres~\cite{Narayanaswamy08,Sasihithlu11} and a sphere in front a plate~\cite{Kruger11,Otey,McCauley}. 
Numerical studies for objects whose scattering properties are not known analytically include transfer for periodic structures~\cite{Rodriguez11}, as well as a cone or finite cylinder in front a plate~\cite{McCauley}, and very recently, numerical scattering techniques were implemented more generally \cite{Rodriguez}. (See also Refs.~\cite{Chapuis08,Biehs11,Ben-Abdallah11,Gamble11,Biehs,Abdallah,Guerout12} for recent studies of various aspects of heat transfer.) 

The radiation of single spheres and plates has theoretically been studied by many
authors~\cite{Rytov3,Rytovc,Kattawar70,Bohren,Hansen98,Bimonte09b,Kruger11,Golyk11}. Radiation of single cylinders, for which an early calculation by Rytov exists~\cite{Rytovc}, was only recently formulated in terms of scattering theory~\cite{Bimonte09b,Kruger11,Golyk11}, and has been investigated experimentally~\cite{Ohman61,li03,Fan09, Singer11, Singer11b}, mostly focussing on polarization effects in the emission of thin tubes (cylinders). Recently, rotating objects were considered, which emit spontaneously at zero temperature~\cite{Manjavacas10, Maghrebi11b}.  Reference~\cite{Beenakker99} provides a basis independent trace formula for the radiation of an isolated object (see also~\cite{Maghrebi11b}).

The main result of this paper is the derivation of
 general trace formulae for heat radiation, transfer and non-equilibrium forces
 (including contributions from the environment) for arbitrary objects.
The trace formulae do not refer to any particular (wave) basis and hence can be employed in rather general situations.
 We  give proofs for the positivity of radiation and transfer. To demonstrate the power of our general results, we provide analytical as well as numerical examples.
The paper is divided into three main parts: The first part (Secs.~\ref{sec:radiation} - \ref{sec:force}) 
formulates the problem in terms of scattering operators (e.g. $\mathbb{T}(\vct{r},\vct{r}')$). 
In the second part (Sec.~\ref{sec:waves}), we transform these expressions into 
matrix forms in arbitrary partial wave basis. 
The third part (Secs.~\ref{sec:hr} - \ref{sec:Appl}) is devoted to specific analytical and numerical examples. 

In particular, in Sec.~\ref{sec:noneq} we describe the model and derive the non-equilibrium correlation function of the electric field, followed by an introduction of the $\mathbb{T}$ operator. In Sec.~\ref{sec:radiation}, we derive  formulae for both the electric field correlation as well as the emitted energy for a single object in terms of $\mathbb{T}$, and prove that the emitted energy is positive for any object made of passive material. In Sec.~\ref{sec:EFneq}, we provide the non-equilibrium  field correlator for $N$ objects. This correlator is then used to derive a trace formula for heat transfer in Sec.~\ref{sec:heat}, where we also prove the positivity of the transfer as well as its symmetry with respect to a permutation of temperatures. Trace formulae for the Casimir force are derived in Sec.~\ref{sec:force}. In Sec.~\ref{sec:waves}, we introduce partial wave expansions and express the trace formulae in terms of the corresponding matrix expressions. A short discussion of differences between equilibrium and non-equilibrium calculations follows in Sec.~\ref{sec:asy}. In Sec.~\ref{sec:hr}, we (re)derive the radiation of a plate, a sphere and a cylinder and analyze the radiation of a sphere in detail. Then we give asymptotic expansions for the cases of two spheres and a sphere in front of a plate for heat transfer (Sec.~\ref{sec:ht}) and forces (Sec.~\ref{sec:fo}). Sec.~\ref{sec:Appl} provides numerical examples for the sphere--plate case, demonstrating that the non-equilibrium force can lead to stable levitation points. In App.~\ref{app:eqf} we show that the results for equilibrium forces can be derived from our trace formula for non-equilibrium forces. Appendices~\ref{app:plane}-\ref{app:conv} present the partial wave expansions and conversion matrices used in Secs.~\ref{sec:hr}-\ref{sec:fo}.

We note that during submission of this manuscript, we became aware of an independent, but partly related work on a trace formula for heat transfer, its symmetry and positivity \cite{Rodriguez}.

\section{Non-equilibrium Fluctuations -- general concepts}
\label{sec:noneq}

We consider an arrangement of $N$ objects labelled by $\alpha=1\dots N$, 
in vacuum at time-independent, homogeneous temperatures $\{T_\alpha\}$, and embedded in an
environment at temperature $T_{\rm env}$. The objects are characterized by their electric and magnetic response $\bbeps(\omega;\vct{r},\vct{r}')$ and $\bbmu(\omega;\vct{r},\vct{r}')$, which can in general be nonlocal complex tensors, $\bbeps(\omega;\vct{r},\vct{r}')=\varepsilon_{ij}(\omega;\vct{r},\vct{r}')$, depending on frequency $\omega$. In this non-equilibrium stationary state, each object is assumed to be at local equilibrium, such that the current fluctuations within the object 
satisfy the fluctuation dissipation theorem at the appropriate temperature~\cite{Rytov3}.

For any two   field operators  ${\hat A}$ and ${\hat B}$, we consider the symmetrized expectation value,
\be
\langle {\hat A}(t,{\bf r})\, {\hat B}(t',{\bf r}')\rangle_{\rm s} \equiv \frac{1}{2}\! 
\left\langle\!  {\hat A}(t,{\bf r}) {\hat B}(t',{\bf r}') +
 {\hat B}(t',{\bf r}') {\hat A}(t,{\bf r}) \!\right\rangle .
\ee
Symmetrization ensures the reality of $\langle  {\hat A}(t,{\bf r}) {\hat B}(t',{\bf r}') \rangle_{\rm s}$ for generally non-commuting quantum operators. In stationary conditions, the expectation value depends only on the time difference $t-t'$, and we can define the spectral density $\langle A({\bf r} ) B^*( {\bf r'} )\rangle_{\omega} $ by
\be
 \left\langle  {\hat A}(t,{\bf r})\, {\hat B}(t',{\bf r}')  \right\rangle_{\rm s}= \int_{-\infty} ^{\infty}\frac{d \omega}{2 \pi} \,e^{-i \omega (t-t')} \, \langle A({\bf r} ) B^*( {\bf r'} )\rangle_{\omega}\;.  
\label{sd}
\ee
Due to the reality of the correlation function on the left hand side, the real part of the spectral density is an even function of frequency, while its imaginary part is odd.

The relevant quantity for our considerations is the spectral density $\mathbb{C}\equiv C_{ij}$ of the electric field $\vct{E}$ at
 points $\vct{r}$ and $\vct{r}'$, from which heat radiation and Casimir forces can then be extracted (see Secs.~\ref{sec:heat} and \ref{sec:force} below), defined by 
\begin{equation}
 C_{ij}({\bf r},{\bf r} ')\equiv \langle E_i({\bf r} ) E_j^*( {\bf r'} )\rangle_{\omega}. 
\end{equation}
In the following we shall not make explicit the dependence on $\omega$ in frequency dependent quantities. 
 In order to derive this correlation in the considered non-equilibrium state, we start with the equilibrium case where all temperatures are equal, $T_\alpha=T_{\rm env}=T$. Then, $\mathbb{C}$ is well known, and can be expressed in terms of $G_{ij}$, the dyadic retarded Green's function of the system~\cite{Eckhardt83}. This relation is a variant of the fluctuation dissipation theorem, and reads~\footnote{$\mathbb{C}$ has an extra factor of $1/(2\pi)$ compared to Ref.~\cite{Kruger11}.}
\begin{align}
C_{ij}^{eq}(T;\vct{r},\vct{r}')=\left[{a}(T)+a_0\right] \Im G_{ij}(\vct{r},\vct{r}').\label{eq:1}
\end{align}
Here~\footnote{$a(T)$ and $a_0$ carry extra factors of $c^2/(2\pi\omega^2)$ compared to Ref.~\cite{Kruger11}.}
\begin{align}
a(T)\equiv {\rm sgn}(\omega)\frac{8\pi\hbar\omega^2}{c^2}[\exp(\hbar |\omega|/k_BT)-1]^{-1} ,
\end{align}
contains the occupation
number of modes with frequency $\omega$, $c$ is the speed of light, and
$\hbar$ is Planck's constant. Zero point fluctuations have amplitude $a_0\equiv {\rm sgn}(\omega) \frac{4\pi\hbar\omega^2}{2c^2}$, but play no role in non-equilibrium phenomena-- they are independent of temperature and present everywhere in space. 
They do contribute to the total force, as they are also responsible for the zero point Casimir effect.

The dyadic  Green's function obeys the Helmholtz equation
\begin{equation}
\left[ \mathbb{H}_0-\mathbb{V}-\frac{\omega^2}{c^2}\mathbb{I}\right]\mathbb{G}(\vct{r},\vct{r}')=\mathbb{I}\delta^{(3)}(\vct{r}-\vct{r}'),\label{eq:H}
\end{equation}
where $\mathbb{H}_0=\boldsymbol{\nabla}\times\boldsymbol{\nabla}\times$, and~\footnote{$\mathbb{V}$ is defined with a minus sign compared to Ref.~\cite{Rahi09}.}
\begin{equation}
\mathbb{V}=\frac{\omega^2}{c^2}(\bbeps-\mathbb{I})+\boldsymbol{\nabla}\times\left(\mathbb{I}-\frac{1}{\bbmu}\right)\boldsymbol{\nabla}\times\label{eq:V}
\end{equation}
is the potential introduced by the objects.
We can separate the term $\langle \vct{E}\otimes\vct{E}^*\rangle_\omega^{0}\equiv a_0\Im \mathbb{G}$ (where we introduce the dyadic vector notation $(\vct{E}\otimes\vct{E}^*)_{ij}=E_iE_j^*$) involving the zero point contribution, and concentrate on the remaining $T$-dependent terms. These can be split up according to their originating thermal sources~\cite{Kruger11}, yielding $N+1$ terms, including the contribution from the environment. The equilibrium correlation in Eq.~\eqref{eq:1} can thus be rewritten as
\begin{align}
\mathbb{C}^{eq}=\langle \vct{E}\otimes\vct{E}^*\rangle_\omega^{0}+\sum_\alpha \mathbb{C}_\alpha^{sc}(T) +\mathbb{C}^{\rm env}(T).\label{eq:4}
\end{align}
This form shows the different contributions to the electric field correlations in equilibrium. The term $\mathbb{C}_\alpha^{sc}(T)$ represents the radiation from the sources in object $\alpha$, and is given by 
\begin{equation}
\mathbb{C}_\alpha^{sc}(T)=a(T)\mathbb{G}\Im[\mathbb{V}_\alpha]\mathbb{G}^*.\label{eq:ros}
\end{equation}
$\mathbb{V}_\alpha$ is the potential of object $\alpha$, i.e., $\bbeps$ and $\bbmu$ in Eq.~\eqref{eq:V} are replaced by $\bbeps_\alpha$ and $\bbmu_\alpha$. Equation~\eqref{eq:ros} is identified with the original definition of the field correlator in Rytov's formalism~\cite{Rytov3, Eckhardt83}.  For the case of $\Im[\bbmu_\alpha]=0$, we have the more familiar expression $\mathbb{C}_\alpha^{sc}(T)=a(T)\frac{\omega^2}{c^2}\mathbb{G}\Im[\bbeps_\alpha]\mathbb{G}^*$. One difference to the heat radiation of object $\alpha$ as described in Refs.~\cite{Rytov3, Eckhardt83} is that in Eq.~\eqref{eq:ros} the radiation is scattered {\it by all objects}, such that $\Im[\mathbb{V}_\alpha]$ is multiplied from both sides by the full Green's function $\mathbb{G}$. For a single object in isolation [see Eq.~\eqref{eq:single} below], $\mathbb{G}$ is replaced by $\mathbb{G}_\alpha$, the Green's function of object $\alpha$ in isolation.

Comparing Eqs.~\eqref{eq:1} and~\eqref{eq:4}, we derive the last term in Eq.~\eqref{eq:4} 
which can be identified as the contribution of the sources in the environment,
\begin{equation}
\mathbb{C}^{\rm env}(T)=- a(T)\mathbb{G} \Im \left[\mathbb{G}_0^{-1}\right] \mathbb{G}^* \, .
\end{equation}
This can be evaluated further by using 
\begin{equation}
\Im[\mathbb{V}]=-\Im[\mathbb{G}^{-1}-\mathbb{G}_0^{-1}],
\end{equation} 
which follows directly from Eq.~\eqref{eq:H} where $\mathbb{G}_0$ is the free Green's function solving the wave equation for $\mathbb{V}=0$. 
Fluctuations in the vacuum are taken into account by the nontrivial term $\mathbb{G}_0^{-1}$,
which can be attributed to infinitesimal environmental ``dust"~\cite{Eckhardt83}. 
Integrating such infinitesimal ``dust'' sources over the infinite space of the environment yields a finite result. 

The intuitive result that the equilibrium field in Eq.~\eqref{eq:4} is the sum of the radiation emitted by the sources in the objects and the sources in the environment can also be corroborated by deriving $\mathbb{C}^{\rm env}$ in Eq.~\eqref{eq:4} along another, straight forward route. We start from the field $\vct{E}$ sourced by the environment, without any objects present. It has the correlator 
\begin{equation}
\langle \vct{E}\otimes \vct{E}^*\rangle^{\rm{free}}_\omega=a(T)\Im \mathbb{G}_0.\label{eq:env0}
\end{equation}
Then, adding the cold, i.e., non-radiating objects, generates scattered fields according to the  Lippmann-Schwinger equation. If the field $\vct{E}$ solves the Helmholtz equation in free space, then the following $\vct{E}^{sc}$ solves it with the objects present,
\begin{equation}
\vct{E}^{sc}=\mathbb{G}\mathbb{G}_0^{-1}\vct{E}.\label{eq:LS}
\end{equation}
Applying the operator $ \mathbb{G}\mathbb{G}_0^{-1}$ to both fields in Eq.~\eqref{eq:env0}, one finds for the environment contribution
\begin{align}
\mathbb{C}^{\rm env}(T)&= \mathbb{G}\mathbb{G}_0^{-1} \langle \vct{E}\otimes \vct{E}^*\rangle^{\rm free}_\omega \mathbb{G}_0^{-1*} \mathbb{G}^*\\&= -a(T)\mathbb{G} \Im\left[ \mathbb{G}_0^{-1} \right]\mathbb{G}^*,
\end{align}
reproducing the last term of Eq.~\eqref{eq:4}. 

Having identified the contributions from the different sources, we can now change the temperature of these sources independently, denoting by $T_\alpha$  the temperature of object $\alpha$ and by $T_{\rm env}$ the temperature of the environment. The field correlator in the non-equilibrium situation is then a simple modification of Eq.~\eqref{eq:4} to
\begin{align}
\mathbb{C}^{\rm neq}(\{T_\alpha\},T_{\rm env})&=\langle \vct{E}\otimes\notag\vct{E}^*\rangle_\omega^{0}+\sum_\alpha \mathbb{C}_\alpha^{sc}(T_\alpha) \\&+\mathbb{C}^{\rm env}(T_{\rm env}).\label{eq:5}
\end{align}
Equation~\eqref{eq:5} gives the general field correlator for arbitrary combinations of the temperatures of the $N$ body system. It contains $N+1$ unknown terms, due to the $N+1$ sources in the system. We note however that one of the sources can be eliminated by introducing the equilibrium correlation function at finite temperature. We chose to eliminate the environment contribution and obtain~\cite{Kruger11}
\begin{equation}
\mathbb{C}^{\rm neq}(\{T_\alpha\},T_{\rm env})=\mathbb{C}^{\rm eq} (T_{\rm env})+\sum_\alpha \left[ \mathbb{C}_\alpha^{\rm sc}(T_\alpha) -\mathbb{C}_\alpha^{\rm sc}(T_{\rm env})\right].\label{eq:6}
\end{equation}
This form shows that we only have to  evaluate the $N$ terms $\mathbb{C}_\alpha^{sc}$ (assuming $\mathbb{C}^{\rm eq}$ is known), in order to compute heat transfer and forces depending on $N+1$ temperatures. For Casimir forces, $\mathbb{C}^{\rm eq} (T_{\rm env})$ will give the equilibrium force at temperature $T_{\rm env}$. 
However, due to its equilibrium origin, this term does not contribute to the heat transfer. In Sec.~\ref{sec:EFneq}, we give the final formula for $\mathbb{C}^{\rm neq}(\{T_\alpha\},T_{\rm env})$ in terms of the scattering operators of the objects. In Secs.~\ref{sec:heat} and \ref{sec:force}, we shall derive the resulting heat transfer and forces, respectively. 

We conclude this section by introducing the classical $\mathbb{T}$-operator which provides a convenient way of rewriting the Helmholtz equation as a Lippmann-Schwinger equation~\cite{Lippmann50}. Starting from 
\begin{equation}
\vct{E}^{sc}=\vct{E}+\mathbb{G}_0\mathbb{V}\vct{E}^{sc},
\end{equation}
we can formally write $\vct{E}^{sc}$ in terms of the $\mathbb{T}$-operator~\footnote{$\mathbb{T}$ has a minus sign compared to Refs.~\cite{Rahi09,Kruger11}.} as
\begin{equation}
\vct{E}^{sc}=\vct{E}+\mathbb{G}_0\mathbb{T}\vct{E}\label{eq:LS3}.
\end{equation}
Solving for $\mathbb{T}$, we obtain
\begin{equation}
\mathbb{T}=\mathbb{V}\frac{1}{1-\mathbb{G}_0\mathbb{V}},\label{eq:TV}
\end{equation}
which to the lowest order equals $\mathbb{V}$, as in the Born approximation.  
Comparing Eq.~\eqref{eq:LS3} to Eq.~\eqref{eq:LS}, we find the following relation between $\mathbb{T}$ and $\mathbb{G}$~\cite{Rahi09},
\begin{equation}
\mathbb{G}=\mathbb{G}_0+\mathbb{G}_0\mathbb{T}\mathbb{G}_0.\label{eq:GT}
\end{equation}
Note that $\mathbb{T}$ is the scattering operator of the entire collection of  objects, whereas 
we shall use $\mathbb{T}_\alpha$ for object $\alpha$ in isolation.

\section{Radiation of one object in isolation}
\label{sec:radiation} 

In this section we  derive the correlator $\mathbb{C}_\alpha$ of a single warm object in  a cold environment, a prerequisite for the derivation of $\mathbb{C}_\alpha^{sc}$ needed in Eq.~\eqref{eq:6}. We shall also compute the energy emitted  of the isolated object and prove that it is positive. 

\subsection{Field correlations}

Equation~\eqref{eq:6} requires the expression for $\mathbb{C}_\alpha^{sc}$,  
the field sourced by object $\alpha$ and scattered by all objects.  
In the absence of other objects,  the thermal field correlator (neglecting zero point terms)  satisfies
\begin{equation}
\mathbb{C}_\alpha(T_\alpha)=a(T_\alpha)\mathbb{G}_\alpha\Im[\mathbb{V}_\alpha]\mathbb{G}_\alpha^*~;\label{eq:single}
\end{equation}
$\mathbb{C}_\alpha(T_\alpha)$ differing from $\mathbb{C}_\alpha^{sc}$ in Eq.~\eqref{eq:ros} by the appearance of the Green's function $\mathbb{G}_\alpha$ instead of $\mathbb{G}$. There are different ways to evaluate Eq.~\eqref{eq:single} for a specific geometry. The most straightforward approach is to start with the Green's function $\mathbb{G}_\alpha$ with one point inside the object and one point outside, as this is the structure of Eq.~\eqref{eq:single}: $\Im[\mathbb{V}_\alpha]$ is only nonzero inside the object and we are interested in the field outside. Once this is accomplished,  Eq.~\eqref{eq:single} can be directly  evaluated by an integration over the volume of the object. Such an approach was used, e.g., in Refs.~\cite{Antezza08,Narayanaswamy08} to find the non-equilibrium Casimir force for parallel plates and the heat transfer between two spheres, respectively. 

Since we aim to describe the non-equilibrium effects through the scattering formalism, 
we would like to express desired observables -- starting from the heat radiation of a single object -- in terms of the scattering operators $\{\mathbb{T}_\alpha\}$. 
The single object's radiation can indeed be expressed in terms of $\mathbb{T}_\alpha$ by starting from the equilibrium situation of $T_\alpha=T_{\rm env}$, where the field correlator $\mathbb{C}^{eq}_\alpha(T_\alpha)$ (not containing zero point fluctuations)  can be split [in a manner similar  to Eq.~\eqref{eq:4}] into contributions of heat sources from the object and from the environment, as
\begin{align}
\mathbb{C}^{eq}_\alpha(T_\alpha)&=\mathbb{C}_\alpha(T_\alpha)+\mathbb{C}_\alpha^{\rm env}(T_\alpha)=a(T_\alpha)\Im\mathbb{G}_\alpha\label{eq:Ceq},\\
\mathbb{C}_\alpha^{\rm env}(T_\alpha)&=-a(T_\alpha)\mathbb{G}_\alpha \Im \left[\mathbb{G}_0^{-1}\right] \mathbb{G}_\alpha^*.\label{eq:envr}
\end{align}
The last equality in Eq.~\eqref{eq:Ceq} follows from Eq.~\eqref{eq:1} after reduction  to one object~\cite{Eckhardt83}.
In order to arrive at the desired heat radiation of the object, we solve Eq.~\eqref{eq:Ceq} for $\mathbb{C}_\alpha(T_\alpha)$, yielding
\begin{align}
\mathbb{C}_\alpha(T_\alpha)=-\mathbb{C}_\alpha^{\rm env}(T_\alpha)+\mathbb{C}^{eq}_\alpha(T_\alpha) \, .
\label{eq:Crad}
\end{align}
$\mathbb{C}^{eq}_\alpha(T_\alpha)$ is readily expressed in terms of the $\mathbb{T}$-operator via Eq.~\eqref{eq:GT}. 
It remains to express the radiation sourced by the environment, see Eq.~\eqref{eq:envr}, in terms of $\mathbb{T}$. This can be achieved along different routes. We want to first present the way introduced in Ref.~\cite{Kruger11} and then present a general formula for the radiated field. 

\subsubsection{Integration over environment ``dust''}

Following the interpretation introduced in Ref.~\cite{Eckhardt83}, the environment can be regarded as composed of ``dust'' characterized by a homogeneous dielectric response $\varepsilon_{\rm env}$ in the infinite space complimentary to the object. The  Green's function of the system [for $\mathbb{V}=\frac{\omega^2}{c^2}\mathbb{I}(\varepsilon_{\rm env}-1)$ outside, and $\mathbb{V}=\mathbb{V}_\alpha$ inside the object] is denoted $\tilde{\mathbb{G}}_\alpha$. It is a simple modification of $\mathbb{G}_\alpha$ as a uniform $\varepsilon_{\rm env}-1$ only changes the speed of light outside the object, replacing $c$ with $c/\sqrt{\varepsilon_{\rm env}}$. Formally interpreting the environment as an additional object, we can use Eq.~\eqref{eq:ros} to get the fields sourced by it. Taking $\varepsilon_{\rm env}\to1$ yields a well defined radiation from the environment,
\begin{equation}
\mathbb{C}_\alpha^{\rm env}(T_\alpha)=\lim_{\varepsilon_{\rm env}\to1}a(T_\alpha)\frac{\omega^2}{c^2}\tilde{\mathbb{G}}_\alpha\Im[\varepsilon_{\rm env}]\tilde{\mathbb{G}}_\alpha^*~.
\end{equation}
Writing out this equation explicitly, it is possible to see more clearly the operator products involved, as
\begin{align}
\notag&C_{\alpha,ij}^{\rm env}(T_\alpha;\vct{r},\vct{r}')=a(T_\alpha)\frac{\omega^2}{c^2}\\&\times \lim_{\varepsilon_{\rm env}\to1} \sum_k\int_{\rm env} d^3 r'' \tilde{G}_{\alpha,ik} (\vct{r},\vct{r}'')\Im [\varepsilon_{\rm env}] \tilde{G}^*_{\alpha,kj} (\vct{r}'',\vct{r}')\label{eq:out}.
\end{align}
First, we note that all the arguments of the Green's functions in Eq.~\eqref{eq:out} lie outside the object, such that $\tilde{\mathbb{G}}_{\alpha}$ can be found by use of scattering theory (see Sec.~\ref{sec:waves}). One additional simplification occurs since the limit $\varepsilon_{\rm env}\to1$ allows to neglect any finite region of integration.  We can hence restrict the integration range to the region with $\xi_1(\vct{r}'')>{\rm max}[\xi_1(\vct{r}),\xi_1(\vct{r}')]$, where $\xi_1$ is the radial component which distinguishes the two expansions of $\mathbb{G}_0$ in Eq.~\eqref{onshell} below. This practical simplification which holds for any finite $\vct{r}$, $\vct{r}'$, 
allows  restriction to one of the cases in Eq.~\eqref{onshell}, such that finally 
\begin{align}
\notag&C_{\alpha,ij}^{\rm env}(T_\alpha;\vct{r},\vct{r}')= \lim_{\varepsilon_{\rm env}\to1} \sum_k\int\limits_{\xi(\vct{r}'')>\text{max}\{\xi(\vct{r}),\xi(\vct{r}')\}} \!\!\!\!d^3 r'' \\&\tilde{G}_{\alpha,ik} (\vct{r},\vct{r}'')\Im [\varepsilon_{\rm env}] \tilde{G}^*_{\alpha,kj} (\vct{r}'',\vct{r}')\label{eq:out2}.
\end{align}
Equation~\eqref{eq:out2} was presented in Ref.~\cite{Kruger11}, and its application for a cylindrical object was worked out in detail in Ref.~\cite{Golyk11}. 
It can be evaluated in a straightforward manner if the matrix elements $\mathcal{T}_{\mu\mu'}$ [see Eq.~\eqref{Tmat}] are known.

\subsubsection{A general formula for the field correlations}

A simpler method for expressing the radiation of the environment, presented in Ref.~\cite{Golyk11},  allows to give Eq.~\eqref{eq:Crad} in closed form. 
First, we rewrite the expression for the environment radiation, Eq.~\eqref{eq:envr}, using Eq.~\eqref{eq:GT}, and (after a few steps)  find~\cite{Golyk11}
\begin{align}
\mathbb{C}_\alpha^{\rm env}(T_\alpha)=a({T_\alpha})[1+\mathbb{G}_0\mathbb{T}_\alpha]\Im \mathbb{G}_0 [\mathbb{T}^*_\alpha\mathbb{G}^*_0+1].\label{eq:envT}
\end{align}
Here, we have used that the $\mathbb{T}$ operator is symmetric [because of Eq.~\eqref{eq:GT} and since $\mathbb{G}$~\cite{Eckhardt83} is symmetric].
Using this form, and additionally writing $\mathbb{C}_\alpha^{eq}$ in terms of $\mathbb{T}_\alpha$ via Eq.~\eqref{eq:GT}, the heat radiation in Eq.~\eqref{eq:Crad} can be given in closed form for an arbitrary object. After some manipulations we find,
\begin{align}
\mathbb{C}_\alpha(T_\alpha)&=
a(T_\alpha) \mathbb{G}_0\bigg[\frac{i}{2}\left(
\mathbb{T}_\alpha^* -\mathbb{T}_\alpha \right)-\mathbb{T}_\alpha \Im[\mathbb{G}_0]\mathbb{T}_\alpha^* \bigg]\mathbb{G}_0^*.\label{eq:radf}
\end{align}
This describes the radiation of an arbitrary object in a basis-independent representation in terms of two well known quantities, the free Green's function and the $\mathbb{T}$-operator. Such a basis-independent representation can be of advantage when numerical methods are employed~\cite{McCauley} to find $\mathbb{T}_\alpha$.  We emphasize again that it holds for any material properties, with magnetic or electric losses. In its derivation, which requires only Eqs.~\eqref{eq:Ceq} and~\eqref{eq:envT}, we even do not have to be cautious about (magnetic or electric) material losses as this information is contained in the   $\mathbb{T}$-operator. In case of lossless materials, Eq.~\eqref{eq:radf} does not correspond to a field configuration that supports energy transport, as will be demonstrated below. To simplify notation, we define the radiation operator $\mathbb{R}_\alpha$ and write 
\begin{align}
\notag\mathbb{C}_\alpha(T_\alpha)&= a(T_\alpha)\mathbb{R}_\alpha,\quad{\rm with}\\
\mathbb{R}_\alpha&\equiv \mathbb{G}_0\bigg[\Im[\mathbb{T}_\alpha] -\mathbb{T}_\alpha \Im[\mathbb{G}_0]\mathbb{T}_\alpha^* \bigg]\mathbb{G}_0^*.\label{eq:radf1}
\end{align}

\subsection{Trace formula for the emitted energy}
\label{sec:traces}

In the previous subsection we  derived the correlation function of the electric field for a single object in an environment at zero temperature, described by the radiation operator $\mathbb{R}_\alpha$ in Eq.~\eqref{eq:radf}. We now consider the heat emitted by this object, obtained by integrating the normal component of the Poynting vector $\vct{S}$ over a surface $\Sigma_\alpha$ enclosing the object, as
\begin{equation}
H_\alpha=\oint_{\Sigma_\alpha} \vct{S}\cdot \vct{n}_{\alpha} , \label{eq:singleobject}
\end{equation}
with 
\begin{align}
\vct{S}(\vct{r})&=\frac{c}{4\pi}\int_{-\infty}^{\infty}\frac{d\omega}{2\pi}\left\langle\vct{E}(\vct{r})\times\vct{B}^*(\vct{r})\right\rangle_\omega\,.
\end{align}
It is  straightforward  to evaluate the integral in Eq.~\eqref{eq:singleobject}  for simple objects such as a sphere or a plane. Nevertheless, a more convenient and illuminating  form can be achieved by reconsidering the derivation of the Poynting theorem~\cite{Jackson}, starting from the work done by the fluctuating fields in a volume element located at $\vct{r}$. This work is given by the electric field times the total electric current $\bf J$ at $\vct{r}$. Hence the total work done on the object (where we include a minus sign to get the emitted energy) is given by an integral over the {\it volume} of the object,
\begin{align}
H_\alpha=- \int_{-\infty}^{\infty}\frac{d\omega}{2\pi}   \int_{V_\alpha} d^3r\left\langle \vct{E}(\vct{r})\cdot\vct{J}^*(\vct{r}) \right\rangle_\omega. \label{eq:volints}
\end{align}
To proceed, we use the free Green's function to convert between the total field and the total current,
\begin{align}
\vct{E}=4\pi i\frac{\omega}{c^2}\mathbb{G}_0 \vct{J}.\label{eq:GJ}
\end{align}
Applying this to Eq.~\eqref{eq:volints}, leads to
\begin{align}
H_\alpha= \frac{c^2}{2\pi}\int_0^\infty\frac{d\omega}{2\pi} \frac{1}{\omega} \Im \int_{V_\alpha} d^3r\left\langle \vct{E}(\vct{r})\cdot (\mathbb{G}_0^{-1*} \vct{E}^*)(\vct{r}) \right\rangle_\omega.\label{eq:Hr} 
\end{align}
From Eq.~\eqref{eq:radf} we note that  the correlation function of the electric field carries the Green's function $\mathbb{G}_0^*$ on its rightmost position. Hence the operation of $\mathbb{G}_0^{-1*}$ in Eq.~\eqref{eq:Hr} is easily performed, leaving $\mathbb{T}_\alpha$ on the  rightmost position. We now note that the range of integration can be extended to {\it all space}, as $\mathbb{R}_\alpha \mathbb{G}_0^{-1*}$ is nonzero within the volume $V_\alpha$ only. This is because $\mathbb{T}_\alpha(\vct{r},\vct{r}')$ is only nonzero if both arguments are located within the volume $V_\alpha$. Thus, the integral in Eq.~\eqref{eq:Hr} together with the scalar product turns into a trace of the operator $\mathbb{R}_\alpha \mathbb{G}_0^{-1*} $, and we get for the energy emitted by an arbitrary object with scattering operator $\mathbb{T}_\alpha$,
\begin{widetext}
\begin{align}
H_\alpha(T_\alpha)=\frac{2\hbar}{\pi}\int_0^\infty d\omega \frac{\omega}{e^{\frac{\hbar\omega}{k_BT_\alpha}}-1} \Tr \left\{\Im[\mathbb{G}_0]\Im[\mathbb{T}_\alpha] - \Im[\mathbb{G}_0] \mathbb{T}_\alpha \Im[\mathbb{G}_0]\mathbb{T}_\alpha^* \right\} \, . 
\label{eq:tracerad}
\end{align}
\end{widetext}
The heat transfer $H$ from the object to the environment at finite temperature $T_{\rm env}$ follows directly from detailed balance, as
\begin{align}
H=H_\alpha (T_\alpha)-H_\alpha(T_{\rm env}) \, .
\end{align} 
Note that the trace in Eq.~\eqref{eq:tracerad} of the operator $\mathbb{R}_\alpha\mathbb{G}_0^{-1*}=(\mathbb{R}_\alpha\mathbb{G}_0^{-1*})_{ij}(\vct{r},\vct{r}')$ is both over the vector indices $i$ and $j$, as well as the positions $\vct{r}$ and $\vct{r}'$~\footnote{The equilibrium Casimir force is initially expressed similarly, i.e., by a trace of an operator, see, e.g., Eq.~(5.6) in Ref.~\cite{Rahi09}.}. This can be converted into a more familiar trace in a partial wave basis, see Eq.~\eqref{eq:trel} in  Sec.~\ref{sec:waves} below. However, when $\Im[\mathbb{G}_0]$ is expanded in wave functions it contains only propagating waves, such that the trace in Eq.~\eqref{eq:trel} is restricted accordingly.

\subsection{Positivity of heat radiation}
\label{sec:pos}

Causality implies that the potential $\mathbb{V}(\omega)$  in Eq.~\eqref{eq:V} is an  analytic function in the upper half of the complex frequency plane, with the  property
\begin{equation}
\mathbb{V}(-z^*)=\mathbb{V}^*(z) \, .
\end{equation}
For real frequencies, the above condition implies that the real part of $\mathbb{V}(\omega)$ is an even function of $\omega$, while its imaginary part is odd. By virtue of this symmetry, we may consider only  positive frequencies.  The imaginary part of the potential $\mathbb{V}(\omega)$ for a body made  of a passive material must be positive semi-definite, i.e.
\begin{equation}
{\rm Im} [\mathbb{V}] \ge 0 \, .
\label{pos0}
\end{equation}
For any positive semi-definite operator $A$, the product $BAB^\dagger$ is also positive semi-definite, and we have 
\begin{equation}
\mathbb{G}^{-1}_0 \mathbb{G} \, {\rm Im} [\mathbb{V} ]\, \mathbb{G}^{*} \mathbb{G}^{*-1}_0  \ge 0\;.
\label{pos1}
\end{equation}
If we furthermore use Eq.~\eqref{eq:TV}, we directly find 
\begin{equation}
{\rm Im} [\mathbb{T} ]-\mathbb{T}\, {\rm Im} [\mathbb{G}_{0}]\, \mathbb{T}^{*}  \ge 0\;.\label{pos}
\end{equation}
As ${\rm Im} [\mathbb{G}_{0}]$ is a  positive semi-definite hermitian operator, and the product of two positive semi-definite hermitian operators is also positive semi-definite, the operator to be traced in Eq.~\eqref{eq:tracerad} is positive semi-definite as well. This shows that the emitted energy $H_\alpha$ is a nonnegative number,
\begin{equation}
H_\alpha\geq 0 \, .
\label{eq:HRpos}
\end{equation}
While this is expected on physical grounds, to our best knowledge, it has not been proven for an object of arbitrary shape before.
Furthermore, as $\mathbb{T} {\rm Im} [\mathbb{G}_{0}] \mathbb{T}^{*}  \ge 0$, we conclude from Eq.~\eqref{pos} 
\begin{equation}
{\rm Im} [\mathbb{T} ] \ge \mathbb{T}\, {\rm Im} [\mathbb{G}_{0}] \,\mathbb{T}^{*}  \ge 0\;,\label{eq:Tpos}
\end{equation}
which proves that ${\rm Im} [\mathbb{T} ]$ is a positive semi-definite operator.

\section{Radiation from multiple objects at different temperatures}
\label{sec:EFneq}

In Sec.~\eqref{sec:radiation} we derived  $\mathbb{C}_\alpha(T_\alpha)$ for the field radiated by an isolated object $\alpha$, in terms of the radiation operator $\mathbb{R}_\alpha$. This radiation is scattered  at all other objects in the system, leading to the modified correlator $\mathbb{C}_\alpha^{sc}$, which is the unknown term in the total non-equilibrium correlator  of Eq.~\eqref{eq:ros}. In the following, $\mathbb{V}_{\bar{\alpha}}$ will denote the composite potential of all objects except object $\alpha$. For two objects, $\mathbb{V}_{\bar{\alpha}}$ is the potential of the second object.  

We place the cold objects described by $\mathbb{V}_{\bar{\alpha}}$ into the field radiated by object $\alpha$. If  the solution to the Helmholtz equation for object $\alpha$ alone is denoted by ${\bf E}_{\alpha,iso}$, then the solution ${\bf E}_{\alpha}$ for all objects can be expressed through the Lippmann-Schwinger equation as~\cite{Lippmann50}
\begin{equation}
{\bf E}_{\alpha}={\bf E}_{\alpha,iso}+\mathbb{G}_\alpha \mathbb{V}_{\bar{\alpha}} {\bf E}_{\alpha} \, .
\label{eq:LS2}
\end{equation}
Writing the potential $\mathbb{V}_{\bar{\alpha}}$ in terms of the Green's function $\mathbb{G}_{\bar{\alpha}}$, we arrive at~\cite{Kruger11}
\begin{align}
{\bf E}_{\alpha}&= \mathbb{O}_{\alpha} {\bf E}_{\alpha,iso} \, ,\\
\mathbb{O}_{\alpha} &=(1+\mathbb{G}_0\mathbb{T}_{\bar{\alpha}})\frac{1}{1-\mathbb{G}_0\mathbb{T}_\alpha\mathbb{G}_0\mathbb{T}_{\bar{\alpha}}}.\label{eq:O}
\end{align}
The multiple scattering operator $\mathbb{O}_{\alpha}$ depends on the composite $T$-operator $\mathbb{T}_{\bar{\alpha}}$ describing scattering by the other objects, as well as on $\mathbb{T}_\alpha$.
Expanding the denominator of $\mathbb{O}_{\alpha}$ leads to
\begin{equation}
\mathbb{O}_{\alpha}=(1+\mathbb{G}_0\mathbb{T}_{\bar{\alpha}})\left[1+\mathbb{G}_0\mathbb{T}_\alpha\mathbb{G}_0\mathbb{T}_{\bar{\alpha}}+\dots \right],\label{eq:exp}
\end{equation}
where the terms in square brackets correspond to an increasing number of back and forth scatterings between the objects. The expansion of Eq.~\eqref{eq:exp} can be useful in order to get simplified analytical results in certain cases, as will be shown below. We note, however, that this expansion does not necessarily converge at close separations,~\footnote{For example $|r^P|$, the magnitude of the Fresnel coefficient can exceed unity for evanescent waves, such that e.g. Eq.~(23) in Ref.~\cite{Volokitin01} for the heat transfer between parallel plates cannot be expanded in multiple reflections.} and a multiple scattering expansion might not be as useful as in equilibrium situations where a fast convergence of such  series is observed~\cite{Maghrebi11}.

 Applying the multiple scattering operator to both sides of the field correlator describing the radiation emitted by object $\alpha$, we arrive at the final formula for the correlator  as
\begin{align}
\mathbb{C}_\alpha^{sc}(T_\alpha)&=\mathbb{O}_{\alpha}\,\mathbb{C}_\alpha(T_\alpha) \,\mathbb{O}_{\alpha}^\dagger
=a({T_\alpha})\mathbb{O}_{\alpha}\,\mathbb{R}_\alpha \,\mathbb{O}_{\alpha}^\dagger \, .
\label{eq:finalC}
\end{align}
In the last line, we have used the definition of the radiation operator in Eq.~\eqref{eq:radf1}. 
Equation~\eqref{eq:finalC}, together with Eqs.~\eqref{eq:1} and~\eqref{eq:6} yields
\begin{equation}
\mathbb{C}^{\rm neq}(\{T_\alpha\},T_{\rm env})=\mathbb{C}^{\rm eq} (T_{\rm env})+\sum_\alpha \left[ \mathbb{C}_\alpha^{\rm sc}(T_\alpha) -\mathbb{C}_\alpha^{\rm sc}(T_{\rm env})\right]\notag.
\end{equation}
This constitutes our final result for the field correlator in the considered non-equilibrium situation, expressed in a basis-independent representation in terms of the T-operators of the objects. 
\section{Trace formula for heat transfer}
\label{sec:heat}

\subsection{Two objects}

Let us consider two objects labelled $1$ and $2$ at temperatures $T_1$ and $T_2$, respectively. One can define different energy fluxes, which in general also depend on the temperature of the environment. For example, an experimental setup could measure the total energy absorbed by object $2$  close to  object $1$ in an environment at a yet different temperature $T_{\rm env}$. While we derive the result for all possible cases in Eq.~\eqref{eq:HT2f} below, we first focus on the usual definition in literature~\cite{Volokitin01,Narayanaswamy08,Sheng09}, where $T_{\rm env}$ is assumed to be zero (or irrelevant in the near field regime), and one considers only the energy exchanged between two objects. The component of radiation emitted by object $1$ and absorbed by object $2$, will be indicated by {\it heat transfer rate} $H_1^{(2)}(T_1)$. In turn, the emission by $2$ which is partly absorbed by $1$ is quantified by $H_2^{(1)}(T_2)$. 

In the previous section we  derived the correlation function of the electric field. For computing the transfer rate $H_1^{(2)}$, the standard method is to integrate the normal component of 
the  Poynting vector [related to $\mathbb{C}_1^{sc}$ in Eq.~\eqref{eq:finalC}]
over a surface $\Sigma_2$ enclosing only object $2$. 
As in Sec.~\ref{sec:traces} for the heat emitted by a single object, we prefer to recast the transfer in terms of a volume integral, which can then be turned into a trace. 
We have to evaluate again the expression $\left\langle \vct{E}(\vct{r})\cdot (G_0^{-1*} \vct{E}^*)(\vct{r}) \right\rangle_\omega$. But now, aiming at the absorption by object 2, the integral must be restricted to the volume of object 2, and the contribution from $\mathbb{C}_1^{sc}$ becomes
\begin{align}
\notag H_1^{(2)}=&- \frac{2\hbar}{\pi} \int\limits_0^\infty d\omega \frac{\omega}{e^{\frac{\hbar\omega}{k_BT_1}}-1}\\
&\Im \sum_i\int\limits_{V_2} d^3r (\mathbb{O}_1\mathbb{R}_1\mathbb{O}_1^\dagger\mathbb{G}_0^{-1*})_{ii}(\vct{r},\vct{r}) \, .
\label{eq:Hf}
\end{align}
We included a minus with respect to Eq.~\eqref{eq:volints}, to indicate the energy {\it absorbed} by object 2. 
In the second line, we have replaced the correlator by $\mathbb{C}_1^{sc}$, resulting in the operator $\mathbb{O}_1\mathbb{R}_1\mathbb{O}_1^\dagger\mathbb{G}_0^{-1*}$, which has to be traced over vector components $i$ [because of the dot product in Eq.~\eqref{eq:volints}] and integrated over the volume $V_2$. It can be split up as
\begin{align}
\mathbb{O}_1\mathbb{R}_1\mathbb{O}_1^\dagger \mathbb{G}_0^{-1*}=\mathbb{O}_1\mathbb{R}_1\mathbb{O}^\dagger_{1,i}+\mathbb{O}_1\mathbb{R}_1\mathbb{O}^\dagger_{1,s},\label{eq:split} 
\end{align}
with the two parts
\begin{subequations}\label{eq:op}
\begin{align}
\mathbb{O}_1\mathbb{R}_1\mathbb{O}^\dagger_{1,i} &=\mathbb{O}_1\mathbb{R}_1 \frac{1}{1- \mathbb{T}^*_2 \mathbb{G}^*_0\mathbb{T}^*_1\mathbb{G}^*_0} \mathbb{T}_2^* \, ,\\
\mathbb{O}_1\mathbb{R}_1\mathbb{O}^\dagger_{1,s} &= \mathbb{O}_1\mathbb{R}_1\frac{1}{1- \mathbb{T}^*_2 \mathbb{G}^*_0\mathbb{T}^*_1\mathbb{G}^*_0} \mathbb{G}_0^{-1*} \, ,
\end{align}
\end{subequations}
where $i$ and $s$ stand for ``interaction" and ``self," respectively. The splitting into the two operators in Eq.~\eqref{eq:op} is done because they differ precisely by the operator on the most right position, 
\begin{align}
\mathbb{O}_{1}\mathbb{R}_1\mathbb{O}_{1,i}^\dagger&=\dots \mathbb{T}_2^* \, ,\\
\mathbb{O}_{1}\mathbb{R}_1\mathbb{O}_{1,s}^\dagger&=\dots \mathbb{T}_1^* \mbox{ (or }\dots \mathbb{T}_1) \, ,
\end{align}
where the dots stand for the remaining parts of these terms. Now we note that $\mathbb{O}_{1}\mathbb{R}_1\mathbb{O}_{1,i}^\dagger(\vct{r},\vct{r})$ is only nonzero if $\vct{r}$ is located \emph{inside} object $2$, whereas $\mathbb{O}_{1}\mathbb{R}_1\mathbb{O}_{1,s}^\dagger(\vct{r},\vct{r})$ is only nonzero if $\vct{r}$ is located inside \emph{object 1}. Thus, the integral over $V_2$ in Eq.~\eqref{eq:Hf} can be extended over all space (without changing the result), if we restrict to $\mathbb{O}_{1}\mathbb{R}_1\mathbb{O}_{1,i}^\dagger(\vct{r},\vct{r})$. Then, the integral turns into a trace of this operator, and we finally have the exact result 
\begin{align}
\notag H_1^{(2)}=&-\frac{2\hbar}{\pi} \int\limits_0^\infty d\omega \frac{\omega}{e^{\frac{\hbar\omega}{k_BT_1}}-1}\Im \Tr \left[\mathbb{O}_1\mathbb{R}_1\mathbb{O}_{1,i}^\dagger\right].
\label{eq:Hf22}
\end{align}
Rewriting $\mathbb{O}_{1}\mathbb{R}_1\mathbb{O}_{1,i}^\dagger$ in terms of the $\mathbb{T}$ operators, we have
\begin{widetext}
\begin{align}
H_1^{(2)}= \frac{-2\hbar}{\pi} \int_0^\infty d\omega \frac{\omega}{e^{\frac{\hbar\omega}{k_BT_1}}-1}\Im \mbox{Tr} \left\{(1+\mathbb{G}_0\mathbb{T}_{2})\frac{1}{1-\mathbb{G}_0\mathbb{T}_1\mathbb{G}_0\mathbb{T}_{2}}\mathbb{G}_0\left[\Im[\mathbb{T}_1] - \mathbb{T}_1 \Im[\mathbb{G}_0]\mathbb{T}_1^*\right]\mathbb{G}^*_0 \frac{1}{1-\mathbb{T}^*_2 \mathbb{G}^*_0\mathbb{T}^*_1\mathbb{G}^*_0} \mathbb{T}_2^*  \right\},\\
=  \frac{2\hbar}{\pi} \int_0^\infty d\omega \frac{\omega}{e^{\frac{\hbar\omega}{k_BT_1}}-1} \mbox{Tr} \left\{ \left[\Im[\mathbb{T}_2] - \mathbb{T}_2^{ *} \Im[\mathbb{G}_0]\mathbb{T}_2\right]\frac{1}{1-\mathbb{G}_0\mathbb{T}_1\mathbb{G}_0\mathbb{T}_{2}} \mathbb{G}_0\left[\Im[\mathbb{T}_1] - \mathbb{T}_1 \Im[\mathbb{G}_0]\mathbb{T}_1^*\right]\mathbb{G}^*_0  \frac{1}{1- \mathbb{T}^{ *}_2 \mathbb{G}^{ *}_0\mathbb{T}^{ *}_1\mathbb{G}^{ *}_0}  \right\}.
\label{eq:Hff}
\end{align}
\end{widetext}
The trace makes no reference to a specific basis, and the transfer is completely determined by the scattering properties of the two objects and the free Green's functions. This form may also be useful for cases where the $\mathbb{T}$-operator is not known explicitly and has to be computed numerically~\cite{McCauley, Reid}, as such methods are most powerful if reference to a specific basis can be avoided. As described below Eq.~\eqref{eq:tracerad}, the trace in Eq.~\eqref{eq:Hff} can be transformed into a trace over partial waves, in analogy to the procedure for equilibrium Casimir forces~\cite{Rahi09}. However, the resulting traces will have restrictions with respect to propagating or evanescent waves [see, e.g. Eq.~\eqref{eq:sptrace2}]. In the following subsections, we  shall prove the symmetry as well as the positivity of Eq.~\eqref{eq:Hff}.
\\

\subsection{Symmetry of transfer}

The result in Eq.~\eqref{eq:Hff} can be used to prove the symmetry of heat transfer: It is intuitively clear that $H_1^{(2)}(T)=H_2^{(1)}(T)$ has to hold since at equal temperatures the objects should not exchange energy. While this is commonly accepted, and has been shown numerically for the case of two spheres~\cite{Narayanaswamy08}, there is to our best knowledge no fundamental principle guaranteeing its validity. Detailed balance cannot be invoked as even for $T_1=T_2$ the system is out of equilibrium if the environment is at a different temperature $T_{\rm env}$. For the case of two parallel plates, this symmetry is apparent from the formula for heat transfer~\cite{Volokitin01,Bimonte09} (in this case, it does follow from detailed balance as $T_{\rm env}$ plays no role). In order to prove the symmetry, we rewrite Eq.~\eqref{eq:Hff} as
\begin{equation}
H_1^{(2)}= \frac{2\hbar}{\pi} \int_0^\infty d\omega \frac{\omega}{e^{\frac{\hbar\omega}{k_BT_1}}-1} \mbox{Tr} \left\{\mathbb{R}_2^* \mathbb{W}_{12}\mathbb{R}_1\mathbb{W}^{*}_{21}  \right\},\label{eq:HTsym}
\end{equation}
where the radiation operator $\mathbb{R}_\alpha$ is defined in Eq.~\eqref{eq:radf1}, and
\begin{equation}
\mathbb{W}_{\alpha\beta}\equiv \mathbb{G}_0^{-1}\frac{1}{1-\mathbb{G}_0\mathbb{T}_\alpha\mathbb{G}_0\mathbb{T}_{\beta}}.
\label{eq:HTW}
\end{equation}
The required symmetry is now apparent from Eq.~\eqref{eq:HTsym}, as the trace allows a cyclic permutation of the operators, and furthermore we can take the complex conjugate of the expression since it is real. We have thus shown that  
\begin{equation}
H_1^{(2)}(T)=H_2^{(1)}(T),
\end{equation}
indeed holds for arbitrary objects. This allows to write the total heat transferred from object $1$ to object $2$,
$H^{1\rightarrow2}=H_1^{(2)}(T_1)-H_2^{(1)}(T_2)$, simply in terms of one (e.g., the first) function, as
\cite{Narayanaswamy08, Sheng09, Rousseau09, Kruger11,Otey},
\begin{align}
H^{1\rightarrow2}=H_1^{(2)}(T_1)-H_1^{(2)}(T_2) \, .
\label{eq:transtot}
\end{align}

\subsection{Positivity of transfer}

We have shown in Section~\ref{sec:pos} that $\mathbb{R}_\alpha$ is a positive semi-definite operator, and the same  holds for $\mathbb{R}_\alpha^*$. With the property $\mathbb{W}_{21}^*=\mathbb{W}_{12}^\dagger$ for the operator in Eq.~\eqref{eq:HTW}, we can write 
\begin{align}
H_1^{(2)}&= \frac{2\hbar}{\pi} \int_0^\infty d\omega \frac{\omega}{e^{\frac{\hbar\omega}{k_BT_1}}-1} \mbox{Tr} \left\{\mathbb{R}_2^* \mathbb{W}_{12}\mathbb{R}_1\mathbb{W}^{\dagger}_{12} \right\}.\label{eq:Hp}
\end{align}
It is clear that $\mathbb{W}_{12}\mathbb{R}_1\mathbb{W}^{\dagger}_{12}$ is positive semi-definite, and thus also $\mathbb{R}_2^* \mathbb{W}_{12}\mathbb{R}_1\mathbb{W}^{\dagger}_{12}$, because it is the product of two semi-definite operators. Thus the integrand in Eq.~\eqref{eq:Hp} is non-negative for any $\omega$ and the  heat transfer $H_1^{(2)}$ is non-negative,
\begin{align}
H_1^{(2)}\geq0 \, .
\label{eq:HTpos} 
\end{align}
This important proof has also to our knowledge not been presented before. 
As naturally expected, it shows that energy is always transferred from the warmer object to the colder one. 

{\subsection{``Self" emission and absorption, and the 
influence of other objects and environment}}

Another important quantity is the heat emitted by object $1$ in the proximity of object $2$, which is for example relevant to the  cooling rate of object $1$. It is given by Eq.~\eqref{eq:Hf}, with the integral taken over $V_1$, and hence determined by the trace of $\mathbb{O}_{1}\mathbb{R}_1\mathbb{O}_{1,s}^\dagger$ in Eq.~\eqref{eq:split}, as
\begin{widetext}
\begin{align}
H_1^{(1)}= \frac{-2\hbar}{\pi} \int_0^\infty d\omega \frac{\omega}{e^{\frac{\hbar\omega}{k_BT_1}}-1}\Im \mbox{Tr} \left\{(1+\mathbb{G}_0\mathbb{T}_{2})\frac{1}{1-\mathbb{G}_0\mathbb{T}_1\mathbb{G}_0\mathbb{T}_{2}}\mathbb{G}_0\left[\Im[\mathbb{T}_1] -\mathbb{T}_1 \Im[\mathbb{G}_0]\mathbb{T}_1^*\right] \frac{1}{1- \mathbb{G}^*_0\mathbb{T}^*_2\mathbb{G}^*_0\mathbb{T}^*_1} \right\}.\label{eq:Hfs}
\end{align}
\end{widetext}
Note that $-H_1^{(1)}$ [as $H_\alpha$ in Eq.~\eqref{eq:tracerad}] is positive if the object emits energy.  As it was the case for $H_1^{(2)}$ in Eq.~\eqref{eq:HTpos}, the sign of $H_1^{(1)}$ is fixed.  Interestingly, we can rewrite Eq.~\eqref{eq:Hfs} similarly to Eq.~\eqref{eq:HTsym}, where now the Green's function of object 1 appears,
\begin{align}
H_1^{(1)}= -\frac{2\hbar}{\pi} \int_0^\infty d\omega \frac{\omega}{e^{\frac{\hbar\omega}{k_BT_1}}-1}\mbox{Tr} \left\{\Im[\mathbb{G}_1]\mathbb{W}_{12}\mathbb{R}_1\mathbb{W}^{\dagger}_{12}\right\}.
\end{align}
As $\Im[\mathbb{G}_1]$ is a positive semi-definite operator, we have proven that 
\begin{align}
H_1^{(1)}\leq0
\end{align}
holds for arbitrary objects. Having derived $H_1^{(2)}$ and $H_1^{(1)}$, we can now write down the {\em total} heat absorbed by object 2 for arbitrary temperatures $T_1$, $T_2$ and $T_{\rm env}$. It is  given by
\begin{align}
H^{(2)} (T_1,T_2,T_{\rm env})=H_1^{(2)}(T_1)+H_2^{(2)}(T_{2})+H_{\rm env}^{(2)}(T_{\rm env}),\label{eq:HT2sum}
\end{align}
where $H_{\rm env}^{(2)}(T_{\rm env})$ (which we do not give explicitly) is the radiation of the environment absorbed by object $2$. Using Eq.~\eqref{eq:6}, we can express $H^{(2)}$ solely in terms of $H_1^{(2)}$ and $H_2^{(2)}$,
\begin{align}
H^{(2)} (T_1,T_2,T_{\rm env})=\sum_{\alpha=1,2} H_\alpha^{(2)}(T_\alpha)-H_{\alpha}^{(2)}(T_{\rm env}) \, .
\label{eq:HT2f}
\end{align}
We stress that, using  Eq.~\eqref{eq:HT2f}, the functions $H_\alpha^{(\beta)}$ are sufficient to describe any heat balance for two objects, including the temperature of the environment.

\subsection{Generalization to $N$ objects}

The generalization to $N>2$ objects is straightforward-- assuming that the composite $\mathbb{T}$-operator of a collection of objects is known. We recall that the correlator $\mathbb{C}_\alpha^{sc}$ in Eq.~\eqref{eq:finalC} is the radiation of object $\alpha$ scattered at all objects. Also for $N>2$, this correlator carries $\mathbb{G}_0^*$ on its rightmost  position;
after the application of its inverse $\mathbb{G}_0^{-1*}$ in Eq.~\eqref{eq:Hf}, the operator on the rightmost  position will be the $\mathbb{T}$-operator of one of the objects~\footnote{The operator $\mathbb{T}_{\bar{ \alpha}}$ can be expanded in terms of the individual $\mathbb{T}$ operators, such that  $\mathbb{T}_{\bar{ \alpha}}$ also carries one of these operators on its rightmost  position}. Again, this is precisely because the final field correlator is always written in terms of the total currents on the objects, and  expressing the correlation in terms of scattering operators allows to identify the sources with individual objects. In other words, $\vct{E}\cdot (\mathbb{G}_0^{-1*} \vct{E}^*)$ in Eq.~\eqref{eq:Hf} can always be decomposed as
\begin{align}
\mathbb{O}_\alpha \mathbb{R}_\alpha \mathbb{O}_\alpha^\dagger \mathbb{G}_0^{-1*}=\sum\limits_{\beta}\mathbb{F}_\alpha^{(\beta)},\label{eq:manysplit}
\end{align}
where $\mathbb{F}_\alpha^{(\beta)}$ is the part which contains $\mathbb{T}^*_\beta$ on the rightmost  position, and is also the only term that contributes to the integral over the volume $V_\beta$. As before, the range of integration can now be extended to {\it all space} since $\mathbb{F}_\alpha^{(\beta)}(\vct{r},\vct{r})$ is only nonzero within $V_\beta$. Thus, also for $N>2$ objects, we can write the heat absorbed by object $\beta$ due to the sources in object $\alpha$ as a trace,  
\begin{align}
H_\alpha^{(\beta)}(T_\alpha)=-\frac{2\hbar}{\pi}\int_0^\infty{d\omega} \frac{\omega}{e^{\frac{\hbar\omega}{k_BT_\alpha}}-1} \Im\Tr \, \mathbb{F}_\alpha^{(\beta)} \, . \label{eq:traceh}
\end{align}
The total heat absorbed by object $\beta$ is then given by a sum over $\alpha$, analogously to Eq.~\eqref{eq:HT2sum},
\begin{align}
H^{(\beta)}(\{T_\alpha\},T_{\rm env})=\sum_\alpha H_\alpha^{(\beta)}(T_\alpha)+H_{\rm env}^{(\beta)}(T_{\rm env}) \, ,
\end{align}
where again we do not need to specify $H_{\rm env}^{(\beta)}(T_{\rm env})$ because we use  Eq.~\eqref{eq:6} to get
\begin{equation}
H^{(\beta)}(\{T_\alpha\},T_{\rm env})=\sum_\alpha \left(H_\alpha^{(\beta)}(T_\alpha)- H_\alpha^{(\beta)}(T_{\rm env})\right)\, .
\label{eq:transfer}
\end{equation}
We emphasize again that $\mathbb{F}_\alpha^{(\beta)}$ can be expressed in terms of $\{\mathbb{T}_\alpha\}$, and Eq.~\eqref{eq:traceh} is free of references to any specific basis.

\section{Trace formula for non-equilibrium force}
\label{sec:force}

\subsection{Two objects}

Let us again start with two objects 1 and 2 at temperatures $T_1$ and $T_2$, respectively, in an environment at $T_{\rm env}$. We have shown in Ref.~\cite{Kruger11b} that the forces on the two objects are not equal and opposite in non-equilibrium, and have to be derived separately.
The force on one of the objects (say 2) is derived in close analogy to the heat transfer in Section \ref{sec:heat}. This force can be found from the surface normal component of the Maxwell stress tensor $\sigma$, integrated over the surface of object $2$~\cite{Jackson},
\begin{align}
\vct{F}^{(2)}=\Re\oint_{\Sigma_2} \mathbb{\sigma}\cdot \vct{n} 
\label{eq:surface}
\end{align}
with
\begin{align}
\sigma_{ij}(\vct{r})=\int_{-\infty}^\infty \frac{d\omega}{8\pi^2}\left\langle E_iE^*_j+B_i B^*_j-\frac{1}{2}\left(|E|^2+|B|^2\right)\delta_{ij}\right\rangle_{\!\!\omega}\!\!\!,\notag
\end{align}
where all fields are evaluated at $\vct{r}$. 
As for the radiation of a single object in Eq.~\eqref{eq:tracerad} and the heat transfer in Eq.~\eqref{eq:Hff}, we can also derive a trace formula for the force by rewriting the surface integral in Eq.~\eqref{eq:surface} as a volume integral. 
Physically, the volume integral  describes the Lorentz force acting on the fluctuating charges and currents inside the object.
This leads  by a straightforward calculation to the following expression for the $j^\text{th}$ component of the force, 
\begin{align}
{\bf \hat{j}}\cdot{\bf F}^{(2)}=\frac{c^2}{4\pi}\int\limits_{-\infty}^\infty\frac{d\omega}{2\pi} \frac{1}{\omega^2}  \int\limits_{V_2} d^3r\left\langle \left[\partial_j\vct{E}(\vct{r})\right]\cdot \left[G_0^{-1*} \vct{E}^*\right](\vct{r}) \right\rangle_\omega \, .
\label{eq:Ff}  
\end{align}
The force on object 2 has contributions due to all sources in the system~\cite{Kruger11b,Bimonte11}. We first consider $\vct{F}_1^{(2)}$ which is  due to the sources in object $1$. It is determined by the correlator $\mathbb{C}_1^{sc}$. We note that the splitting of the operator $\mathbb{O}_1 \mathbb{R}_1\mathbb{O}_1^\dagger$ in Eq.~\eqref{eq:split} is helpful here too. Here only $\mathbb{O}_{1,i}^\dagger$ contributes to the integral in Eq.~\eqref{eq:Ff} in which case the range of integration can be extended over all space. We hence find for the force on object $2$ due to the sources in object $1$
\begin{widetext}
\begin{align}
\vct{F}_1^{(2)}=\frac{2\hbar}{\pi} \int_0^\infty d\omega \frac{1}{e^{\frac{\hbar\omega}{k_BT_1}}-1}\Re \mbox{Tr} \left\{\boldsymbol{\nabla}(1+\mathbb{G}_0\mathbb{T}_{2})\frac{1}{1-\mathbb{G}_0\mathbb{T}_1\mathbb{G}_0\mathbb{T}_{2}}\mathbb{G}_0\left[\Im[\mathbb{T}_1] - \mathbb{T}_1 \Im[\mathbb{G}_0]\mathbb{T}_1^*\right]\mathbb{G}^*_0 \frac{1}{1- \mathbb{T}^*_2 \mathbb{G}^*_0\mathbb{T}^*_1\mathbb{G}^*_0} \mathbb{T}_2^*  \right\}.\label{eq:forcetra} 
\end{align}
We refer to this force in the following as \emph{interaction} force~\cite{Kruger11b} (this is the reason for the subscript $i$ in $\mathbb{O}_{1,i}^\dagger$).
The force on object 1 due to the sources in object 1, i.e., $\vct{F}_1^{(1)}$, is referred to as the \emph{self} force. $\vct{F}_2^{(2)}$ is found by exchanging indices 1 and 2 in the equation below. It is given by the self-part in Eq.~\eqref{eq:op}, and we have
\begin{align}
\vct{F}_1^{(1)}=\frac{2\hbar}{\pi} \int_0^\infty d\omega \frac{1}{e^{\frac{\hbar\omega}{k_BT_1}}-1}\Re \mbox{Tr} \left\{\boldsymbol{\nabla}(1+\mathbb{G}_0\mathbb{T}_{2})\frac{1}{1-\mathbb{G}_0\mathbb{T}_1\mathbb{G}_0\mathbb{T}_{2}}\mathbb{G}_0\left[ \Im[\mathbb{T}_1] - \mathbb{T}_1 \Im[\mathbb{G}_0]\mathbb{T}_1^*\right] \frac{1}{1- \mathbb{G}^*_0\mathbb{T}^*_2 \mathbb{G}^*_0\mathbb{T}^*_1}   \right\}.\label{eq:sforcetr} 
\end{align}
\end{widetext}
The total force on  object 2 is then given by a sum over all contributing (thermal and quantum) sources, 
\begin{equation}
\vct{F}^{(2)}(\{T_\alpha\},T_{\rm env})=\sum_{\alpha=1,2} \vct{F}_\alpha^{(2)}(T_\alpha)+\vct{F}_{\rm env}^{(2)}(T_{\rm env})+\vct{F}_0^{(2)}.\label{eq:totnf}
\end{equation}
$\vct{F}_{\rm env}^{(2)}(T_{\rm env})$ is the force on object 2 due to thermal fluctuations of the environment and $\vct{F}_0^{(2)}$ is the contribution from zero point fluctuations, i.e., the usual zero-temperature Casimir force.  Using Eq.~\eqref{eq:6} the  total force can be expressed in terms of $\{\vct{F}_\alpha^{(2)}(T_\alpha)\}$ and the equilibrium force $\vct{F}^{(2,eq)}$ as~\cite{Kruger11, Kruger11b}
\begin{align}
\vct{F}^{(2)}(\{T_\alpha\},T_{\rm env})\notag&=\vct{F}^{(2,eq)}(T_{\rm env}) \\&+ \sum_{\alpha=1,2} \left[\vct{F}_\alpha^{(2)}(T_\alpha)- \vct{F}_\alpha^{(2)}(T_{\rm env}) \right]. \label{eq:totf}
\end{align}
Equation~\eqref{eq:totf} has the advantage over Eq.~\eqref{eq:totnf} that the evaluation of the term $\vct{F}_{\rm env}^{(2)}$ is not necessary (although possible with slightly more effort). Equations~\eqref{eq:forcetra},~\eqref{eq:sforcetr} and~\eqref{eq:totf} allow the computation of the non-equilibrium force between arbitrary objects in an arbitrary basis, and constitute another of our  main results.

\subsection{Generalization to $N$ objects}

The generalization to $N>2$ objects follows by considering the composite $\mathbb{T}$ operator of the collection of objects. Using the decomposition of $\mathbb{O}_\alpha \mathbb{R}_\alpha \mathbb{O}_\alpha^\dagger \mathbb{G}_0^{-1*}$ in Eq.~\eqref{eq:manysplit}, we  note that only $\mathbb{F}_\alpha^{(\beta)}$ contributes to the force on object $\beta$, because it has  $\mathbb{T}^*_\beta$ on its rightmost  position. After extending the range of integration to all space, we find 
\begin{align}
\vct{F}_\alpha^{(\beta)}(T_\alpha)=\frac{2\hbar}{\pi}\int_0^\infty{d\omega} \frac{1}{e^{\frac{\hbar\omega}{k_BT_\alpha}}-1} \Re\Tr \left[\boldsymbol{\nabla}\mathbb{F}_\alpha^{(\beta)}\right]. \label{eq:tracef}
\end{align}
The total force on object $\beta$ is given by Eq.~\eqref{eq:totf} with the upper index 2 replaced by $\beta$ and the sum running over all objects in the system.

\section{Partial wave representation}
\label{sec:waves}

\subsection{Partial wave expansions of the free Green's function and the $\mathbb{T}$ operator}

\begin{figure*}
\begin{center}\includegraphics[width=0.6\textwidth]{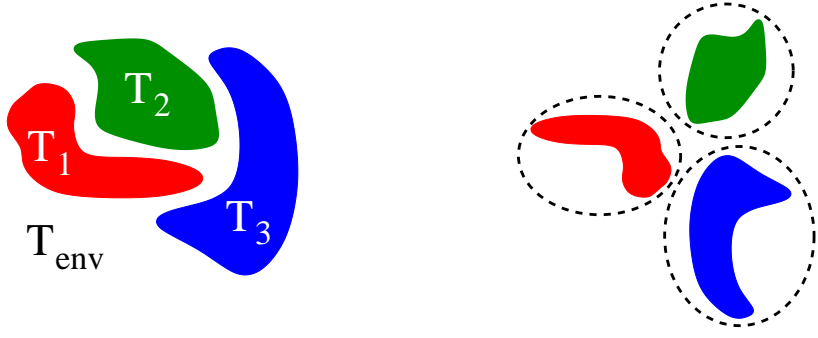}
\end{center}\caption{\label{fig:system}The results derived in the previous sections (Secs. \ref{sec:radiation} - \ref{sec:force}) are completely general, and apply e.g. to the configuration on the left hand side. The right hand side shows a configuration that allows representation in partial waves as derived in this section, because the enclosing spheres or ellipsoids do not overlap.} 
\end{figure*}

\subsubsection{Free Green's function}

In Sections \ref{sec:radiation}, \ref{sec:heat} and \ref{sec:force}, we derived trace formulae for  heat radiation, transfer and  non-equilibrium forces, in terms of operators involving $\mathbb{G}_0$ and $\mathbb{T}_\alpha$. These formulae hold for any geometry. In this section, we present the derivation of the corresponding formulae in partial wave bases. As is the case for equilibrium Casimir forces (see, e.g., Refs.~\cite{Emig08,Rahi09}), the traces of operators in the previous sections will turn into sums over matrix elements with respect to partial wave indices, providing simple closed form equations for specific geometries. The matrix expressions below will have the general restrictions discussed in Refs.~\cite{Emig08,Rahi09}. For example, two objects can only be described in a spherical basis if their enclosing spheres do not overlap, and a plane waves basis can only be used for two objects if they can be separated by a plane, see Fig.~\ref{fig:system}. 

In Ref.~\cite{Rahi09}, the relevant functions (Green's functions, translation matrices and scattering amplitudes), were given for imaginary frequencies, as needed for equilibrium computations. In the present case of non-equilibrium, we have to evaluate them for real frequencies, which leads to some differences in the definitions as outlined in Appendices \ref{app:plane} to \ref{app:conv}. In contrast to Ref.~\cite{Rahi09}, we choose to write the free Green's function without complex conjugations of waves, which yields an expansion that is manifestly analytic in the upper complex frequency plane, 
\begin{widetext}
\be \mathbb{G}_0({\bf r},{\bf r}')= i \sum_{\mu}
  \left\{ \begin{array}{cc} {\bf
E}_{\mu}^{\rm out}({\bf r}) \otimes {\bf
E}_{\sigma(\mu)}^{{\rm reg}}({\bf r}')  &\;\;\;\; {\rm if}
\;\xi_1({\bf r})>\xi_1'({\bf r}')\\{\bf E}_{\sigma(\mu)}^{{\rm
reg}}({\bf r}) \otimes  {\bf E}_{  \mu}^{{\rm
out}}({\bf r}')  &\;\;\;\; {\rm if} \;\xi_1({\bf r})<\xi_1'({\bf
r}')
\end{array}\right. \label{onshell}.
\ee
\end{widetext}
Here  $\mu$ runs over polarizations (electric and magnetic) as well as indices of vector-functions. $\xi_1$ is the `radial' coordinate~\cite{Rahi09} which gives rise to the two different pieces of the expansion in Eq.~\eqref{onshell}. $\vct{E}^{reg}$ denote  waves which  are regular at the origin, and $\vct{E}^{out}$ denote outgoing waves that are typically singular at the origin. The function $\sigma(\mu)$ is a permutation among the indices, which fulfills $\sigma(\sigma(\mu))=\mu$; along the lines of Ref.~\cite{Tsang}, where, for example, in spherical waves $\sigma(\{l,m\})=\{l,-m\}$, i.e., the permutation changes the multipole index $m$ to $-m$.
See Appendices \ref{app:plane}, \ref{app:sphere} and \ref{app:cyl} for the expansions in plane, spherical and cylindrical waves.

Another important quantity that appears in the non-equilibrium formulae is the imaginary part of $\mathbb{G}_0$. It is regular everywhere in space and hence can be expanded in regular waves
\be {\rm Im} [ \mathbb{G}_0({\bf r},{\bf
r}')]= \sum_{\mu \in  {\rm pr}}   {\bf E}^{\rm
reg}_{\mu}({\bf r}) \otimes {\bf E}^{{\rm reg}}_{\sigma(\mu)
}({\bf r}')
. \label{imGbis} 
\ee
Here, the sum runs only over propagating waves.

\subsubsection{Definition of matrix elements of the $\mathbb{T}$-operator and their relation  to  the scattering amplitude}

We define the matrix elements in accordance with the definition of the $\mathbb{T}$ operator in Eq.~\eqref{eq:LS3}. The homogeneous solution of the wave equation is the regular wave ${\bf
E}_{\mu '}^{ \rm reg}( {\bf r})$, and the scattering solution ${\bf E}^{sc}$ is then 
\begin{align}
{\bf E}^{\rm sc}_{\mu'}({\bf r})&=(1+\mathbb{G}_0\mathbb{T}){\bf E}_{\mu '}^{ \rm reg}\notag\\
&= {\bf E}_{\mu '}^{ \rm reg}( {\bf
r})+\sum_{\mu} \, {\bf E}_{\mu }^{\rm
out}({\bf r}) \,{\cal T}_{\mu \mu '},\label{eq:scatter}
\end{align}
where the second term is the scattered field.
From this equation the  matrix elements are obtained as~\footnote{${\cal T}_{\mu \mu'}$ in Eq.~\eqref{Tmat} is different from $\mathcal{F}_{\mu\mu'}$ in Ref.~\cite{Rahi09} (the form of Eq.~\eqref{Tmat} is analytic in the upper complex frequency plane). Actual elements are identical to Ref.~\cite{Rahi09} in most cases.}
\be
 {\cal T}_{\mu \mu'}= i \int d^3 {\bf r} \int d^3 {\bf r}' {\bf E}_{ \sigma(\mu) }^{{\rm reg}}({\bf r})\,\mathbb{T}({\bf r},{\bf r}')\,{\bf E}_{\mu '}^{\rm reg}( {\bf r}')\;.\label{Tmat}
 \ee
 Note that the integrand  involves no complex conjugations (in contrast to e.g. Ref.~\cite{Rahi09}). This ensures manifest analyticity of the matrix element ${\cal T}_{\mu \mu'}$ in the upper complex frequency plane. However, in most cases, the elements defined in Eq.~\eqref{Tmat} are identical to those defined previously (e.g., for spheres~\cite{Kruger11b} or for cylinders in Ref.~\cite{Golyk11}).~\footnote{More precisely, taking into account the prefactor $C_\mu$ in Ref.~\cite{Rahi09} as well as the expansion of $\mathbb{G}_0$, the definition of the elements in \eqref{Tmat} and Ref.~\cite{Rahi09} are identical if $\mathcal{T}$ is diagonal for the indices on which $C_\mu$ depends}

The symmetry of $\mathbb{T}$ implies that the matrix elements $ {\cal T}_{\mu \mu'}$ satisfy the condition
 \be
  {\cal T}_{\mu \mu'}= {\cal T}_{\sigma(\mu') \sigma(\mu)}\;.\label{simT}
 \ee
The $\mathcal{S}$ matrix [which we employ in Eq.~\eqref{eq:trel2} below] is  defined as 
\be
 {\cal T}=\frac{{\cal S}-{\cal I}}{2}\;.
\label{SF}
 \ee

\subsubsection{Properties of partial waves}

The formulae for heat radiation, transfer and forces in Secs.~\ref{sec:radiation}, \ref{sec:heat} and \ref{sec:force} involve complex conjugates or imaginary parts of the operators $\mathbb{T}_\alpha$ and $\mathbb{G}_0$. This is in contrast to the equilibrium force formula, see App.~\ref{app:eqf} and Ref.~\cite{Rahi09}, which involves no complex conjugation or imaginary parts. In order to relate, e.g., $\Im[\mathbb{T}]$ to the matrix elements ${\cal T}_{\mu\mu'}$, we have to know the behavior of partial waves under complex conjugation. We note that the expansion of Eq.~\eqref{onshell} is not unique, as are the properties below. However, for any basis, the Green's function can be written in the form of Eq.~\eqref{onshell} and the partial waves can be assumed to fulfill the properties below. Under complex conjugation we have for propagating modes 
\begin{align}
 {\bf E}_{\sigma(\mu)}^{{\rm reg}}({\bf
r})={\bf E}_{ \mu}^{{\rm reg}*}({\bf r})\;\;\;\mu \in {\rm
pr} \, .
\label{conjpr1} 
\end{align}
For evanescent waves, complex conjugation  involves a phase $e^{i \phi_{\mu}}$~\footnote{One could also choose the definitions $\tilde{\bf E}_{\sigma(\mu)}^{{\rm reg}}\equiv {\bf E}_{\sigma(\mu)}^{{\rm reg}}e^{-i \phi_{\mu}/2}$ and $\tilde{\bf E}_{\sigma(\mu)}^{{\rm out}}\equiv {\bf E}_{\sigma(\mu)}^{{\rm out}}e^{i \phi_{\mu}/2}$, in which case the waves $\tilde{\bf E}$ would obey a relation like Eq.~\eqref{conjpr1}, simplifying the analysis. However, we choose not to do this because our  waves are analytic in the upper complex frequency plane, which is not necessarily the case for the waves $\tilde{\bf E}$.},
\begin{align} 
 {\bf E}_{\sigma(\mu)}^{{\rm
reg}}({\bf r})&= e^{i \phi_{\mu}} {\bf E}_{ \mu}^{{\rm reg}*}({\bf r})\;\;\;\mu \in {\rm
ev},\label{conjev1} \\  {\bf
E}_{\sigma(\mu)}^{{\rm out}}({\bf r})&=- e^{-i \phi_{\mu}}
{\bf E}_{ \mu}^{{\rm out}*}({\bf r})\;\;\;\mu \in
{\rm ev}.
\label{conjev2} 
\end{align} 
The phase $e^{i\phi_{\mu}}=e^{i\phi_{\sigma(\mu)}}$ can be easily found for any specific basis, as in Apps.~\ref{app:plane},~\ref{app:sphere} and~\ref{app:cyl}. Equations~\eqref{conjpr1} and~\eqref{conjev1} are identical for propagating modes where $e^{i \phi_{\mu}}=1$. Using Eqs.~\eqref{onshell}, \eqref{imGbis}, \eqref{Tmat} as well as the relations \eqref{conjpr1} and \eqref{conjev1}, we can in a straightforward manner evaluate the operator expressions for radiation, transfer and forces, as demonstrated in the following subsections.

\subsection{Heat radiation}

In this subsection we omit for brevity the index $\alpha$ keeping in mind that all quantities refer to object $\alpha$.

\subsubsection{Field correlator}

The partial wave representation of the field correlator $\mathbb{C}(T)$ for a single object in a cold environment  in Eq.~\eqref{eq:radf1} is now easily found  and conveniently expressed in terms of outgoing waves, as 
\begin{align}
\notag \mathbb{G}_0 (\mathbb{T}^*-\mathbb{T}) \,\mathbb{G}_0^{*}=& i\sum_{\mu,\mu '}\left(e^{i \phi_{\mu }} {\cal T}^\dagger_{\mu \mu'}+ e^{-i \phi_{\mu '}} {\cal T}_{\mu  \mu '} \right)\times\\&\times{\bf E}_{\mu}^{\rm out} \otimes {\bf E}_{\mu '}^{{\rm out}*},
\end{align} 
and also 
\begin{align} 
\mathbb{G}_0 \mathbb{T} {\rm Im} [\mathbb{G}_0]
\mathbb{T}^{*} \mathbb{G}_0^{*} =\sum_{\mu, \mu '}
\sum_{\rho \in {\rm pr}}  { {\cal
T}}_{ \mu \rho} {  {\cal T}}^{\dagger}_{  \rho \mu'}
{\bf E}_{\mu}^{\rm out} \otimes {\bf E}_{\mu '}^{{\rm out}*}.\label{eq:qt}
\end{align}
Note that $\mu$ and $\mu'$ run over all waves because the (near-) field correlator also contains evanescent waves. However, the index $\rho$ in Eq.~\eqref{eq:qt} is restricted to propagating waves. The origin of this restriction is mathematically due to the restriction in Eq.~\eqref{imGbis}, and  physically due to the fact that the environment radiation that enters the derivation of $\mathbb{C}$ via Eq.~\eqref{eq:Crad}, contains only propagating waves. It is useful to introduce the projector on propagating waves,
\begin{equation}
\Pi^{\rm pr}_{\mu \mu'}=\delta_{\mu\mu'}\delta_{\mu\,\rm pr}.\label{eq:proj}
\end{equation}
With this definition the net correlator is
\begin{align}
\notag\mathbb{C}(T)=-a(T)&\sum_{\mu\mu '} 
\left[\frac{1}{2} \left( {\cal T} e^{-i  \Phi} +
e^{i \Phi}{\cal T}^{\dagger}   \right)+   {\cal T}  \Pi^{\rm pr} {\cal T}^{\dagger}\right]_{\mu \mu'}\\
&\times {\bf E}_{\mu}^{\rm out} \otimes {\bf E}_{\mu '}^{{\rm out}*}.\label{eq:Rad}
\end{align}
We also define the matrix version of the radiation operator $\mathbb{R}$ in Eq.~\eqref{eq:radf1},
\begin{align}
R_{\mu\mu'}\equiv-\left[\frac{1}{2} \left( {\cal T} e^{-i  \Phi} +
e^{i \Phi}{\cal T}^{\dagger}   \right)+   {\cal T}  \Pi^{\rm pr} {\cal T}^{\dagger}\right]_{\mu \mu'},\label{eq:radop}
\end{align}
which yields for the emitted field 
\be \mathbb{C}(T) ={a}(T)\, \sum_{\mu,\mu '} {\cal R}_{\mu \mu'} {\bf E}_{\mu}^{\rm out} \otimes {\bf E}_{\mu '}^{{\rm out}*}\,. \label{genem}
\ee

\subsubsection{Emitted energy}

The operator-trace for the emitted energy in Eq.~\eqref{eq:tracerad} is readily written as a  trace of an (infinite) matrix. For example, the first term in Eq.~\eqref{eq:tracerad} reads
\begin{align}
&\notag\Tr \{\Im[\mathbb{G}_0] \Im[\mathbb{T}]\}=\\&\int\int d^3rd^3r'\sum_{\mu\in pr} \notag
\vct{E}^{reg}_{\sigma(\mu)}(\vct{r}')\Im[\mathbb{T}(\vct{r}',\vct{r})]\vct{E}^{reg}_{\mu}(\vct{r})\label{eq:conv}\\
&=-\sum_{\mu\in pr} \Re\mathcal{T}_{\mu\mu}\equiv-\Tr_{pr}\Re\mathcal{T},
\end{align}
where we have also used the symmetry of $\mathbb{T}$. In the last equality, we have defined the trace over propagating waves of the matrix $\mathcal{T}_{\mu\mu'}$. The second term in Eq.~\eqref{eq:tracerad} is treated analogously, and we obtain for the emitted energy in a partial wave basis
\begin{align}
H=-\frac{2\hbar}{\pi}\int d\omega \frac{\omega}{e^{\frac{\hbar\omega}{k_BT}}-1} \Tr_{pr} \left\{\Re[\mathcal{T}] + \mathcal{T} \mathcal{T}^\dagger \right\}.\label{eq:trel}
\end{align}
In the second term, we defined $\Tr_{pr}[\mathcal{T} \mathcal{T}^\dagger]=\sum_{\mu\mu'} |\mathcal{T}_{\mu\mu'}|^2$, where both $\mu$ and $\mu'$ run over propagating modes. 

Given our definition of the matrix form of the radiation operator in Eq.~\eqref{eq:radop}, we can also write the emitted energy as
\begin{align}
H=\frac{2\hbar}{\pi}\int d\omega \frac{\omega}{e^{\frac{\hbar\omega}{k_BT}}-1} \Tr_{pr} \mathcal{R}.\label{eq:trel1}
\end{align}
There is an alternative representation of the emitted energy in terms of the scattering matrix $\cal{S}$ known in literature~\cite{Beenakker99,Maghrebi11b}, which also follows from Eq.~\eqref{SF} as
\begin{align}
H=\frac{\hbar}{2\pi}\int d\omega \frac{\omega}{e^{\frac{\hbar\omega}{k_BT}}-1} \Tr_{pr} \left[{\cal I}-\mathcal{S}\mathcal{S}^\dagger\right].\label{eq:trel2}
\end{align}
In Ref.~\cite{Maghrebi11b}, Eq.~\eqref{eq:trel2}  was derived by an entirely different route, starting directly from partial waves expansions and using identities for vector waves.

\subsection{Heat transfer}

Let us transform the heat transfer expression $H_1^{(2)}$ in Eq.~\eqref{eq:Hff} into a partial waves basis. First we note that Eq.~\eqref{eq:Hff} contains $\mathbb{G}_0$ in two different ways.
(i) $\Im[\mathbb{G}_0]$ connecting $\mathbb{T}$ operators of the same object. While the free Green's function  is singular at the origin,  its imaginary part is regular and can be expanded using Eq.~\eqref{imGbis}. This case where $\mathbb{G}_0$ is sandwiched by $\mathbb{T}$ operators of the same object does not occur in the equilibrium Casimir formula~\cite{Rahi09}.
(ii) The remaining $\mathbb{G}_0$ in Eq.~\eqref{eq:Hff} are similar to equilibrium as they connect $\mathbb{T}_1$ and $\mathbb{T}_2$. To expand the latter we use the techniques presented in detail in Ref.~\cite{Rahi09}, expanding the outgoing waves in Eq.~\eqref{onshell} in the coordinate system of the other object as
\begin{align}
{\bf E}^{\rm out}_{\mu}({\bf r}_{\beta})= \sum_{\mu'} {\cal U} _{\mu' \mu}^{\alpha \beta}({\bf X}_{\alpha \beta}) {\bf E}^{\rm reg}_{\mu'}({\bf r}_{\alpha}),
\end{align}
where ${\bf r}_{\alpha}$ and ${\bf r}_{\beta}$ denote the same position measured relative to the origin of systems $\alpha$ and $\beta$ respectively, and  ${\bf X}_{\alpha \beta}={\bf r}_{\alpha}-{\bf r}_{\beta}$ is the vector connecting the two coordinate origins~\cite{Wittmann88,Emig08,Rahi09}. We can then write the free Green's function as
\begin{align}
\mathbb{G}_0(\vct{r}_2,\vct{r}_1)=i\sum_{\mu\mu'}  \mathcal{U}^{21}_{\mu'\mu}\vct{E}^{reg}_{\mu'}(\vct{r}_2)\otimes\vct{E}^{reg}_{\sigma(\mu)}(\vct{r}_1).\label{eq:G0ex}
\end{align}
Here, $\vct{r}_1$ and $\vct{r}_2$ are measured in the coordinate system of objects $1$ and $2$ respectively, and are located on the corresponding object. Applying the expansions in Eqs.~\eqref{imGbis} and~\eqref{eq:G0ex}  to Eq.~\eqref{eq:Hff}, we arrive at the  expression for the heat transfer rate in matrix form.

\subsubsection{Spherical Basis}

We start with the spherical basis, as it is most useful for studies of compact objects. It also allows for the most concise representation as it does not contain evanescent modes. 
The expression for heat transfer is
\begin{widetext}
\begin{equation}
H_1^{(2)}= \frac{2\hbar}{\pi} \int_0^\infty d\omega \frac{\omega}{e^{\frac{\hbar\omega}{k_BT_1}}-1}\mbox{Tr}\left\{\left[\frac{\mathcal{T}^\dagger_2+\mathcal{T}_2}{2}+\mathcal{T}^\dagger_2\mathcal{T}_{2}\right]\frac{1}{\mathcal{I}-\mathcal{U}\mathcal{T}_1\mathcal{U}\mathcal{T}_{2}}\mathcal{U}\left[\frac{\mathcal{T}^\dagger_1+\mathcal{T}_1}{2} + \mathcal{T}_1 \mathcal{T}_1^\dagger\right] \frac{1}{\mathcal{I}- \mathcal{U}^\dagger\mathcal{T}^\dagger_2 \mathcal{U}^\dagger\mathcal{T}^\dagger_1}\mathcal{U}^\dagger \right\}\label{eq:sptrace},
\end{equation}
\end{widetext}
where the trace involves summing over all  indices $\{l, m,P\}$, of spherical waves;
and the adjoint of $\mathcal{U}$ is to be taken with respect to all these indices, i.e. $\mathcal{U}_{\mu'\mu}^{21\dagger}=\mathcal{U}_{\mu\mu'}^{12*}$. For example, one of  the terms (linear in the $\mathbb{T}$ matrices) reads explicitly $\Tr[ \mathcal{T}^\dagger_2\mathcal{U}\mathcal{T}_1 \mathcal{U}^\dagger]=\mathcal{T}_{2,\mu'\mu}^*\mathcal{U}^{21}_{\mu'\mu''}\mathcal{T}_{1,\mu''\mu'''}\mathcal{U}^{21*}_{\mu\mu'''}$ with sums over all indices. 
Equation~\eqref{eq:sptrace} is valid in spherical basis, but the objects described  can be of any shape. For homogeneous spheres, the matrix $\mathcal{T}_{l'm'lm}^{P'P}$ is diagonal in $l$, $m$ and $P$, and Eq.~\eqref{eq:sptrace} simplifies further (see Sec.~\ref{sec:Hsp}).

\subsubsection{Arbitrary Basis}

In general, the wave expansion in Eq.~\eqref{onshell} contains also evanescent waves, which behave differently under complex conjugation, see Eq.~\eqref{conjev2}. The consequence is that Eq.~\eqref{eq:sptrace} is modified and contains factors of $e^{i\phi}$ in some places [compare to Eq.~\eqref{eq:Rad}]. The result does however take a simple form in terms of the redefined matrices 
\begin{equation}
\tau_{\mu\mu'}\equiv e^{-i\phi_\mu}\mathcal{T}_{\mu\mu'}, \hspace{1cm} \upsilon_{\mu\mu'}\equiv \mathcal{U}_{\mu\mu'}e^{i\phi_{\mu'}}.\label{eq:Tsmall}
\end{equation}
Because $\Im[\mathbb{G}_0]$ contains only propagating waves, the projection $\Pi^{\rm pr}$ of Eq.~\eqref{eq:proj}  appears between the matrices of the same object, such that the heat transfer in an arbitrary basis is
\begin{widetext}
\begin{equation}
H_1^{(2)}= \frac{2\hbar}{\pi} \int_0^\infty d\omega \frac{\omega}{e^{\frac{\hbar\omega}{k_BT_1}}-1}\mbox{Tr}\left\{\left[\frac{\tau^\dagger_2+\tau_2}{2}+\tau^\dagger_2\Pi^{\rm pr}\tau_{2}\right]\frac{1}{\mathcal{I}-\upsilon\tau_1\upsilon\tau_{2}}\upsilon\left[\frac{\tau^\dagger_1+\tau_1}{2} + \tau_1 \Pi^{\rm pr}\tau_1^\dagger\right] \frac{1}{\mathcal{I}- \upsilon^\dagger\tau^\dagger_2 \upsilon^\dagger\tau^\dagger_1}\upsilon^\dagger \right\}\label{eq:sptrace2}.
\end{equation}
\end{widetext}
We note that the matrix $\tau$ is in general non-analytic in the upper complex frequency plane, whereas $\mathcal{T}$ is manifestly analytic. Equation~\eqref{eq:sptrace2} includes Eq.~\eqref{eq:sptrace} as a special case since in the spherical basis $\tau=\mathcal{T}$, $\upsilon=\mathcal{U}$ and $\Pi^{\rm pr}=\mathcal{I}$.

\subsection{Non-equilibrium Force}

The traces in Eqs.~\eqref{eq:forcetra} and~\eqref{eq:sforcetr} can also be transformed to sums over partial waves. The discussion before Eq.~\eqref{eq:sptrace} holds in close analogy here, too. Note that  Eqs.~\eqref{eq:forcetra} and \eqref{eq:sforcetr} are directly for the force, whereas in equilibrium one starts with a free energy~\cite{Rahi09}. In App.~\ref{app:eqf} we demonstrate that the force obtained in our non-equilibrium formalism, if applied to equilibrium, can be integrated to yield the free energy; while more generally Eqs.~\eqref{eq:forcetra} and \eqref{eq:sforcetr} cannot be thus integrated. In order to evaluate the gradients in Eqs.~\eqref{eq:forcetra} and~\eqref{eq:sforcetr}, we note that they act in two different ways on $\mathbb{G}_0$. In the first case, the gradient acts on a $\mathbb{G}_0$ that connects $\mathbb{T}_1$ and $\mathbb{T}_2$. After using Eq.~\eqref{eq:G0ex},  this action turns into the derivative of $\mathcal{U}(\vct{d})$,
\begin{align}
\notag&\boldsymbol{\nabla}_{r_2}\mathbb{G}_0(\vct{r}_2,\vct{r}_1)=\\
\notag &=-i\sum_{\mu\mu'}  \boldsymbol{\nabla}_{{ X_{21}}_{}}\mathcal{U}^{21}_{\mu'\mu}({\bf X}_{21})\vct{E}^{reg}_{\mu'}(\vct{r}_2)\otimes\vct{E}^{reg}_{\sigma(\mu)}(\vct{r}_1),\\
&=i\sum_{\mu\mu'}  \left({\bf p}\mathcal{U}^{21}\right)_{\mu'\mu}({\bf X}_{21})\vct{E}^{reg}_{\mu'}(\vct{r}_2)\otimes\vct{E}^{reg}_{\sigma(\mu)}(\vct{r}_1)
.\label{eq:G0exd}
\end{align}
In the last expression, we have introduced the infinitesimal translation operator
\begin{equation}
{\bf p}_{\mu\mu'}= - \boldsymbol{{ \nabla}}_{a}{\cal V}^{} _{\mu \mu'}({\bf a} )\vert_{{\bf a}={\bf 0}},\label{defp}
\end{equation} 
where ${\cal V}^{} _{\mu \mu'}$ describes the translation of regular waves in the same basis by
\begin{align}
{\bf E}^{\rm reg}_{\mu}({\bf r}_{1})=\sum_{\mu} {\cal V} _{\mu' \mu}^{}(\vct{a}) {\bf E}^{\rm reg}_{\mu'}({{\bf r}_1+\bf a }).
\end{align}
The two representations of Eq.~\eqref{eq:G0exd} can be equivalently used; we present formulae below in terms of the second representation which allows for a more compact notation.  In the second case, the gradient acts on the free Green's function connecting $\mathbb{T}_2$ with itself (a combination which does not appear in calculations of equilibrium forces). This  gradient can also be easily expressed in terms of {\bf p}, [where in contrast to Eq.~\eqref{eq:G0exd}, both points are measured in the same coordinate system] as
\begin{align}
{\bf \nabla}_{ r}{\rm Im}[\mathbb{G}_0]({\bf r},{\bf r}')= \sum_{\mu' \in {\rm pr}}  \sum_{\mu \in {\rm pr}} {\bf p}^{}_{\mu' \mu} {\bf E}^{\rm
reg}_{\mu'}( {\bf r}) \otimes {\bf E}^{{\rm reg}}_{\sigma(\mu)}( {\bf r}').
\end{align}
Apart from these derivatives, the result is in close analogy to the heat transfer in Eq.~\eqref{eq:sptrace2}. Here, for brevity, we give only the result for arbitrary basis, see Eq.~\eqref{eq:Tsmall}, [In particular, in the spherical basis, $\tau=\mathcal{T}$, $\upsilon=\mathcal{U}$ and $\Pi^{\rm pr}=\mathcal{I}$ hold, further simplifying the expression]  
\begin{widetext}
\begin{align}
{\bf F}_1^{(2)}=\frac{2 \hbar}{\pi} \int_{0}^{\infty} d \omega \frac{1}{e^{\frac{\hbar\omega}{k_BT_1}}-1}\, \Im\mbox{Tr} \left\{\left[\tau_2^{\dagger}  {\bf p}+\tau_2^{\dagger} \Pi^{\rm pr} {\bf p} \,\Pi^{\rm pr} \tau_2\right]\;\frac{1}{\mathcal{I}- \upsilon \tau_1 \upsilon \tau_2 }  \upsilon  \left[\frac{\tau^\dagger_1+\tau_1}{2} + \tau_1 \Pi^{\rm pr}\tau_1^\dagger\right] \upsilon^{\dagger}    \frac{1}{\mathcal{I}-  \tau_2^{\dagger}  \upsilon^{\dagger}  \tau_1^{\dagger}  \upsilon^{\dagger}  } \right\}\;.\label{eq:spftrace}
\end{align}
The self contribution of Eq.~\eqref{eq:sforcetr} is also readily written in a partial wave basis as 
\begin{align}
{\bf F}_1^{(1)}=\frac{2 \hbar}{\pi} \int_{0}^{\infty} d \omega \frac{1}{e^{\frac{\hbar\omega}{k_BT_1}}-1}\, \Im\mbox{Tr} \left\{\left[ {\bf p}\,\upsilon\, \tau_{2}\upsilon+\Pi^{\rm pr}{\bf p} \Pi^{\rm pr}  \right]\; \frac{1}{\mathcal{I}-  \tau_1 \upsilon \tau_2  \upsilon}    \left[\frac{\tau^\dagger_1+\tau_1}{2} + \tau_1 \Pi^{\rm pr}\tau_1^\dagger\right]    \frac{1}{\mathcal{I}- \upsilon^{\dagger} \tau_2^{\dagger}  \upsilon^{\dagger}  \tau_1^{\dagger}    }   \right\}\;.\label{eq:spsftrace}
\end{align}
\end{widetext}
For the meaning of the matrix multiplications in Eqs.~\eqref{eq:spftrace} and \eqref{eq:spsftrace}, see  the discussion below Eq.~\eqref{eq:sptrace}.
For, e.g.,  homogeneous spheres, the expressions simplify further as described in Sec.~\ref{sec:fsp}. 

\section{Key differences between equilibrium and non-equilibrium} \label{sec:asy}

In contrast to equilibrium force calculations, the expressions for heat transfer and non-equilibrium interactions are non-analytic in the upper complex frequency plane (e.g. due to the presence of adjoint quantities) and have to be evaluated for real frequencies. This restriction reflects the fact that non-equilibrium quantities can be strongly influenced by small changes in the resonances of the dielectric functions~\cite{Bimonte11, Kruger11b}, effects of which are marginal along the imaginary frequency axis. While non-equilibrium effects are hence richer in their phenomenology, they are also harder to evaluate numerically due to oscillatory behavior of the functions involved.

Setting aside the issue of convergence of the series in Eq.~\eqref{eq:exp} at close proximity, the evaluation of heat transfer and interactions can be simplified in the limit where the  separation $d$ is much larger than the size of the objects, where the following one reflection approximation
\begin{equation}
\mathbb{O}_\alpha\approx1+\mathbb{G}_0\mathbb{T}_{\bar \alpha},
\end{equation}
becomes asymptotically exact. This is because most of the waves involved will not scatter twice at the same object.
We will use this approximation in Secs.~\ref{sec:ht} and~\ref{sec:fo} to derive analytic expressions for the cases of a sphere in front a plate, and for two spheres.
In these  cases we observe a difference in convergence of multipole expansions between equilibrium and non-equilibrium situations. For equilibrium Casimir interactions (see e.g. Refs.~\cite{Zandi10,Emig07}), the convergence of the multipole expansion is governed by the ratio $R/d_s$, because the dominant wavelength is of the order of the surface to surface separation $d_s$. For heat transfer~\cite{Narayanaswamy08} and non-equilibrium forces~\cite{Kruger11b}, the convergence is governed by both $R/d_s$ and $R/\lambda_T$, and the maximum multipole order to be included is roughly given by the larger of the two numbers (see Refs.~\cite{Narayanaswamy08, Sasihithlu11a} for  more detailed discussions regarding heat transfer). Even if $R/d_s$ is small, one might still need many multipoles if $R/\lambda_T$ is not small. For $R\ll\lambda_T$, however, there is a well defined asymptotic large $d$ expansion, equivalent to the Casimir Polder limit plus higher order corrections, in equilibrium~\cite{Emig07}. For example, Eqs.~\eqref{eq:tsmalls} or \eqref{eq:f1} (and in general all our equations for $R\ll\lambda_T$), are  asymptotic large $d$ expansions, in the sense that the given orders in inverse $d$ have no corrections from higher order reflections.

\section{Examples for heat radiation}\label{sec:hr}

\subsection{Radiation of a plate}\label{sec:plate}

The radiation of a plate, i.e., a semi-infinite planar body occupying the space $z<0$, has been extensively studied by many authors~\cite{Rytov3, Eckhardt83}. The main emphasis is on the radiated energy, whereas the correlator of the emitted field, as discussed in Sec.~\ref{sec:radiation}, can be found, e.g., in Ref.~\cite{Bimonte09}. 
The simplicity of the plate geometry (labeled by $p$) enables easy derivation of formulae, e.g. from  Eqs.~\eqref{eq:tracerad} and~\eqref{eq:radf}. 
We will  use this geometry to demonstrate two things: 
First, we show that Eqs.~\eqref{eq:Rad} and~\eqref{eq:trel} can indeed be applied to planar geometries, which is slightly less obvious than for spheres as below. 
Second, we show that it is not strictly necessary to use a basis that satisfies the conditions of Eqs.~(\ref{conjpr1})--(\ref{conjev1}). It is sufficient to have a transformation from the desired basis to a basis that does fulfill Eqs.~(\ref{conjpr1})--(\ref{conjev1}).

We first consider the emission of a plane-parallel dielectric slab of {\it finite-thickness} in the region $-l \le z \le 0$. We compute the correlator of the electric field for any two points ${\bf r}$ and ${\bf r}'$ outside the slab using the general formula of Eq.~(\ref{eq:Rad}). For simplicity, we consider the case when both ${\bf r}$ and ${\bf r}'$ are on the side $z>0$, so that the field can be expressed in terms of right traveling waves ${{\bf E}}^{\rm out}_{R,P,{\bf k}_{\perp}}$. As described in App.~\ref{app:plane}, Eq.~(\ref{eq:Rad}) is not applicable to such waves, but  holds for the waves of definite parity, ${{\bf E}}^{\rm out}_{s,P,{\bf k}_{\perp}}$. However, we can still use Eq.~(\ref{eq:Rad}), noting that the two sets of waves are related by the unitary transformation
 \begin{equation}
 {\bf E}^{\rm out}_{s,P,{\bf k}_{\perp}}=\sum_{j=L,R}{{\bf E}}^{\rm out}_{j,P,{\bf k}_{\perp}}\, O_{js}.
 \end{equation}
This transformation follows immediately from Eq.~(\ref{pwout}),
\begin{align}
O_{js}=\frac{1}{\sqrt{2}}(\delta_{s,+}\delta_{j,R}+\delta_{s,+}\delta_{j,L}+i\delta_{s,-}\delta_{j,R}-i\delta_{s,-}\delta_{j,L}).
\end{align}
Because ${{\bf E}}^{\rm out}_{L,P,{\bf k}_{\perp}}({\bf r})=0$ for $z>0$ [see Eq.(\ref{inpla})], we have 
\begin{align}
\mathbb{C}_p (T_p )
={ a}(T_{p})\,  \sum_{ P,P'=N,M} \int \frac{d^2 { k}_{\perp}}{(2 \pi)^2}\int \frac{d^2 { k}'_{\perp}}{(2 \pi)^2}\notag\\
\times \,  {\tilde {\cal R}}_{R,P,{\bf k}_{\perp},R,P',{\bf k}'_{\perp}}\,  { {\bf E}}_{R,P,{\bf k}_{\perp}}^{\rm out} ({\bf r})\otimes { {\bf E}}_{R,P',{\bf k}'_{\perp}}^{{\rm out}*}({\bf r}')\,, \label{emslab}
\end{align}
where we have introduced the transformed radiation operator 
\begin{align} 
{\tilde {\cal R}} =O {  {\cal R}} O^{\dagger}=-\frac{1}{2}\left({\tilde {\cal T}}   {e^{-i\tilde\phi}}+
 {e^{i\tilde\phi}} {\tilde {\cal T}}^{\dagger}   \right)-   {\tilde {\cal T}}  \Pi^{\rm pr} {\tilde {\cal T}}^{\dagger}\;,
\end{align}
given in terms of the transformed T-operator, ${\tilde {\cal T}}=O {\cal T}O^{\dagger}$, and $e^{i\tilde\phi}=O e^{i \phi} O^{\dagger}$. In contrast to $e^{i \phi}$, $e^{i\tilde\phi}$ is not diagonal, and we find by  explicit computation that
\begin{align}
\notag(e^{i\tilde\phi})_{j,P,{\bf k}_{\perp},j',P',{\bf k}_{\perp}'}=\delta_{PP'}(2 \pi)^2 \delta^{(2)}({\bf k}_{\perp}-{\bf k}'_{\perp}) \times\\\times[\Theta_{\rm pr} \delta_{jj'}-i (1-\delta_{jj'}) \,\Theta_{\rm ev}]\;,
\label{Psimat}
\end{align}
 where  we have used the step functions
\begin{align}
\Theta_{\rm pr}=\Theta(\omega/c-k_\perp),\hspace{1cm} \Theta_{\rm ev}=\Theta(k_\perp-\omega/c) \, .  
\label{eq:step}
\end{align}

According to Eqs.~(\ref{eq:scatter}) and~(\ref{Tmat}), the matrix $\tilde{\cal T}$ is related to the Fresnel reflection and transmission coefficients, $r^{(R)}_P$  and $t^{(R)}_P$ for outgoing waves to the right of the slab, by  
\begin{align}
{\tilde {\cal T}}_{R,P,{\bf k}_{\perp},R,P',{\bf k}'_{\perp}}&=\delta_{PP'}(2 \pi)^2 \delta^{(2)}({\bf k}_{\perp}-{\bf k}'_{\perp})\; \frac{ t^{(R)}_P -1}{2},\\
{\tilde {\cal T}}_{R,P,{\bf k}_{\perp},L,P',{\bf k}'_{\perp}}&=  \delta_{PP'}(2 \pi)^2 \delta^{(2)}({\bf k}_{\perp}-{\bf k}'_{\perp}) \;\frac{r^{(R)}_P}{2} \, .
\end{align}
Both $t^{(R)}_P$ and $r^{(R)}_P$ (see e.g. Ref.~\cite[p.299]{Landauel}) depend on the thickness $l$ of the slab.
Substitution of the above matrix elements in Eq.~(\ref{emslab}) gives
\begin{align}
\mathbb{C}_p (T_p)
\notag={a}(T_{p})  \sum_{ P}  \int \frac{d^2 { k}_{\perp}}{(2 \pi)^2}  \left[\left(1-|r^{(R)}_P|^2-|t^{(R)}_P|^2\right) \Theta_{\rm pr}\right.\\\left.+ 2 {\rm Im} r^{(R)}_P \Theta_{\rm ev}\right]\,  {{\bf E}}_{R,P,{\bf k}_{\perp}}^{\rm out} ({\bf r})\otimes {{\bf E}}_{R,P,{\bf k}_{\perp}}^{{\rm out}*}({\bf r}'). \label{emslabfin}
\end{align}
 The first term in Eq.~\eqref{emslabfin} describes propagating waves which carry energy emitted by the slab. The second term corresponds to evanescent waves, which do not contribute to the energy emitted. In the situation of heat transfer between multiple objects, it is the evanescent waves which lead to a strong increase of transfer at close separations~\cite{Volokitin01}.
Equation~(\ref{eq:trel}) for the total emitted energy per surface area $A$ is also readily evaluated to give
\begin{align}
\notag\frac{H_p(T_p)}{A}=  \frac{\hbar}{2\pi} \int_0^{\infty} d \omega  \frac{\omega}{e^{\frac{\hbar \omega}{k_B T_p}}-1} \sum_P  \int_{ k_{\perp} \le \frac{\omega}{c}} \frac{d^2 { k}_{\perp}}{(2 \pi)^2}\times\\ \times \left[\left(  1-|r^{(R)}_P|^2-|t^{(R)}_P|^2  \right)+\left(  1-|r^{(L)}_P|^2-|t^{(L)}_P|^2  \right)\right].\label{emis2}
\end{align}
In view of Eq.~(\ref{emslabfin}), $H_{p}$ represents the sum of the emissions of the two faces of the slab, the emission of the right (left) face being given by the first (second) pair of round brackets.  
For an infinitely  thick plate, the transmission vanishes and $r^{(R)}_P$ approaches $r^P$ given in Eq.~\eqref{eq:Fresnel}. The emission to the right is then
\begin{align}
\frac{H_p}{A}=\frac{\hbar}{2\pi}\int\limits_0^\infty d\omega\frac{\omega}{e^{\frac{\hbar\omega}{k_BT_p}}-1} \int_{k_\perp<\frac{\omega}{c}}\frac{d^2k_\perp}{(2\pi)^2} \sum_{P}\left[1-|r^P|^2\right].\label{eq:radp}
\end{align}
A slab  made of a perfectly reflecting material ($|\varepsilon|\to\infty$) does not emit energy as in this case the Fresnel coefficients approach unity (and the transmission vanishes). This is a manifestation of Kirchhoff's law: since a plate with perfect reflectivity does not absorb electromagnetic waves, it can also not radiate. 
The blackbody limit is obtained by letting $r^P \to 0$ in Eq.~\eqref{eq:radp}, in which case $H_p$ approaches the Stefan--Boltzmann law~\cite{Rytov3,Eckhardt83}.

\subsection{Radiation of a sphere}

Another known result is for heat emitted by a sphere~\cite{Kattawar70}, which we re-derive here in our notation for completeness. (We also provide the field correlator, which seems not available in the literature.) 
The spherical basis (see Appendix \ref{app:sphere} for details) is naturally the most appropriate, and Eqs.~\eqref{eq:Rad} and~\eqref{eq:trel} are readily evaluated.
For simplicity, we consider a homogeneous sphere of radius $R$. The matrix  $\mathcal{T}_{\mu\mu'}$ is diagonal [see Eq.~\eqref{eq:Ts}], 
\begin{align}
\mathbb{C}_s(T)=&-a(T)\sum_{P,l,m}\notag\left[ \Re\mathcal{T}^{P}_{l}+|\mathcal{T}^{P}_{l}|^2\right]\\
&\vct{E}^{\rm out}_{Plm}(r,\theta,\phi)\otimes\vct{E}^{\rm out*}_{Plm}(r',\theta',\phi') \label{eq:sphererad}.
\end{align}
The trace over the index $\mu$ now involves sums over $P$, $l$, and $m$. The  rate of energy emission by the sphere is found from Eqs.~\eqref{eq:tracerad} or~\eqref{eq:trel}, as
\begin{equation}
H_s=-\frac{2\hbar}{\pi}\int_0^\infty d\omega  \frac{\omega}{e^{\frac{\hbar\omega}{k_BT}}-1} \sum_{P,l,m}\left[\Re\mathcal{T}^{P}_{l}+|\mathcal{T}^{P}_{l}|^2\right].\label{eq:rads}
\end{equation}
Equation~\eqref{eq:rads} can also be obtained from Eq.~\eqref{eq:sphererad} by integrating the Poynting vector   over the surface of the sphere $\Sigma_s$, where the following identity is useful,
\begin{equation}
\Im\int_{\Sigma_s}\left(\vct{E}^{\rm out}_{Plm}(R,\theta,\phi)\times\boldsymbol{\nabla}\times\vct{E}^{\rm out*}_{Plm}(R,\theta,\phi)\right)\cdot\hat{\bf r}\notag=1.
\end{equation}
In the limit of perfect conductivity (or reflectivity), the emission of the sphere vanishes because one has 
\begin{equation}
\lim_{|\varepsilon| \to\infty}\Re[\mathcal{T}^P_{l}]= -\lim_{|\varepsilon| \to\infty}|\mathcal{T}^P_{l}|^2\label{eq:limit}.
\end{equation}
Equation~\eqref{eq:limit} is a general property of scattering operators, which is connected to the fact that the $\mathcal{S}$-matrix in Eq.~\eqref{SF} becomes unitary in this limit. Objects with unitary $\mathcal{S}$ do not absorb energy and hence cannot radiate heat~\cite{Tsang,Bohren,Beenakker99,Maghrebi11b}.

\begin{figure}
\includegraphics[angle=270,width=1\linewidth]{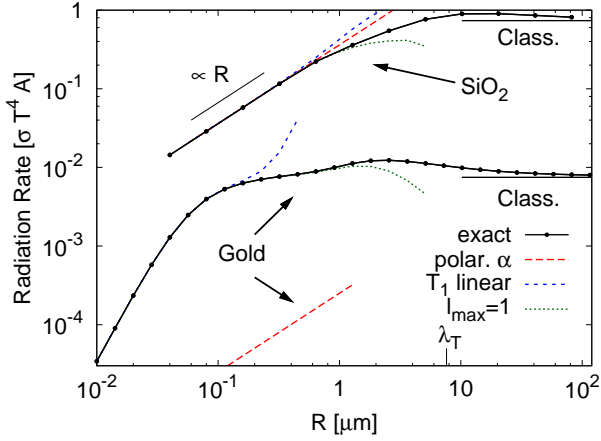}
\caption{\label{fig:1} Energy emitted by SiO$_2$ (upper curves) and Gold (lower curves)
spheres at $T=300$K in a cold environment as a function of radius $R$, normalized by the Stefan--Boltzmann result (Eq.~\eqref{eq:Hslimit} with $\epsilon(T)=1$). 
Solid lines correspond to the exact result from Eq.~\eqref{eq:rads} (for SiO$_2$, the data are  from~\cite{Kruger11}); different approximations valid for small spheres are depicted by: Long dashes correspond to Eq.~\eqref{eq:Hslimit2}, short dashes to Eq.~\eqref{eq:rads1} omitting the term $|\mathcal{T}^{P}_{1}|^2$, and dots to Eq.~\eqref{eq:rads1}.}
\end{figure}

Figure~\ref{fig:1} shows the radiation of a sphere for two different materials, SiO$_2$ and gold, as examples for dielectrics and conductors, respectively. The SiO$_2$ sphere (where we have used optical data) has been analyzed in Ref.~\cite{Kruger11}: If $R$ is much larger than the thermal wavelength and the skin depth $\delta=c/(\Im\sqrt{\varepsilon\mu}\omega)$, the emitted heat becomes proportional to the surface area of the sphere, and can be written as
\begin{equation}
\lim_{ R\gg\{\lambda_T,\delta\}}H_s=4\pi R^2\sigma T^4\epsilon(T)=4\pi R^2 \frac{H_p}{A}.\label{eq:Hslimit}
\end{equation} 
with $\sigma=\pi^2 k_B^4 / (60 \hbar^3 c^2)$. 
Here, $\epsilon(T)$ is the emissivity of the plate, which can be derived from Eq.~\eqref{eq:radp}  as shown by the last equality in Eq.~\eqref{eq:Hslimit} ($\epsilon(T)\to 1$ corresponds to a black body).

In the opposite limit, where $R$ is the smallest scale, the emission is proportional to the volume of the sphere, and the normalized curve in Fig.~\ref{fig:1} is linear in $R$. In this limit, one can use the expansion of the T-matrix for small $R^*=\frac{\omega R}{c}$, as
\begin{align}
\nonumber \mathcal{T}_{1}^{N}&=i\frac{2(\varepsilon-1)}{3(\varepsilon+2)}{R^*}^3+2i\frac{2-3\varepsilon+\varepsilon^2(1+\mu)}{5(2+\varepsilon)^2} {R^*}^5 \\&-\frac{4(\varepsilon-1)^2}{9(2+\varepsilon)^2}{R^*}^6 +\mathcal{O}\left({R^*}^7\right),\label{eq:Fi}
\end{align}
and accordingly for $T_{1}^{M}$ with $\varepsilon\leftrightarrow\mu$. The first term in Eq.~\eqref{eq:Fi} is commonly attributed to the dipole polarizability
\begin{equation}
\alpha\equiv\frac{\varepsilon-1}{\varepsilon+2}R^3.\label{eq:pola}
\end{equation}
  For $\mu=1$, the radiation is then given by
\begin{equation}
\lim_{ R\ll\{\lambda_T,\delta\}}H_s=\frac{4\hbar}{c^3\pi}\int_0^\infty d\omega  \frac{\omega^4}{e^{\frac{\hbar\omega} {k_BT}}-1} \Im\alpha\label{eq:Hslimit2}.
\end{equation}
While this is a good approximation for small SiO$_2$ spheres, it can have a very limited range of validity for other materials, as can be seen from comparison to gold in Fig.~\ref{fig:1}.
For the dielectric response of gold we have used the Drude model
\begin{align}
\varepsilon_{Au}(\omega)=1-\frac{\omega_p^2}{\omega(\omega+i\omega_\tau)},\label{eq:Drude}
\end{align}
with $\omega_p=9.03$~eV and $\omega_\tau=2.67\times 10^{-2}$~eV. For large $R$, 
 the blackbody limit is approached, but with  a much smaller emissivity compared to SiO$_2$ [$\epsilon(T)$ vanishes as $1/\sqrt{\varepsilon}$, as can be seen from expanding $H_p$ in Eq.~\eqref{eq:radp} for large $\varepsilon$]. However, the limit of Eq.~\eqref{eq:Hslimit2} is not approached for the physically accessible radii, as can be seen in Fig.~\ref{fig:1}. The relevant skin depth for gold is roughly 20 nm, and the exact curve approaches the result of Eq.~\eqref{eq:Hslimit2} for $R\approx 1$ nm. However the radiation from a sphere of $R\ll \lambda_T$ \emph{for all materials} can be approximated by restricting the sum in Eq.~\eqref{eq:rads} to  $l=1$, as
\begin{equation}
\lim_{ R\ll\lambda_T}H_s=-\frac{6\hbar}{\pi}\int_0^\infty d\omega  \frac{\omega}{e^{\frac{\hbar\omega}{k_BT}}-1} \sum_{P}\left[\Re\mathcal{T}^{P}_{1}+|\mathcal{T}^{P}_{1}|^2\right].\label{eq:rads1}
\end{equation}
As indicated by the dotted lines in Fig.~\ref{fig:1}, this result holds for both  dielectric and conducting spheres, accurately describing the emission up to roughly $R/\lambda_T\approx10\%$. In Fig.~\ref{fig:1}, we also show the result after omitting the quadratic term in Eq.~\eqref{eq:rads1}, an approximation used in Ref.~\cite{Kruger11b} in order to obtain short equations for the non-equilibrium interactions between spheres and for a sphere in front of a plate. While we expect this additional approximation to have a smaller range of validity compared to Eq.~\eqref{eq:rads1}, we find that it is as good as Eq.~\eqref{eq:Hslimit2} for SiO$_2$, and much better than Eq.~\eqref{eq:Hslimit2} for gold. Additionally, in contrast to Eq.~\eqref{eq:Hslimit2}, it can be extended to magnetic materials. See Table~\ref{table:1} for a summary of the various approximations and their limits of validity.

Lastly, we note that when normalized by the volume of the sphere, the radiation of SiO$_2$ decreases monotonically, while that of gold has a sharp maximum around $R=100$~nm, a feature which is not captured by the dipole approximation of Eq.~\eqref{eq:Hslimit2}. 

\begin{table}
\begin{ruledtabular}
\begin{tabular}{|c|c|}
\hline
$-(2l+1)\sum_{l,P}(\Re\mathcal{T}^{P}_{l}+|\mathcal{T}^{P}_{l}|^2) \approx$&Range of validity\\
\hline\hline
$-3 \sum_{P}(\Re\mathcal{T}^{P}_{1}+|\mathcal{T}^{P}_{1}|^2)$ &$R\ll\lambda_T$\\
\hline
$-3 \sum_{P}\Re\mathcal{T}^{P}_{1}$&$|\sqrt{\mu\varepsilon}|R\ll\lambda_T$\\
\hline
$2 \Im\frac{\varepsilon-1}{\varepsilon+2}\left(\frac{\omega R}{c}\right)^3$&$|\sqrt{\varepsilon}|R\ll\lambda_T$, $\mu=1$\\
\hline
\end{tabular}
\end{ruledtabular}
\caption{\label{table:1}Range of validity of different approximations for the radiation of a sphere.}
\end{table}

\subsection{Radiation of a cylinder}

An infinite cylinder is the last shape for which heat radiation can be computed analytically. 
The emitted energy  was originally described in Ref.~\cite{Rytovc}, and more recently re-derived, and numerically analyzed in detail, in Refs.~\cite{Bimonte09b,Kruger11,Golyk11}, paying special attention to the polarized nature of the radiation. The correlation of the emitted fields was also recently  studied in Ref.~\cite{Golyk12}. 

In cylindrical vector basis (see App.~\ref{app:cyl} for details), the matrix $\mathcal{T}_{\mu\mu'}$ for a cylinder is not diagonal in polarizations~\cite{Bohren, Noruzifar11,Golyk11}, but symmetric.
With Eq.~\eqref{eq:Rad}, the correlation of the emitted field is then readily evaluated as
\begin{align}
&\mathbb{C}_c(T_c)=-a(T_c)\sum_{n=-\infty}^{\infty}\int\limits_{-\infty}^{\infty}\frac{dk_z}{2\pi}\sum_{P,P'}\notag\Biggl[e^{i\phi_{k_z}} \Re\mathcal{T}^{PP'}_{n,k_z}+\\&\sum_{P''}\mathcal{T}^{PP''}_{n,k_z}\mathcal{T}^{P'P''*}_{n,k_z}\Theta\left(\frac{\omega}{c}-|k_z|\right)\Biggr]\vct{E}^{\rm out}_{Pnk_z}(\vct{r})\otimes\vct{E}^{\rm out*}_{P'nk_z}(\vct{r}') .\label{eq:radcylf}
\end{align}
The quadratic term carries the step function $\Theta$ as it only spans propagating waves with $-\frac{\omega}{c}$ to $\frac{\omega}{c}$,  whereas the term linear in $\mathcal{T}$ contains both propagating and evanescent waves. The phase factor of the linear term is the real function
\begin{align}
e^{i \phi_{k_z}}=[\Theta (\omega/c-| { k}_{z}|) \,+(-1)^{(n+1)} \Theta (|k_z|-\omega/c)].
\end{align}
The energy radiated by the cylinder~\cite{Rytovc, Kruger11,Golyk11} follows from Eq. \eqref{eq:trel} [or Eq.~\eqref{eq:tracerad}], and reads
\begin{align}
\notag\frac{|H_c|}{L}=-\frac{2\hbar}{\pi}\int d\omega \frac{\omega}{e^{\frac{\hbar\omega}{k_BT}}-1} \sum_{P}\sum_{n=-\infty}^{\infty}\\\int\limits_{-\omega/c}^{\omega/c}\frac{dk_z}{2\pi} \left(\Re[\mathcal{T}_{n,k_z}^{PP}]+|\mathcal{T}_{n,k_z}^{PP}|^2+|\mathcal{T}_{n,k_z}^{P\bar P}|^2\right) \, ,
\label{eq:radcyl}
\end{align}
where $\bar{P}=M$ if $P=N$ and vice versa.
We note that Eq.~\eqref{eq:radcyl} can also be found from Eq.~\eqref{eq:radcylf} by computing the Poynting vector, where the following relation is useful 
\begin{equation}
\Im\frac{1}{L}\int_{\Sigma_c}\left[\vct{E}^{\rm out}_{P,n,k_z}(\vct{r})\times\boldsymbol{\nabla}\times\vct{E}^{\rm out*}_{P',n,k_z}(\vct{r}')\right]\cdot\hat{\boldsymbol{\rho}}=\delta_{PP'}. 
\end{equation}
In Ref.~\cite{Golyk11}, the emission of a gold wire was found to be very large at $R\approx20$~nm, where it exceeds the Stefan--Boltzmann value by a factor of 10. 
This feature is not found for the gold sphere in Fig.~\ref{fig:1}, which we attribute to the fact that the gold wire allows for unrestricted electric fields along the wire, but the sphere does not. 

\section{Examples for Heat transfer} \label{sec:ht}

\subsection{Two spheres}\label{sec:Hsp}

\begin{figure}
\includegraphics[width=0.8\linewidth]{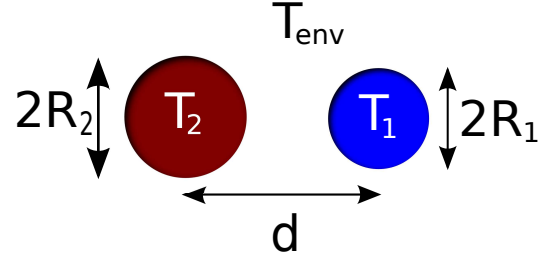}
\caption{\label{fig:twospheres}
The system of two spheres in thermal non-equilibrium.}
\end{figure}

The heat transfer between two spheres was studied numerically in great detail in Refs.~\cite{Narayanaswamy08,Sasihithlu11}, and analytically in Ref.~\cite{Volokitin01}. Our Eq.~\eqref{eq:sptrace} provides a compact  exact representation for this  transfer in a spherical basis. Here, we present a more explicit evaluation for the case of homogeneous spheres, where the matrices $\mathcal{T}_{\mu\mu'}$ are diagonal [see Eq.~\eqref{eq:Ts}]. For simplicity, we consider only the one reflection approximation. 
Assuming that the spheres are small compared to the thermal wavelengths, we then provide an asymptotic expansion valid at large separations. This extends the results in Ref.~\cite{Volokitin01} to spheres of arbitrary material. The spheres have radii $R_j$ ($j=1,2$),  complex dielectric and magnetic permeabilities $\varepsilon_j$ and $\mu_j$, at temperatures $T_j$, with their centers  separated by a distance $d$, as in Fig.~\ref{fig:twospheres}. 
The trace in Eq.~\eqref{eq:sptrace} simplifies further since $\mathcal{T}_{\mu\mu'}$ is diagonal, and the  expression for transfer from sphere 1 to sphere 2 reads (in the one reflection approximation) 
\begin{widetext}
\begin{align}
&\lim_{d\gg R}H_{}^{1\rightarrow2}(T_1,T_2)=\frac{2\hbar}{\pi}\int_0^\infty d\omega\left[\frac{\omega}{e^{\frac{\hbar\omega}{k_BT_1}}-1} -\frac{\omega}{e^{\frac{\hbar\omega}{k_BT_2}}-1} \right]\sum_{PP'll'm}
\left(\Re[\mathcal{T}^{P}_{1,l}]+|\mathcal{T}^{P}_{1,l}|^2\right) \left(\Re[\mathcal{T}^{P'}_{2,l'}]+|\mathcal{T}^{P'}_{2,l'}|^2\right) |\mathcal{U}^{21}_{l'lP'Pm} |^2.
\label{eq:heattransfer}
\end{align}
\end{widetext}
Equation~\eqref{eq:heattransfer} contains the translation matrices $\mathcal{U}$ for spherical waves (see App.~\ref{sec:U}). 
If the radii are small compared to the separation as well as the thermal wavelengths, the above expression can be evaluated using $l_{max}=1=l'_{max}$ (see also Fig.~\ref{fig:1} and Table~\ref{table:1}). This corresponds to an asymptotic large $d$ expansion of the transfer (denoting $\mathcal{T}_{j}^P=\mathcal{T}_{j,l=1}^P$),
\begin{align}
& \lim_{\{d,\lambda_{T_j}\}\gg R_j}H^{1\rightarrow2}=\frac{2\hbar}{\pi}\int_0^\infty \!\!d\omega \left[\frac{\omega  }{e^{\frac{\hbar\omega}{k_BT_1}}-1} -\frac{\omega }{e^{\frac{\hbar\omega}{k_BT_2}}-1}\right]\notag\\
& \times \sum_{P,P'}\left(\Re[{\cal T}^P_1]+|{\cal T}^P_1|^2\right)\notag \left(\Re[{\cal T}^{P'}_2]+|{\cal T}^{P'}_2|^2\right)\\
& \times \left(\frac{9c^2}{2\omega^2d^2}+\frac{9c^4}{2\omega^4d^4}+\frac{27c^6}{2\omega^6d^6}\delta_{P,P'}\right) \, .
\label{eq:tsmalls}
\end{align}
This equation has a common limit with Eq.~(62) in Ref.~\cite{Volokitin01}, if we restrict to the first term in Eq.~\eqref{eq:Fi} and linearize Eq.~\eqref{eq:tsmalls} in $\alpha$. As we have shown in Fig.~\ref{fig:1} and Table~\ref{table:1}, such a replacement is only valid if the spheres are non-magnetic and small compared to the skin-depth of the material; it is in general not valid for conductors for which the skin-depth is of the order of a few nanometers. Equation~\eqref{eq:tsmalls} requires only $\{d,\lambda_{T_j}\}\gg R_j$, and thus holds for all materials. Also the limit of perfect reflectivity ($|\varepsilon|\to\infty$) is only captured correctly when including the quadratic terms in Eq.~\eqref{eq:tsmalls}, as only then the transfer asymptotically approaches zero [compare to Eq.~\eqref{eq:limit}]~\footnote{We note that Eq.~\eqref{eq:tsmalls} has an additional factor of $8\pi$ compared to Eq.~(62) in Ref.~\cite{Volokitin01}.}.

\subsection{Sphere and plate}
 \label{sec:SPH}

\begin{figure}
\includegraphics[width=0.8\linewidth]{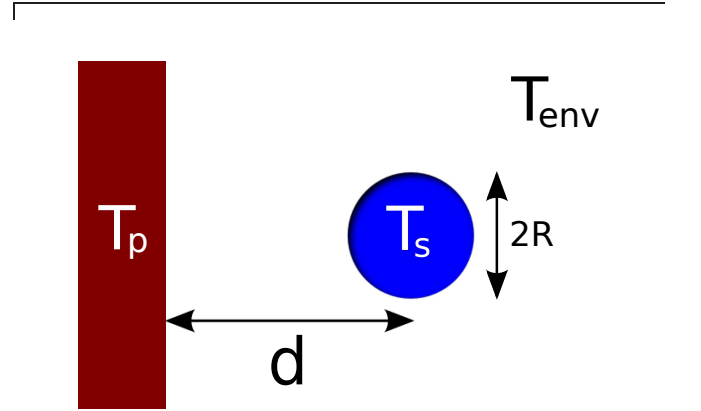}
\caption{\label{fig:sphereplate}
The system of a sphere in front a plate in thermal non-equilibrium.}
\end{figure}

The heat transfer between a sphere and a plate was considered numerically in Refs.~\cite{Kruger11,Otey,McCauley} and analytically in Ref.~\cite{Volokitin01}. Here, we provide the result  in a one-reflection approximation, including the large $d$ expansion for small $R/\lambda_T$. The exact transfer given in Eq.~\eqref{eq:sptrace2} is more complicated  because every reflection involves an integration over wave-vectors. We consider the system shown in Fig.~\ref{fig:sphereplate} with the sphere described by $R$, $\varepsilon_s$ and $\mu_s$, and the plate by $\varepsilon_p$ and $\mu_p$. The sphere-center to surface separation is denoted by $d$.

The heat transfer can be split into contributions from propagating ($H_{pr}$) and evanescent ($H_{ev}$) waves,  $H_{p}^{(s)}=H_{pr}+H_{ev}$ as in Eq.~\eqref{emslabfin}. We have
\begin{widetext}
\begin{align}
\lim_{d\gg R }&\notag H_{}^{p\rightarrow s}=\frac{2\hbar }{\pi}\int_0^\infty d\omega\left[\frac{\omega}{e^{\frac{\hbar\omega}{k_BT_p}}-1}-\frac{\omega}{e^{\frac{\hbar\omega}{k_BT_s}}-1}\right] \\&\frac{c^2}{\omega^2}\int\frac{d^2k_\perp}{(2\pi)^2}\frac{1}{2} \sum_{P,P'} \left(\frac{1}{2}(1-|r^P|^2)\Theta_{\rm pr}+\Im[r^P]e^{2ik_zd}\Theta_{\rm ev}\right) \sum_{l,m}  \left(\Re[\mathcal{T}_l^{ P'}]+|\mathcal{T}_l^{P'}|^2 \right)  |D_{lmP'P\vct{k}_\perp}|^2,
\label{eq:Hpr}
\end{align}
\end{widetext}
with $\Theta_{\rm pr}$  and $\Theta_{\rm ev}$ given in Eq.~\eqref{eq:step}. If the sphere is small compared to the thermal wavelength, we can restrict to the terms with $l=1$ (see Fig.~\ref{fig:1} and Table~\ref{table:1}), to get the asymptotic result for large $d$  as
\begin{align}
&\sum_{m=-1}^1|D_{1mP'P\vct{k}_\perp}|^2=\notag\\&\left\{\begin{array}{cc}
6\pi\frac{\omega}{ck_z}&k_\perp<\omega/c\\
6\pi \frac{\omega}{c|k_z|}\left[1+ (2 \frac{k_\perp^2c^2}{\omega^2}-2) \delta_{PP'} \right]& k_\perp>\omega/c
\end{array}.
\right.
\end{align}
The distance independent term due to propagating waves now becomes  (denoting $\mathcal{T}^{ P'}=\mathcal{T}_{l=1}^{ P'}$),
\begin{align}
\lim_{\{\lambda_{T_\alpha},d\}\gg R} H_{pr}^{p\rightarrow s}=\frac{3\hbar }{2\pi}\int_0^\infty d\omega\left[\frac{\omega}{e^{\frac{\hbar\omega}{k_BT_p}}-1} -\frac{\omega}{e^{\frac{\hbar\omega}{k_BT_s}}-1}\right]\notag\\\frac{c}{\omega}\int_0^{\frac{\omega}{c}}k_\perp dk_\perp\frac{1}{k_z}\sum_{P,P'}(1-|r^P|^2) \left(\Re[\mathcal{T}^{ P'}]+|\mathcal{T}^{P'}|^2 \right)\label{eq:Hplim}.
\end{align}
Keeping only the terms linear in $T^{P}$ in Eq.~\eqref{eq:Hplim}, and simplifying these by restricting to electric and magnetic dipole polarizabilities, Eq.~(9) of Ref.~\cite{Kruger11} is reproduced. The evanescent part can be analyzed in the two limits of $d\ll\lambda_T$ and $d\gg\lambda_T$. If $d$ is the largest scale, i.e., for $d\gg\lambda_{T_\alpha}\gg R$, the evanescent part of the transfer decays as $d^{-2}$ (setting $\mu_p=1$),  
\begin{align}
 H^{p\rightarrow s}_{ev}=\frac{3\hbar c^2}{2\pi d^2}\int_0^\infty \frac{d\omega}{\omega}\left[\frac{1}{e^{\frac{\hbar\omega}{k_BT_p}}-1}-\frac{1}{e^{\frac{\hbar\omega}{k_BT_s}}-1}\right]\notag\\ 
\times \Im\left[\frac{\varepsilon_p-1}{\sqrt{1-\varepsilon_p}}\right] \sum_{P'} \left(\Re[\mathcal{T}^{ P'}]+|\mathcal{T}^{P'}|^2 \right). \label{eq:Hev1}
\end{align}
In the opposite limit, where the separation is much smaller than $\lambda_{T}$ but still much larger than $R$, i.e., for $\lambda_{T_\alpha}\gg d \gg R$, we find a  $\sim d^{-3}$ fall-off, and
\begin{align}
H^{p\rightarrow s}_{ev}=\frac{3\hbar c^3}{2\pi d^3}\int_0^\infty \frac{d\omega}{\omega^2}\left[\frac{1}{e^{\frac{\hbar\omega}{k_BT_p}}-1}-\frac{1}{e^{\frac{\hbar\omega}{k_BT_s}}-1}\right]\notag\\ \Im\left[\frac{\varepsilon_p-1}{1+\varepsilon_p}\right]  \left(\Re[\mathcal{T}^{N}]+|\mathcal{T}^{N}|^2 \right). \label{eq:Hev2}
\end{align}
Equation~\eqref{eq:Hev2} is consistent with the first term in Eq.~(28) of Ref.~\cite{Volokitin01} (after setting ${\cal T}^N=i\frac{2\omega^3}{3c^3}\alpha$, and restricting to terms linear in the dipole polarizability $\alpha$). We note again that only the expressions in Eqs.~\eqref{eq:Hplim} to \eqref{eq:Hev2}, including quadratic terms, are valid for small spheres ($R\ll\lambda_T$) made of any material.
\\

\section{Examples for Non-equilibrium interactions}\label{sec:fo}

\subsection{Two spheres}
\label{sec:fsp}

Let us consider the Casimir interaction of the two spheres in Fig.~\ref{fig:twospheres} (again, with $R_j$,  $\varepsilon_j$ and $\mu_j$) at temperatures $T_j$ in an environment at $T_{\rm env}$. The centers of the two sphere, $\mathcal{O}_1$  and $\mathcal{O}_2$, are separated by  $\mathcal{O}_1-\mathcal{O}_2=d\hat{\vct{z}}$. We derive the total force ${\bf F}^{(2)}$ acting on sphere 2 (${\bf F}^{(1)}$ is then found by interchanging indices 1 and 2 everywhere). This force has three contributions in Eq.~\eqref{eq:totf}: The equilibrium force for the two spheres evaluated at $T_{\rm env}$, and two contributions due to the deviations of $T_1$ from $T_{\rm env}$ (${\bf F}_1^{(2)}$), and of $T_2$ from $T_{\rm env}$ (${\bf F}_2^{(2)}$). Physically, these forces follow  from the heat radiation of sphere 1 and 2, respectively  [see Eq.~\eqref{eq:rads}]. Equations~\eqref{eq:spftrace} and~\eqref{eq:spsftrace}  give the exact expressions for these forces between arbitrary objects in a basis independent form. For homogeneous spheres, these equations simplify due to the diagonality of the matrix $\mathcal{T}_{\mu\mu'}$ [see Eq.~\eqref{eq:Ts}], as well as the absence of evanescent modes. Here, we give a more explicit form of the result in a one reflection approximation, including all wave indices. The force points from center to center, and we define it to be positive when it is directed towards the other sphere (attractive). From Eq.~\eqref{eq:spftrace}, we have 
\begin{widetext}
\begin{align}
\lim_{d\gg R_j}F_1^{(2)}(T_1)=&-\frac{2\hbar}{\pi}\int_0^\infty d\omega\frac{1}{e^{\frac{\hbar\omega}{k_BT_1}}-1}\sum_{PP'll'm}
\left(\Re[\mathcal{T}^{P}_{1,l}]+|\mathcal{T}^{P}_{1,l}|^2\right)\notag \Im\left[\mathcal{T}^{P'*}_{2,l'} \left(\frac{\partial}{\partial d}\mathcal{U}^{21}_{P'Pl'lm}(d)\right) \mathcal{U}^{21*}_{P'Pl'lm}\right.\\
&+\left.\mathcal{T}^{P'}_{2,l'}\mathcal{T}^{P''*}_{2,l''} \left(\left.\frac{\partial}{\partial d} \mathcal{V}_{P''P',l''l'm}(d)\right|_{d=0}\right)   \mathcal{U}^{21}_{P'Pl'lm}  \mathcal{U}^{21*}_{P''Pl''lm}\right]\label{eq:traceforces1}\\
&\hspace{-1cm}=\frac{2\hbar}{c\pi}\int_0^\infty d\omega\frac{\omega}{e^{\frac{\hbar\omega}{k_BT_1}}-1}\sum_{PP'll'm}\left(\Re \mathcal{T}^{P}_{1,l} + | \mathcal{T}^{P}_{1,l}|^2\right)  \left\{ a(l', m)\Re\left[ \left(\mathcal{T}_{2, l'}^{P'}+2\mathcal{T}_{2,l'}^{M} \mathcal{T}_{2,l'}^{N*}\delta_{P',M}\right)\mathcal{U}^{21}_{ P'P, l' lm }\mathcal{U}^{21*}_{\bar{P'}P, l' lm }\right]\notag\right.\\
&\hspace{-1cm}\left.+ \, b(l', m)\Im\left[  \left(2\mathcal{T}_{2, l'+1}^{P'}  \mathcal{T}_{2, l'}^{ P'*} +\mathcal{T}_{2, l'}^{ P'*}+\mathcal{T}_{2, l'+1}^{ P'}\right)\mathcal{U}^{21}_{P'P, l'+1,lm}\mathcal{U}^{21*}_{ P'P,l' lm} \right]\right\},
\label{eq:f}
\end{align} 
\end{widetext}
with $\bar{P}=N$ if ${P}=M$  and vice versa. In the second expression, we have expressed the derivative of the matrix elements of $\mathcal{U}$ in terms of other elements of it, e.g., by using $\partial_{d} {\cal U}= -{p_z\cal U}$ and $ \partial_{d} {\cal V}(d\,{\hat {\bf z}})\vert_{d=0}=-p_z$, with $p_z$ given in Eq.~\eqref{pzs}. In doing so, the functions 
\begin{align}
\label{eq:a}a(l,m)&= \frac{m}{l(l+1)},\\
b(l,m)&=\frac{1}{l+1}\sqrt{\frac{l(l+2)(l-m+1)(l+m+1)}{(2l+1)(2l+3)}},
\label{eq:b}
\end{align}
appear.
The first representation,  Eq.~\eqref{eq:traceforces1}, can be evaluated numerically or analytically. We show also the second representation, because it is found when performing the calculation as described in Ref.~\cite{Kruger11b}, i.e., by evaluating the Maxwell stress tensor instead of using the trace formula of Eq.~\eqref{eq:spftrace}. The product of two spherical waves gives a net force on the sphere only if they differ in polarization or by 1 in the multipole order $l$~\cite{Crichton00}, as a spherically symmetric field generates no net force. 

For the self force, we evaluate Eq.~\eqref{eq:spsftrace} in the one reflection approximation, also writing the matrices more explicitly, such that
\begin{widetext}
\begin{align}
\lim_{d\gg R_j}F_2^{(2)}(T_2)=&\frac{2\hbar}{\pi}\int_0^\infty d\omega\frac{1}{e^{\frac{\hbar\omega}{k_BT_2}}-1}\sum_{PP'll'm}
\left(\Re[\mathcal{T}^{P}_{2,l}]+|\mathcal{T}^{P}_{2,l}|^2\right) \Im\left[-\mathcal{T}^{P'}_{1,l'} \left(\frac{\partial}{\partial d}\mathcal{U}^{21}_{PP'll'm}(d)\right) \mathcal{U}^{12}_{P'Pl'lm}\right]\label{eq:selffs}\\
=&\frac{2\hbar}{c\pi}\int_0^\infty d\omega\frac{\omega}{e^{\frac{\hbar\omega}{k_BT_2}}-1}\sum_{PP'll'm}\left(\Re \mathcal{T}^{P}_{2,l} + | \mathcal{T}^{P}_{2,l}|^2\right)  \left\{ a(l, m)\Re\left[ \mathcal{T}_{1,  l'}^{ P'}\mathcal{U}^{12}_{P'P, l' lm }\mathcal{U}^{21}_{\bar{P}P', l l'm }\right]\notag\right.\\
&\left.+b(l, m) \Im \left[\mathcal{T}_{1, l'}^{ P'} \mathcal{U}^{12}_{ P'P,l' l m}\mathcal{U}^{21}_{{P} P', l+1, l'm }\right] 
-b(l-1, m) \Im \left[\mathcal{T}_{1, l'}^{ P'} \mathcal{U}^{12}_{ P'P, l' l m}\mathcal{U}^{21}_{{P} P', l-1, l' m}\right]\right\}
\label{eq:fself}.
\end{align}
\end{widetext}
In the second expression, we have again re-expressed the $d$-derivative in order to find the form which naturally arises when considering the stress tensor, as done in Ref.~\cite{Kruger11b}.
We note that in both Eq.~\eqref{eq:selffs} and Eq.~\eqref{eq:fself}, none of the $\mathcal{U}$ matrices is conjugated, in accordance with Eq.~\eqref{eq:spsftrace}. This can lead to an oscillatory behavior of the self force as a function of $d$~\cite{Kruger11b}.

When the spheres are small compared to the thermal wavelength, $R\ll\lambda_T$,  we can restrict the partial wave sum to the dipole moment ($l=1$) in all $\mathcal{T}$ matrices. 
This results in an asymptotic large $d$ expansion which is the non-equilibrium  counterpart of the Casimir Polder limit, as (denoting again $\mathcal{T}_{j}^P=\mathcal{T}_{j,l=1}^P$) 
\begin{align}
& \lim_{\{d,\lambda_{T_1}\}\gg R_j}F_1^{(2)}=-\frac{\hbar}{c\pi}\int_0^\infty d\omega \frac{\omega }{e^{\frac{\hbar\omega}{k_BT_1}}-1}\notag\sum_{P,P'}\\&\left(\Re[{\cal T}^P_1]+|{\cal T}^P_1|^2\right)\Biggl[\frac{9c^2}{\omega^2d^2} \left(\Re[{\cal T}^{P'}_2]+\Re[{\cal T}_2^P{\cal T}_2^{\bar{P*}}]\delta_{PP'}\right)\notag\\
&+\Im[{\cal T}^{P'}_2]\left(\frac{9c^3}{\omega^3d^3}+\frac{81c^7}{\omega^7d^7}\delta_{PP'}\right)\notag\\&+\left(\Im[{\cal T}^{P'}_2]-\frac{1}{2}\Im[{\cal T}^{P}_2{\cal T}^{\bar{P}*}_2]\delta_{PP'}\right)\frac{18c^5}{\omega^5d^5}
\Biggr]\label{eq:f1}.
\end{align}
This expression is identical to Eq.~(6) in Ref.~\cite{Kruger11b}, except that we have here included the terms quadratic in $\mathcal{T}$. As shown in Fig.~\ref{fig:1} and Table~\ref{table:1}, Eq.~\eqref{eq:f1} holds for any material, while Eq.~(6) in Ref.~\cite{Kruger11b} holds only for materials with sufficiently small real and imaginary part of $\varepsilon-1$ and $\mu-1$.

For large separations, $F_1^{(2)}$ decays as $d^{-2}$ and is repulsive. This originates from momentum transfer to the second sphere via absorption or scattering of photons. The remaining terms in Eq.~\eqref{eq:f1}, with higher powers in $1/d$, are (in most cases) attractive.
Similarly, the self force $F_2^{(2)}(T_2)$ is expanded for $\{d,\lambda_{T_2}\}\gg R_j$, as
\begin{align}
&\notag F_2^{(2)}=\frac{\hbar}{c\pi}\int_0^\infty d\omega\frac{\omega}{e^{\frac{\hbar\omega}{k_BT_2}}-1}\sum_{P}\left(\Re[{\cal T}_2^P]+|{\cal T}_2^P|^2\right) \\&\Re\Biggl\{\Biggl[({\cal T}^P_1-{\cal T}^{\bar{P}}_1)\left(\frac{9c^2}{\omega^2d^2}+i\frac{27c^3}{\omega^3d^3}\right)-({\cal T}^P_1-\frac{{\cal T}^{\bar{P}}_1}{2})\frac{72c^4}{\omega^4d^4}\nonumber\\&-({\cal T}^P_1-\frac{{\cal T}^{\bar{P}}_1}{8})i\frac{144c^5}{\omega^5d^5}+{\cal T}^P_1\left(\frac{162c^6}{\omega^6d^6}+i\frac{81c^7}{\omega^7d^7}\right)\Biggr]e^{2i\frac{\omega}{c}d}\Biggr\}.\label{eq:selft}
\end{align}
(This expression is, up to quadratic terms, identical to Eq.~(7) in Ref.~\cite{Kruger11b}.) We emphasize again that, in contrast to $F_1^{(2)}$, this term can oscillate as a function of $d$ at a scale set by material resonances (see Ref.~\cite{Kruger11b}). 

We now repeat the low temperature expansion for dielectrics, which was given in Ref.~\cite{Kruger11b}. The leading low temperature behavior of the force for insulators can be derived by requiring $\lambda_T\gg\lambda_0$, where $\lambda_0$ is the wavelength of the lowest resonance of the material. The dielectric functions and polarizabilities are then expanded as~\cite{Jackson}
\begin{align}\label{eq:insu}
\varepsilon_j&=\varepsilon_{0,j}+i\frac{\lambda_{in,j}\omega}{c}+\mathcal{O}(\omega^2),\\
\alpha_j&=\alpha_{0,j} +i\alpha_{i0,j}\frac{\lambda_{in,j}\omega}{c}+\mathcal{O}(\omega^2),\label{eq:al}
\end{align}
with $\varepsilon_{0,j}$, $\lambda_{in,j}$, $\alpha_{0,j}$  and $\alpha_{i0,j}=3R^3_j/(\varepsilon_{0,j}+2)^2$ real. For  $\lambda_{T_1}\gg\lambda_0$, the interaction term is then given in closed form by
\begin{align}
&\lim_{\{d,\lambda_{T_1}\}\gg R_j}F_1^{(2)}=\frac{\hbar c}{3d^2}\frac{\lambda_{in,1}\alpha_{i0,1}}{\lambda_{T_1}^7}\Biggl[\frac{-32\pi^7\lambda_{in,2}\alpha_{i0,2}}{5\lambda_{T_1}}\notag\\
&+\alpha_{0,2}\left(\frac{32\pi^5\lambda_{T_1}}{21d}+\frac{8\pi^3\lambda_{T_1}^3}{5d^3}+\frac{18\pi\lambda_{T_1}^5}{d^5}\right)\Biggr].\label{eq:FT}
\end{align}
The self force $F_2^{(2)}$ does not oscillate to lowest order in temperature and takes a more complicated form.  In the limit where $d$ is the largest scale, we have
\begin{equation}
\lim_{d\gg\lambda_{T_2}\gg \{R_j,\lambda_0\}}F_2^{(2)}=\frac{60\hbar c}{\pi d^9}\lambda_{in,2}\alpha_{i0,2}\alpha_{0,1}\, . \label{eq:larged}
\end{equation}
While in this range of $d$ the force $F_2^{(2)}$ is independent of temperature, it vanishes as $T_2\to0$ since
with $\lambda_{T_2}$ the largest scale ($\lambda_{T_2}\gg\{d,R_j,\lambda_0\}$), one has
\begin{equation}
\lim_{d\gg R_j}F_2^{(2)}=\frac{6\pi\hbar c}{d^7\lambda_{T_2}^2}\lambda_{in,2}\alpha_{i0,2}\alpha_{0,1},\label{eq:smallT}
\end{equation}
which is identical to $F_1^{(2)}$ in this limit, with indices 1 and 2 interchanged.

\subsection{Sphere and plate}
\label{sec:SP}

The force on a small sphere in front of a plate, (see  Fig.~\ref{fig:sphereplate}), has been studied in Ref.~\cite{Kruger11b}  for various combinations of temperatures (0~K or 300~K).
The forces on the sphere and the plate are not equal and opposite in non-equilibrium.
We focus here on the force experienced by the sphere, which is directed  normal to the plate. 
The force $F^{(s)}$ is defined to be positive when the sphere, centered at $z=d$, is attracted to the plate (occupying the space $z\leq0$). 
The total force on the sphere given in Eq.~\eqref{eq:totf} contains three contributions: The equilibrium force at the temperature of the environment, and two contributions from the deviations of $T_s$ and $T_p$ from $T_{\rm env}$, denoted by $F_s^{(s)}$ and $F_p^{(s)}$. These are most conveniently expressed in terms of the formulae translating plane waves to spherical waves (see App.~\ref{app:D}). 

The interaction force is naturally split into contributions from propagating and evanescent waves, $F_p^{(s)}=F_{p,pr}^{(s)}+F_{p,ev}^{(s)}$. In a one reflection approximation the former is independent  of separation, and
\begin{widetext}
\begin{align}
\lim_{d\gg R}F_p^{(s)}(T_p)=&\frac{2\hbar}{c\pi}\int_0^\infty d\omega\frac{\omega}{e^{\frac{\hbar\omega}{k_BT_p}}-1} \frac{c^2}{\omega^2}\int\frac{d^2k_\perp}{(2\pi)^2}\frac{1}{2} \sum_{P,P'} \left(\frac{1}{2}(1-|r^P|^2)\Theta_{\rm pr}+\Im[r^P]e^{-2|k_z|d}\Theta_{\rm ev}\right) \notag\times\\
&\times\sum_{l,m} \left\{ a(l, m)\Re\left[ \left(\mathcal{T}_l^{ P'}+2\mathcal{T}_l^{P'}\mathcal{T}_l^{\bar{P'}*} \delta_{P'M}\right)  D_{lmP'P\vct{k}_\perp} D^*_{lm\bar{P'}P\vct{k}_\perp}\right]\notag+\right.\\
&\left.b( l, m)\Im\left[ \left(2 \mathcal{T}_{l+1}^{P'} \mathcal{T}_{l}^{P'*}+ \mathcal{T}_l^{ P'*}+\mathcal{T}_{l+1}^{P'}\right) D_{l+1,mP'P\vct{k}_\perp} D^*_{lmP'P\vct{k}_\perp}\right]\right\}\label{eq:Fpr}.
\end{align}
The unit-step functions $\Theta_{\rm pr}$  and $\Theta_{\rm ev}$ defined in Eq.~\eqref{eq:step} project onto propagating and evanescent modes, respectively. The self force due to the sources in the sphere comes from the radiation of the sphere at temperature $T_s$. In the one reflection approximation, we find
\begin{align}
\lim_{d\gg R}F_{s}^{(s)}(T_s)=\frac{-\hbar }{c\pi}\int_0^\infty d\omega\frac{\omega}{e^{\frac{\hbar\omega}{k_BT_s}}-1} \sum_{P,P',l,m} \left( \Re[{\cal T}^P_l]+|{\cal T}^P_l|^2\right) \frac{c^3}{\omega^3}\int\frac{d^2k_\perp}{(2\pi)^2}|k_z|\Re\left[g(k_\perp) r^{P'} e^{2ik_zd} |D_{lmPP'\vct{k}_\perp}|^2 \right],
\end{align}
\end{widetext}
where the function $g(k_\perp)$ contains projectors on propagating and evanescent modes,
\begin{align}
g(k_\perp)=(-1)^{m+l}(1-2\delta_{PP'})\Theta_{\rm pr}+\Theta_{\rm ev}.
\end{align}
$F_{s}^{(s)}$ contains no separation independent term and vanishes for $d\to\infty$. Let us perform the simplifications that arise for the case $R\ll\lambda_{T}$, where the restriction to dipole order, $l_{\rm max}=1$, yields again the asymptotic large $d$ expansion. We obtain for the Casimir Polder limit of the interaction force 
\begin{align}
&\lim_{\{d,\lambda_{T_p}\}\gg R}F_p^{(s)}=\frac{3\hbar}{2c\pi}\int_0^\infty d\omega\frac{\omega}{e^{\frac{\hbar\omega}{k_BT_p}}-1}\left(f_{pr}+f_{ev}\right)\label{eq:spf},
\end{align}
with the functions (we set $\mathcal{T}^P_{l=1}=\mathcal{T}^P$)
\begin{align}
\notag f_{pr}&=\left(\frac{c}{\omega}\right)^2\int_{0}^{\omega/c}k_\perp dk_\perp\sum_{P,P
'}(1- |r^P|^2)\left(\Re[{\cal T}^{P'}]\right.\\&\left.+\Re[{\cal T}^P{\cal T}^{\bar P*}]\delta_
{PP'}\right),\\
\notag f_{ev}&=2\left(\frac{c}{\omega}\right)^2\int_{\omega/c}^{\infty}k_\perp dk_\perp e^{-2d\sqrt{k_\perp^2-\omega^2/c^2}}\\&\sum_{P}\left\{(\Im\left[r^P\right]\left[\left(2 \frac{k^2_\perp c^2}{\omega^2}-1\right) \Im[{\cal T}^P]-\Im[{\cal T}^P{\cal T}^{\bar P*}]\right]\notag\right.\\&+\Im[r^{\bar{P}}]\Im[{\cal T}^P]\biggr\}.\label{eq:Fev}
\end{align}
The interaction force in Eq.~\eqref{eq:spf} is identical to Eq.~(16) from Ref.~\cite{Kruger11b}, except that we have here added the terms quadratic in $\mathcal{T}$. This equation is thus valid for any material as long as $\{d,\lambda_{T_p}\}\gg R$.  In the limit where the separation is the largest scale, i.e. for $d\gg \{R,\lambda_{T_p}\}$, the evanescent contribution in  Eq.~\eqref{eq:spf} decays like $d^{-3}$~\cite{Kruger11b} (compare also Ref.~\cite{Antezza05} for the situation of an atom and a  plate). The self force becomes 
\begin{align}
&\lim_{\{d,\lambda_{T_s}\}\gg R}F_{s}^{(s)}\notag=\frac{-3\hbar c}{\pi}\sum_{P}\int_0^\infty d\omega\frac{\Re[{\cal T}^P]+|{\cal T}^P|^2}{\omega(e^{\frac{\hbar\omega}{k_BT_s}}-1)}\times\\&\times\int_{0}^{\infty}k_\perp dk_\perp \Re\left\{e^{2idk_z}\left[r^P\left(2 \frac{k^2_\perp c^2}{\omega^2}-1\right)+r^{\bar{P}}\right]\right\}.\label{eq:Fss}
\end{align}
This is identical to Eq.~(19) from Ref.~\cite{Kruger11b}, again up to the  terms quadratic in $\mathcal{T}$.  $F_{s}^{(s)}$ behaves in this limit similarly as $F_{2}^{(2)}$ in Eq.~\eqref{eq:selft}, i.e., it can oscillate for dielectrics as a function of $d$, falling off at large separations as $1/d$.

We now repeat the expansion of the forces for small temperatures and dielectric spheres given in Ref.~\cite{Kruger11b}.  For a dielectric sphere and plate, we can employ Eqs.~\eqref{eq:insu} and \eqref{eq:al},
to obtain the leading behavior at low temperatures ($\lambda_{T_p}\gg\{\lambda_0,R\}$, but not necessarily ${\lambda_{T_p}\gg}  d$).
The $d$ independent part now becomes,
\begin{equation} 
\lim_{d\gg R}F_{p,pr}^{(s)}=-\frac{8\pi^5}{63} \frac{\hbar c}{\lambda_{T_p}^6}f_{pr}(\omega=0) \lambda_{in,s}\alpha_{i0} \, .
\label{eq:propf}
\end{equation}
$F_{p,ev}^s$ can be analyzed in the following two limits, corresponding to expansions of the function $f_{ev}$,
\begin{align}
\lim_{d\gg\lambda_{T_p}\gg \{R,\lambda_0\}} F_{p,ev}^{(s)}=\frac{\pi}{6}\frac{\hbar c }{\lambda_{T_p}^2d^3} \Re\left[\frac{1+\varepsilon_{0,p}}{\sqrt{\varepsilon_{0,p}-1}}\right] \alpha_{0}.\label{eq:evf}
\end{align}
In the opposite limit, with $\lambda_{T_p}\gg\{d,R_j,\lambda_0\}$, we have
\begin{align}
\lim_{d\gg R}F_{p,ev}^{(s)}=\frac{\pi}{2}\frac{\hbar c \lambda_{in,p}}{\lambda_{T_p}^2d^4} \frac{1}{(1+\varepsilon_{0,p})^2} \alpha_{0}.\label{eq:evc}
\end{align}
Equation~\eqref{eq:evf} is similar to Eq.~(12) in Ref.~\cite{Antezza05}.
As it is the case for $F_2^{(2)}$, to leading order in temperature the self part $F_s^{(s)}$ does not oscillate. For $d\gg\lambda_{T_s}\gg \{R,\lambda_0\}$, we have $F_s^{(s)}\propto 1/d^6$, while for $\lambda_{T_s}\gg\{d,R_j,\lambda_0\}$ 
\begin{equation}
\lim_{d\gg R}F_{s}^{(s)}=\frac{\pi}{4}\frac{\hbar c}{\lambda_{T_s}^2 d^4}\frac{\varepsilon_{0,p}-1}{\varepsilon_{0,p}+1} \lambda_{in,s}\alpha_{i0},
\end{equation}
which is identical to Eq.~\eqref{eq:evc} when interchanging real and imaginary parts for $r^P$ and $\alpha$.

\section{Applications: Strong 
non-equilibrium forces for nanospheres}
\label{sec:Appl}

Sections~\ref{sec:SP} and \ref{sec:SPH} give asymptotic formulae for forces and transfer for the sphere-plate geometry. Here, we present two examples, focussing on cases where the total forces allow for stable levitation when including the gravitational force acting on the sphere. In all cases, $\{d,\lambda_T\}\gg R$ is assumed, such that we can use the corresponding approximations. The equilibrium force at finite temperature is computed from Eq.~(17) in Ref.~\cite{Antezza04}.

\subsection{Metal sphere in a hot environment}

\begin{figure}
\includegraphics[angle=270,width=0.95\linewidth]{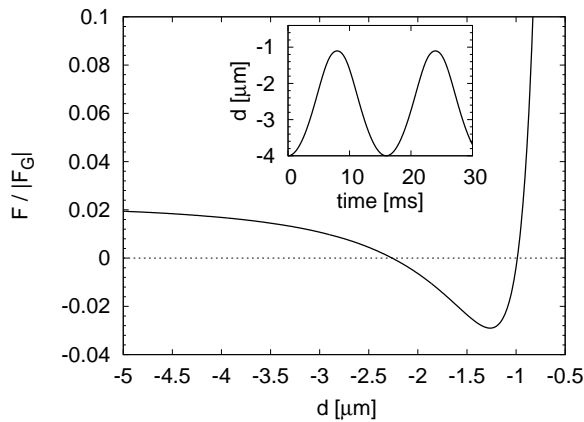}
\caption{\label{fig:AlSiC} The net force (including gravity ) on a spherical aluminum shell (outer and inner radii of 73 and 23~nm) {\it below} a SiC plate, for $T_p=300$~K and $T_{\rm env}=2862$~K ($\lambda_{T_{\rm env}}=800$~nm). ($T_s$ plays no visible role.)
The force is normalized by the gravitational force $F_G$, with positive values pointing up towards the plate (i.e. gravity gives a negative contribution). 
The point of zero force at smaller $d$ is stable;  the inset shows oscillations around this point
when the sphere is initially put at $d=-4$~$\mu$m with zero velocity.} 
\end{figure}

We start with a metal sphere, choosing aluminum for its low density, in front of a SiC plate. The  Drude model of Eq.~\eqref{eq:Drude} is used to model aluminum, with $\omega_p=12.04$~eV and $\omega_\tau=12.87\times 10^{-2}$~eV~\cite{Zeman87}. For the dielectric response of SiC we use,~\cite{Spitzer59}
\begin{equation}
\varepsilon_{SiC}=\varepsilon_\infty\frac{\omega^2-\omega_{LO}^2+i\omega\gamma}{\omega^2-\omega_{TO}^2+i\omega\gamma},
\end{equation}
where $\varepsilon_\infty=6.7$, $\omega_{LO}=0.12$~eV, $\omega_{TO}=0.098$~eV, $\gamma=5.88\times10^{-4}$~eV. Both functions have a sufficient range of validity for our purposes.   We find that the distance independent part of the interaction force in Eq.~\eqref{eq:spf} becomes comparable to the weight of the sphere (aluminum has a mass density of $2.7 \mbox{g}/\mbox{cm}^3$), at roughly $T=2700$~K (for $R=90$~nm). In Fig.~\ref{fig:AlSiC}, we show that when gravity is included a stable point of zero force, i.e., a levitation point, is possible in this regime. The main part of the figure shows the net force on the sphere {\it hanging below the plate}, at $T_p=300$~K and $T_{\rm env}=2862$~K. In this situation, $F_p^{(s)}$ enters Eq.~\eqref{eq:totf} with a minus sign, such that it becomes attractive for $d\to\infty$, and thus almost balances the gravitational force (which in this setup corresponds to a repulsive force). The near field part of $F_p^{(s)}$ [see Eqs.~\eqref{eq:Fev} or \eqref{eq:evc}] is now positive, leading to the repulsive barrier seen in the figure. At the shortest separations $d$, the attractive equilibrium part is dominant. 

The gravitational force can be fine-tuned by using a spherical shell, which has (almost) identical optical properties to a solid sphere as long as the shell thickness is large compared to the skin depth of aluminum (around 20 nm). Figure  \ref{fig:AlSiC}  shows the situation for outer and inner radii of 73 and 23 nm, respectively.  To illustrate a potential experiment, we also show the periodic motion of the sphere (the solution to Newton's equation of motion), starting from $d(t=0)=-4$~$\mu$m with zero initial velocity in the inset of Fig.~\ref{fig:AlSiC}. The sphere oscillates in the potential minimum on a timescale of tens of milliseconds. Finally, we note that the temperature of the sphere plays almost no role, as the self force in Eq.~\eqref{eq:Fss} is negligible due to the small emissivity of a metal sphere (see Fig.~\ref{fig:1}). This is of advantage, as the temperature of the sphere cannot be controlled in such a situation. We note however that due to the proximity to the plate, it is likely to attain a temperature close to the plate temperature of 300~K, which avoids potential melting of the sphere.

\subsection{Bouncing hot dielectric sphere over a room temperature plate}

\begin{figure}
\includegraphics[angle=270,width=0.95\linewidth]{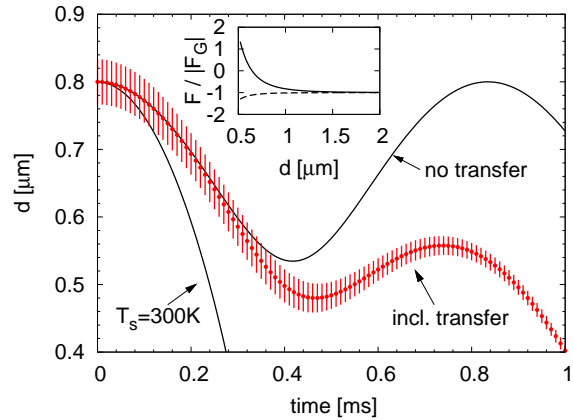}
\caption{\label{fig:Bimontium} Inset: The net force on a dielectric sphere of radius 60~nm {\it above} a dielectric plate, including the gravitational force $F_G$ on the sphere (which is negative), for $T_p=T_{\rm env}=300$~K and $T_s=916$~K. 
The strong non-equilibrium repulsion is due to near field effects, not radiation pressure, and
leads to a stable zero force point (including gravity). The dashed line shows the total force for $T_p=T_{\rm env}=T_s=300$~K. The main figure shows the trajectory of the sphere, when dropped from 800~nm. Points show the trajectory including cooling down of the sphere, where the length of the vertical bars is proportional to $T_s(t)-300~K$ in arbitrary units.}         
\end{figure}

In Ref.~\cite{Bimonte11} it was shown that dielectric materials can support strong non-equilibrium near field forces when their resonances are slightly detuned (see also Refs.~\cite{Cohen03,Henkel02}). Here we demonstrate that this can lead to a repulsive non-equilibrium force on a small sphere that can exceed gravitational forces at moderate temperatures. We use for sphere and plate the following oscillator models, with one resonance each in both  infrared and the optical range, 
\begin{align}
\epsilon_{\alpha}=1+\frac{C_{\alpha}\,\omega_{\alpha}^2}{\omega_{\alpha}^2-\omega^2-i
\gamma_{\alpha}
\omega}+\frac{D_{\alpha}\,\Omega_{\alpha}^2}{\Omega_{\alpha}^2-\omega^2-i
\Gamma_{\alpha} \omega}\;.\label{model} 
\end{align}
The parameters used are given in Table~\ref{table:2}, and resemble realistic values~\cite{Kittel}. Most importantly, we have detuned the infrared resonances, which strongly  changes the non-equilibrium forces (we found the strongest effects for a detuning by 1.19, see Table~\ref{table:2}). The inset of Fig.~\ref{fig:Bimontium} shows the total (Casimir and gravity) force for a sphere of $R=60$~nm above a plate (gravity being attractive) for temperatures $T_p=T_{\rm env}=300$~K and $T_s=916$~K. The mass density of the sphere is assumed to be $2 \mbox{g}/\mbox{cm}^3$.  The total force is equal to the gravitational force at $d\to\infty$ (because all Casimir contributions vanish asymptotically in this case), and becomes repulsive for separations below one micron due to the non-equilibrium force $F_s^{(s)}$ (which is stronger than the gravitational force). For comparison, the dashed line gives the total force for $T_p=T_{\rm env}=T_s=300$~K, which is purely attractive. We emphasize that the non-equilibrium repulsion is not due to radiation pressure but arises from  near field (evanescent) effects.

Again, mimicking a potential experimental realization, we show in the main part of Fig.~\ref{fig:Bimontium} the solution to Newton's equation of motion for a sphere dropped from a height of $d(t=0)=800$~nm. The falling sphere bounces back due the repulsive barrier, oscillating on a timescale of milliseconds. 
For comparison, we include the trajectory in thermal equilibrium, where the sphere just drops onto the plate.  
Of course the sphere will lose  energy due to heat radiation and transfer and cool off in time. 
Interestingly, the (near field) transfer to the plate [where we can use Eq.~\eqref{eq:Hpr} with $l_{max}$=1] is also a strong function of the detuning parameter, and is maximal at around the value of 1.19  (making the transfer to the environment negligible in this case). We have  solved the coupled equations for the time dependent trajectory and temperature.
Suppressing dependencies on time independent parameters $T_p$ and $T_{\rm env}$, the two equations are
\begin{table}
\begin{ruledtabular}
\begin{tabular}{|c||c|c|c|c|c|c|}
\hline
&$C_\alpha$&$\omega_\alpha$&$\gamma_\alpha$&$D_\alpha$&$\Omega_\alpha$&$\Gamma_\alpha$\\
\hline\hline
plate&3&$10^{13}$&$10^{11}$&$1$&$10^{16}$&$5\times10^{14}$\\
\hline
sphere&1.5&$1.19\times10^{13}$&$10^{11}$&$0.5$&$10^{16}$&$5\times10^{14}$\\
\hline
\end{tabular}
\end{ruledtabular}
\caption{\label{table:2} Parameters for the oscillator model of the dielectric functions of the sphere and plate. $\omega_\alpha$, $\Omega_\alpha$, $\gamma_\alpha$ and $\Gamma_\alpha$ are given in rad/sec, the remaining parameters are dimensionless.}
\end{table}
\begin{align}
\dot{T}_s&=-\frac{H_{}^{s\rightarrow p}(T_s(t);d(t))}{\kappa_s},\\
m_s\ddot{d}(t)&=F(T_s(t);d(t)) ,
\end{align}
where $m_s$ is the mass of the sphere. The heat capacity of the sphere is estimated from  $\kappa_s/m_s=800$~J/(kg K), which is a realistic value for solid materials. 
The solution in Fig.~\ref{fig:Bimontium} indicates that the sphere falls slightly further due to cooling down, but still bounces back  before eventually cooling too much to counterbalance gravity. The vertical bars denote the difference $T_s-300K$ in arbitrary units. The distance dependence of the cooling rate (which also depends on $T_s$) is hardly visible. We note that the time to eventual drop of the sphere is roughly four times longer for the hot sphere ($T_s(t=0)=916$~K) compared to $T_s(t=0)=300$~K,  making such an experiment sensitive to  non-equilibrium effects.

While the examples presented in this section may prove demanding to actual experimentation,  we believe that they provide valuable new insights into the physics of non-equilibrium fluctuations. Also, compared to the examples in Ref.~\cite{Kruger11b}, we have increased the ratio of non-equilibrium repulsion to gravity by almost six orders of magnitude, and it is well possible that it can be increased even further by choosing appropriate materials or compounds.

\section{Acknowledgments}

We thank R.~L.~Jaffe, N.~Graham, M.~T.~H.~Reid, M.~F.~Maghrebi, V.~A.~Golyk, A.~P.~McCauley, P.~L.~Sambegoro, G.~Chen and A.~Narayanaswamy for valuable discussions. 
One of us (MKr) especially acknowledges discussions with M.~F.~Maghrebi.
This research was supported by the DFG grant No.~KR 3844/1-1, NSF
Grant No.~DMR-08-03315, DOE grant No. DE-FG02-02ER45977, and the ESF Research Network CASIMIR.

\begin{appendix}

\section{Equilibrium force formula from the field correlator}
\label{app:eqf}

The trace formulae for the non-equilibrium force in Eqs.~\eqref{eq:forcetra} and~\eqref{eq:sforcetr} were derived starting from Eq.~\eqref{eq:Ff}. This starting point can be further  supported by showing that it leads to the known result for the equilibrium Casimir force, as demonstrated in this Appendix for two objects (1 and 2). In equilibrium, the field correlator is given by Eq.~\eqref{eq:1}, and all we need is the Green's function of the system. It is found by starting from object 1 in isolation with $\mathbb{G}_1=(1+\mathbb{G}_0\mathbb{T}_1)\mathbb{G}_0$ (see Eq.~\eqref{eq:GT}), and  inserting object 2 by use of the operator $\mathbb{O}_1$ in Eq.~\eqref{eq:O}, as
\begin{equation}
\mathbb{G}=(1+\mathbb{G}_0\mathbb{T}_2)\frac{1}{1-\mathbb{G}_0\mathbb{T}_1\mathbb{G}_0\mathbb{T}_2}(1+\mathbb{G}_0\mathbb{T}_1)\mathbb{G}_0.\label{eq:Gw2}
\end{equation}
In applying Eq.~\eqref{eq:Ff}, we note that we may as well consider the complex conjugate of the integrand, in which case we have $\mathbb{G}_0^{-1}$, a better fit to Eq.~\eqref{eq:Gw2}. Furthermore, we note that by symmetry, only terms with an even number of $\mathbb{T}$ matrices contribute to the force. Using the arguments given below Eq.~\eqref{eq:Ff}, we find the forces on the two objects to be (note that the field correlator in Eq.~\eqref{eq:1} is real), 
\begin{align}
{\bf F}^{(1,eq)}=&\frac{2\hbar}{\pi} \int_0^\infty d\omega \left[\frac{1}{e^{\frac{\hbar\omega}{k_BT}}-1}+\frac{1}{2}\right]\notag\\&\Im \mbox{Tr} \left\{\boldsymbol{\nabla}\mathbb{G}_0\mathbb{T}_{2}\frac{1}{1-\mathbb{G}_0\mathbb{T}_1\mathbb{G}_0\mathbb{T}_{2}}\mathbb{G}_0\mathbb{T}_1\right\},\label{eq:eq1}\\
{\bf F}^{(2,eq)}=&\frac{2\hbar}{\pi} \int_0^\infty d\omega \left[\frac{1}{e^{\frac{\hbar\omega}{k_BT}}-1}+\frac{1}{2}\right]\notag\\&\Im \mbox{Tr} \left\{\boldsymbol{\nabla}\mathbb{G}_0\mathbb{T}_1\mathbb{G}_0\mathbb{T}_2  \frac{1}{1-\mathbb{G}_0\mathbb{T}_1\mathbb{G}_0\mathbb{T}_{2}}\right\}.\label{eq:eq2}
\end{align}
In Eq.~\eqref{eq:eq2}, we have re-summed the expansion of the inverse operator. The above two  equations satisfy ${\bf F}^{(1,eq)}=-{\bf F}^{(2,eq)}$ as expected, which can be seen by expanding the inverse operators. We can hence write
\begin{equation}
{\bf F}^{(1,eq)}=\frac{1}{2}\left({\bf F}^{(1,eq)}-{\bf F}^{(2,eq)} \right).
\end{equation}
Noting that $\boldsymbol{\nabla}_r \mathbb{G}_0(\vct{r},\vct{r}')=-\boldsymbol{\nabla}_{r'} \mathbb{G}_0(\vct{r},\vct{r}')$, one can  see that subtracting Eq.~\eqref{eq:eq2} from Eq.~\eqref{eq:eq1} yields a result which can be written as a derivative with respect to the position of object 1, as
\begin{align}
{\bf F}^{(1,eq)}=&\frac{\hbar}{\pi} \int_0^\infty d\omega \left[\frac{1}{e^{\frac{\hbar\omega}{k_BT}}-1}+\frac{1}{2}\right]\notag\\
&\Im \mbox{Tr} \left\{\boldsymbol{\nabla}_{\mathcal{O}_1} \log\left[1-\mathbb{G}_0\mathbb{T}_1\mathbb{G}_0\mathbb{T}_{2}\right]\right\}.
\end{align}
This derivative can be taken out of the trace, leading to
\begin{align}
{\bf F}^{(1,eq)}=-\boldsymbol{\nabla}_{\mathcal{O}_1}\mathcal{F}=& \boldsymbol{\nabla}_{\mathcal{O}_1}\frac{\hbar}{\pi} \int_0^\infty d\omega \left[\frac{1}{e^{\frac{\hbar\omega}{k_BT}}-1}+\frac{1}{2}\right]\notag\\
&\Im \mbox{Tr} \left\{\log\left[1-\mathbb{G}_0\mathbb{T}_1\mathbb{G}_0\mathbb{T}_{2}\right]\right\}.\label{eq:eqf}
\end{align}
Here we have introduced the Casimir free energy $\mathcal{F}$~\cite{Rahi09}, and the result in Eq.~\eqref{eq:eqf} is consistent with previous work. Using Eqs.~\eqref{eq:G0ex} and~\eqref{Tmat}, we find the even more familiar result in terms of partial waves,
\begin{align}
\mathcal{F}=&-\frac{\hbar}{\pi} \int_0^\infty d\omega \left[\frac{1}{e^{\frac{\hbar\omega}{k_BT}}-1}+\frac{1}{2}\right]\notag\times\\
&\times\Im \mbox{Tr} \log[\mathcal{I}-{\cal U}^{21}  {\cal T}_1 {\cal U}^{12 }  {\cal T}_2 ].
\end{align}

\section{Plane waves basis}\label{app:plane}
\subsection{Free Green's function}
The plane waves basis is most conveniently chosen with the symmetry axis pointing along the $z$-direction, for plates lying in the $xy$ plane; ${\bf x}_\perp$ and ${\bf k}_\perp$ are the spatial coordinate and the wave-vector perpendicular to this axis, respectively. The plane wave basis is fundamentally different from coordinates with a singular point (such as the spherical basis) and we give two representations of the Green's function with different properties.
We consider the following  vector eigenfunctions,
\begin{align}
&\vct{M}^\pm_{\vct{k}_\perp}(\vct{x}_\perp,z)=\frac{i}{2\sqrt{k_z}|\vct{k}_\perp|}\left(\hat{\bf x} k_y-\hat{\bf y}k_x\right) e^{i\vct{k}\cdot\vct{r}} ,\label{eq:PW1}\\
&\vct{N}^\pm_{\vct{k}_\perp}(\vct{x}_\perp,z)=\frac{\frac{c}{\omega}}{2\sqrt{k_z}|\vct{k}_\perp|}\left(\pm\hat{\bf x}k_xk_z\pm\hat{\bf y}k_yk_z+\hat{\bf z}k_\perp^2\right)e^{i\vct{k}\cdot\vct{r}},\label{eq:PW2}
\end{align}
where $\vct{k}=(\vct{k}_\perp,k_z)^T$, with $k_z=\sqrt{\frac{\omega^2}{c^2}-{k}_\perp^2}$. We consider two sets of eigenfunctions, for reasons discussed below, and start with elementary right and left moving waves,
\begin{align}
 { {\bf E}}_{{R,P,{\bf k}_{\perp}}}^{\rm reg } ({\bf r})&={\bf P}^{-}_{{\bf k}_{\perp}}({\bf x}_{\perp},z),\\
   {{\bf E}}_{{L,P,{\bf k}_{\perp}}}^{\rm reg } ({\bf r})&={\bf P}^{+}_{{\bf k}_{\perp}}({\bf x}_{\perp},-z),\\
  {{\bf E}}_{{R,P,{\bf k}_{\perp}}}^{\rm out}( {\bf r})&= \left\{
\begin{array}{cc}2  { {\bf E}}_{{R,P,{\bf k}_{\perp}}}^{\rm reg } ({\bf r} )&z\geq0 \\
0&z<0
\end{array}\right.\label{inpla},\\
{ {\bf E}}_{{L,P,{\bf k}_{\perp}}}^{\rm out}( {\bf r})&= \left\{
\begin{array}{cc}0&z\geq0\\
2  { {\bf E}}_{{L,P,{\bf k}_{\perp}}}^{\rm reg } ({\bf r} )&z<0 
\end{array}\right..
 \end{align}
We also define waves of definite {parity} under reflections at the $z=0$ plane, that carry an index $s=\pm$,
\begin{align}
{\bf E}_{{s,P,{\bf k}_{\perp}}}^{\rm reg }( {\bf r})&=\frac{ i^{ \frac{1-s}{2}}}{ \sqrt{2 }}\,\left[{ {\bf E}}_{{R,P,{\bf k}_{\perp}}}^{\rm reg }({\bf r} ) +s \, { {\bf E}}_{{L,P,{\bf k}_{\perp}}}^{\rm reg }({\bf r} ) \right],\\
{\bf E}_{{s,P,{\bf k}_{\perp}}}^{\rm out}({\bf r})&=\frac{i^{ \frac{1-s}{2}}}{ \sqrt{2}}\left[{ {\bf E}}^{\rm out}_{{R,P,{\bf k}_{\perp}}}({\bf r} ) +s\, { {\bf E}}^{\rm out}_{{L,P,{\bf k}_{\perp}}}({\bf r} ) \right].\label{pwout}
\end{align}

The free Green's function can now be written as
\begin{widetext}
\begin{align}
\mathbb{G}_0({\bf r},{\bf r}')&= i \sum_{P=N,M} \sum_{j=L,R}\int \frac{d^2 { k}_{\perp}}{(2 \pi)^2} 
\left\{
\begin{array}{ll}
{ {\bf E}}_{{j,P,{\bf k}_{\perp}}}^{\rm out } ({\bf r}) \otimes
{ {\bf E}}_{{{\bar j},P,-{\bf k}_{\perp}}}^{\rm reg } ({\bf r}')& |z|>|z'|,\\
{ {\bf E}}_{{{\bar j},P,-{\bf k}_{\perp}}}^{\rm reg } ({\bf r}) \otimes {{\bf E}}_{{{j},P,{\bf k}_{\perp}}}^{\rm out } ({\bf r}')& |z|<|z'|,\\
\end{array}\right.\\
\mathbb{G}_0({\bf r},{\bf r}')&=i \sum_{P=N,M}\sum_{s=\pm} \int \frac{d^2 { k}_{\perp}}{(2 \pi)^2}
\left\{
\begin{array}{ll}
{{\bf E}}^{\rm out}_{{s,P,{\bf k}_{\perp}}} ({\bf r}) \otimes
   {{\bf E}}^{\rm reg}_{{s,P,-{\bf k}_{\perp}}} ({\bf r}')& |z|>|z'|,\\
 {{\bf E}}^{\rm reg}_{{s,P,-{\bf k}_{\perp}}} ({\bf r}) \otimes
   {{\bf E}}^{\rm out}_{{s,P,{\bf k}_{\perp}}} ({\bf r}') & |z|<|z'|.
\end{array}\right.
\end{align}
\end{widetext}
where ${\bar L}=R$ and ${\bar R}=L$ in the first line.
We note that both representations are of the form of Eq.~\eqref{onshell}, and obey Eq.~\eqref{imGbis}. The first variant is convenient for study the problems involving only one side of a (thick) slab, as done in this manuscript. These waves  have $\mu =(j,P,{\bf k}_{\perp})$ and $\sigma(\mu)=({\bar j},P,-{\bf k}_{\perp})$, but they do not obey the required properties under complex conjugation in Eqs.~(\ref{conjpr1})--(\ref{conjev2}). On the other hand, the waves with definite parity fulfill all properties required, i.e., Eqs.~\eqref{onshell},~\eqref{imGbis},~(\ref{conjpr1})  --(\ref{conjev2}), for our final formulae for transfer and forces to be directly applicable. They have $\mu=(s,P,   {\bf k}_{\perp})$, and $\sigma(\mu) =(s,P,-{\bf k}_{\perp})$, and their phases are 
\begin{equation}
e^ {i \phi_\mu }= [\Theta (\omega/c-| {\bf k}_{\perp}|)   -i\,(-1)^{(1-s)/2}  \Theta (| {\bf k}_{\perp}|-\omega/c) ].
\end{equation}  

\subsection{Fresnel coefficients} 
The Fresnel reflection coefficients $r^P$  for reflection are given in Ref.~\cite{Jackson},
\begin{align}
r^M\left(k_\perp,\omega\right)=\frac{\mu\sqrt{\frac{\omega^2}{c^2}-k_\perp^2}-\sqrt{\varepsilon\mu\frac{\omega^2}{c^2}-k_\perp^2}}{\mu\sqrt{\frac{\omega^2}{c^2}-k_\perp^2}+\sqrt{\varepsilon\mu\frac{\omega^2}{c^2}-k_\perp^2}}\label{eq:Fresnel}.
\end{align}
$r^N$ is obtained from $r^M$ by interchanging $\mu$ and $\varepsilon$. In Sec.~\ref{sec:plate}, we give the relation between the Fresnel coefficients and the $\mathcal{T}$ matrix elements of a plate.

\section{Spherical Basis}\label{app:sphere}
\subsection{Partial waves and free Green's function}
Here we adopt a wave expansion similar to Ref.~\cite{Tsang}, where the waves, depending on spherical coordinates $r$, $\theta$, and $\phi$, are   
\begin{align}
\vct{E}^{\rm reg}_{Mlm}&=\sqrt{\frac{(-1)^m\omega}{c}}\frac{1}{\sqrt{l(l+1)}}j_l\left(\frac{\omega}{c}r\right)\nabla\times\vct{r}Y_l^m(\theta,\phi),\\
\vct{E}^{\rm out}_{Mlm}&=\sqrt{\frac{(-1)^m\omega}{c}}\frac{1}{\sqrt{l(l+1)}}h_l\left(\frac{\omega}{c}r\right)\nabla\times\vct{r}Y_l^m(\theta,\phi),\\
\vct{E}^{\rm reg}_{Nlm} &=\frac{c}{\omega}\nabla\times\vct{E}^{\rm reg}_{Mlm},\\
\vct{E}^{\rm out}_{Nlm} &=\frac{c}{\omega}\nabla\times\vct{E}^{\rm out}_{Mlm}.
\end{align}
$j_l$ is the spherical Bessel function of order $l$, and $h_l$ is the spherical Hankel function of the first kind of order $l$. $Y_l^m(\theta,\phi)$ are the spherical harmonics, where we use the standard definition according to Ref.~\cite{Jackson} (which is different from the one used in Ref.~\cite{Tsang}).
The free Green's function $\mathbb{G}_0$ is then given by Eq.~\eqref{onshell} with $\mu=\{P,l,m\}$ and $\sigma(\mu)=\{P,l,-m\}$, and $\sum_\mu\rightarrow\sum_P\sum_{l=1}^{\infty}\sum_{m=-l}^{l}$. These definitions fulfill Eqs.~\eqref{imGbis} and~(\ref{conjpr1})--(\ref{conjev2}). The phase function is unity throughout, $e^{i\phi_\mu}=1$. 

\subsection{$\mathcal{T}$-matrix of a sphere}
The matrix elements of ${\cal T}$ for a sphere of radius $R$, as defined in Eq.~\eqref{Tmat}, are well known~\cite{Tsang} and sometimes  referred to as Mie coefficients.
Considering for simplicity spheres with isotropic and local $\varepsilon$ and $\mu$, renders the matrix $\mathcal{T}^{P'P}_{l'lm'm}$ diagonal and independent of $m$, $\mathcal{T}^{P'P}_{l'lm'm}=\mathcal{T}^P_{l}\delta_{PP'}\delta_{ll'}\delta_{mm'}$.  The matrix element  $\mathcal{T}^P_{l}$ can be conveniently written in terms of $R^*=R\omega/c$ and $\tilde R^*=\sqrt{\varepsilon\mu}R\omega/c$, as
\begin{align}
\mathcal{T}_{l}^M=-\frac{\mu j_l(\tilde R^*)\frac{d}{d R^*}\left[R^*j_l(R^*)\right]-j_l(R^*)\frac{d}{d \tilde R^*}\left[\tilde R^*j_l(\tilde R^*)\right]}{\mu j_l(\tilde R^*)\frac{d}{d R^*}\left[R^*h_l(R^*)\right]-h_l(R^*)\frac{d}{d \tilde R^*}\left[\tilde R^*j_l(\tilde R^*)\right]}.\label{eq:Ts}
\end{align}
$\mathcal{T}_{l}^N$ follows from $\mathcal{T}_{l}^M$ by interchanging $\mu$ and $\varepsilon$.

\section{Cylindrical Basis}\label{app:cyl}
\subsection{Free Green's function}
The cylindrical coordinates are denoted by $\rho$ (radial), $z$ (along the cylinder axis), and azimuthal angle $\phi$.  The free Green's function is now expanded in terms of cylindrical waves, which, using a notation similar to the one in Ref.~\cite{Tsang}, are defined by
\begin{align}
\vct{E}^{\rm reg}_{M,n,k_z}(\varrho,z,\phi)&=\sqrt{(-1)^n}\frac{1}{2k_\varrho}\nabla\times\hat{\vct{z}}\Psi^{\rm reg}_{n,k_z}(\varrho,z,\phi),\\
\vct{E}^{\rm out}_{M,n,k_z}(\varrho,z,\phi)&= \sqrt{(-1)^n}\frac{1}{2k_\varrho}\nabla\times\hat{\vct{z}}\Psi^{\rm out}_{n,k_z}(\varrho,z,\phi),\\
\vct{E}^{\rm reg}_{N,n,k_z} &=\frac{c}{\omega}\nabla\times\vct{E}^{\rm reg}_{M,n,k_z},\\
\vct{E}^{\rm out}_{N,n,k_z} &=\frac{c}{\omega}\nabla\times\vct{E}^{\rm out}_{M,n,k_z},
\end{align}
with $k_\varrho=\sqrt{\frac{\omega^2}{c^2}-k_z^2}$, and
\begin{align}
\Psi^{\rm reg}_{n,k_z}(\varrho,z,\phi)=&J_n(k_\varrho\varrho)e^{ik_zz+in\phi},\\
\Psi^{\rm out}_{n,k_z}(\varrho,z,\phi)=&H^{(1)}_n(k_\varrho\varrho)e^{ik_zz+in\phi}.
\end{align}
$J_n$ is the Bessel function of order $n$, and $H^{(1)}_n$ is the Hankel function of the first kind of order $n$. The free Green's function $\mathbb{G}_0$ is then given by Eq.~\eqref{onshell} with $\mu=\{P,n,k_z\}$ and $\sigma(\mu)=\{P,-n,-k_z\}$, and $\sum_\mu\rightarrow\sum_P\sum_{n=-\infty}^{\infty}\int\frac{dk_z}{2\pi}$. These definitions fulfill Eqs.~\eqref{imGbis} and~(\ref{conjpr1})--(\ref{conjev2}) with the phase
\begin{align}
e^{i \phi_{\mu}}=[\Theta (\omega/c-| { k}_{z}|) \,+(-1)^{(n+1)} \Theta (|k_z|-\omega/c)],
\end{align}
where the $\Theta$-function is used to stress that the phase is unity for  propagating waves.

\subsection{$\mathcal{T}$-Matrix of a cylinder}
The $\mathcal{T}$-matrix for an infinite homogeneous cylinder is diagonal in $n$ and $k_z$, but not in polarization $P$~\cite{Bohren}. 
The coefficients $\mathcal{T}^{P'P}_{n,k_z}$ take a lengthy form and are not reproduced here. They can be found in Refs.~\cite{Bohren,Noruzifar11,Golyk11}; in Ref.~\cite{Golyk11} the matrix elements are given in precisely the notation used here. Ref.~\cite{Golyk11} also gives $\mathcal{T}^{P'P}_{n,k_z}$ for  uniaxial materials. 

\section{Conversion matrices}\label{app:conv}
\subsection{Plane waves to spherical waves}\label{app:D}
The outgoing plane wave eigenfunctions in Eq.~\eqref{inpla} are expanded in spherical waves in the following way, 
\begin{align}
\frac{\omega}{c}\vct{E}^{\rm out}_{R,P,\vct{k}_\perp}(\vct{x}_\perp,z)&=\sum_{P',l,m} D_{lmP'P\vct{k}_\perp}  \vct{E}^{\rm reg}_{P'lm},
\end{align} 
with the matrix elements expressed in terms of Legendre Polynomials $P_l^m$, as
\begin{subequations}\label{eq:Ds}
\begin{align}
D_{lmMM\vct{k}_\perp}&= \frac{-i^{l+1}}{\sqrt{(-1)^m}} \sqrt{\frac{4\pi(2l+1)(l-m)!}{l(l+1)(l+m)!}}\sqrt{\frac{c}{\omega}}\frac{|\vct{k}_\perp|}{\sqrt{k_z}} \notag\times\\&\times P_l^{'m}\left(\frac{c}{\omega}\sqrt{\frac{\omega^2}{c^2}-k_\perp^2}\right) e^{-im\Phi_{\vct{k}_\perp}},\\
D_{lmNM\vct{k}_\perp}&= \frac{mi^{l+1}}{\sqrt{(-1)^m}} \sqrt{\frac{4\pi(2l+1)(l-m)!}{l(l+1)(l+m)!}}\frac{\omega}{c|\vct{k}_\perp|} \notag\times \\ &\times\sqrt{\frac{\omega}{c{k}_z}}  P_l^{m}\left(\frac{c}{\omega}\sqrt{\frac{\omega^2}{c^2}-k_\perp^2}\right) e^{-im\Phi_{\vct{k}_\perp}},\\
D_{lmNN\vct{k}_\perp}&=D_{lmMM\vct{k}_\perp},\\
D_{lmMN\vct{k}_\perp}&=D_{lmNM\vct{k}_\perp}.
\end{align}
\end{subequations}
$\Phi_{\vct{k}_\perp}$ is the the angle of $\bf k_\perp$ with respect to the $x$-axis.

\subsection{Spherical waves to spherical waves}\label{sec:U}
Outgoing spherical waves can be expanded in regular spherical waves with respect to a different origin, shifted by $\vct{d}$. In the cases considered in this paper, the translation can always be chosen along the $z$-axis, such that
\begin{align}
\vct{E}^{\rm out}_{Plm}(\vct{r})=\sum_{P'l'}\mathcal{U}^{\pm}_{P'P,l'lm}(d)\vct{E}^{\rm reg}_{P'l'm}(\vct{r}\pm d\hat{\bf z}).
\end{align}
For example, if the coordinate system of object 2 is centered at $\mathcal{O}_2=\mathcal{O}_1-d\hat{\bf z}$ as considered in Eqs.~\eqref{eq:traceforces1},~\eqref{eq:f}, \eqref{eq:selffs} and~\eqref{eq:fself}, one has $\mathcal{U}^{21}=\mathcal{U}^{+}$ in Eq.~\eqref{eq:G0ex}. The elements of $\mathcal{U}$ are~\cite{Wittmann88},
\begin{align}
\mathcal{U}^\pm_{P'P,l'lm}=\sum_{\nu}\biggl[\frac{l(l+1)+l'(l'+1)-\nu(\nu+1)}{2}\delta_{PP'}\notag\\
\mp imd\frac{\omega}{c}\left(1-\delta_{PP'}\right)\biggr]A^\pm_{l'l\nu m}(d),\label{eq:Upm}
\end{align}
with the function
\begin{align}
A^\pm_{l'l\nu m}(d)=(-1)^mi^{l-l'\pm\nu}(2\nu+1)\sqrt{\frac{(2l+1)(2l'+1)}{l(l+1)l'(l'+1)}}\notag\\
\left(\begin{array}{ccc}
l&l'&\nu\\
0&0&0
\end{array}
\right)
\left(\begin{array}{ccc}
l&l'&\nu\\
m&-m&0
\end{array}
\right)h_\nu\left(\frac{d\omega}{c}\right).\label{eq:A}
\end{align}
The regular part $\mathcal{V}$, as used in Eq.~\eqref{eq:traceforces1}, is obtained from $\mathcal{U}^+$ by replacing $h_\nu$ in Eq.~\eqref{eq:A} with $j_\nu$~\cite{Wittmann88}. We find the useful relation
 for the matrix $-p_z= \partial_{d} {\cal V}(d\,{\hat {\bf z}})\vert_{d=0}$ (see Eq.~(\ref{defp})), 
\begin{align}
&-p_{z;P'Pl'lm}=\frac{\omega}{c}\left\{i(1-\delta_{P'P})\,\delta_{l',l} \, a(l,m)\right.\notag\\&\left.+ \,\delta_{P'P} [-b(l,m)\,\delta_{l',l+1}+b(l',m)\,\delta_{l'+1,l}]\right\},\label{pzs}
\end{align}
with $a(l,m)$ and $b(l,m)$ defined in Eqs.~\eqref{eq:a} and~\eqref{eq:b}.

\end{appendix}

%

\begin{thebibliography}{91}%
\makeatletter
\providecommand \@ifxundefined [1]{%
 \@ifx{#1\undefined}
}%
\providecommand \@ifnum [1]{%
 \ifnum #1\expandafter \@firstoftwo
 \else \expandafter \@secondoftwo
 \fi
}%
\providecommand \@ifx [1]{%
 \ifx #1\expandafter \@firstoftwo
 \else \expandafter \@secondoftwo
 \fi
}%
\providecommand \natexlab [1]{#1}%
\providecommand \enquote  [1]{``#1''}%
\providecommand \bibnamefont  [1]{#1}%
\providecommand \bibfnamefont [1]{#1}%
\providecommand \citenamefont [1]{#1}%
\providecommand \href@noop [0]{\@secondoftwo}%
\providecommand \href [0]{\begingroup \@sanitize@url \@href}%
\providecommand \@href[1]{\@@startlink{#1}\@@href}%
\providecommand \@@href[1]{\endgroup#1\@@endlink}%
\providecommand \@sanitize@url [0]{\catcode `\\12\catcode `\$12\catcode
  `\&12\catcode `\#12\catcode `\^12\catcode `\_12\catcode `\%12\relax}%
\providecommand \@@startlink[1]{}%
\providecommand \@@endlink[0]{}%
\providecommand \url  [0]{\begingroup\@sanitize@url \@url }%
\providecommand \@url [1]{\endgroup\@href {#1}{\urlprefix }}%
\providecommand \urlprefix  [0]{URL }%
\providecommand \Eprint [0]{\href }%
\@ifxundefined \urlstyle {%
  \providecommand \doi  [0]{\begingroup \@sanitize@url \@doi}%
  \providecommand \@doi [1]{\endgroup \@@startlink {\doibase
  #1}doi:\discretionary {}{}{}#1\@@endlink }%
}{%
  \providecommand \doi  [0]{doi:\discretionary{}{}{}\begingroup
  \urlstyle{rm}\Url }%
}%
\providecommand \doibase [0]{http://dx.doi.org/}%
\providecommand \Doi [0]{\begingroup \@sanitize@url \@Doi }%
\providecommand \@Doi  [1]{\endgroup\@@startlink{\doibase#1}\@@Doi}%
\providecommand \@@Doi [1]{#1\@@endlink}%
\providecommand \selectlanguage [0]{\@gobble}%
\providecommand \bibinfo  [0]{\@secondoftwo}%
\providecommand \bibfield  [0]{\@secondoftwo}%
\providecommand \translation [1]{[#1]}%
\providecommand \BibitemOpen [0]{}%
\providecommand \bibitemStop [0]{}%
\providecommand \bibitemNoStop [0]{.\EOS\space}%
\providecommand \EOS [0]{\spacefactor3000\relax}%
\providecommand \BibitemShut  [1]{\csname bibitem#1\endcsname}%
\bibitem [{\citenamefont {Planck}(1951)}]{Planck}%
  \BibitemOpen
  \bibfield  {author} {\bibinfo {author} {\bibfnamefont {M.}~\bibnamefont
  {Planck}},\ }\href@noop {} {\bibfield  {journal} {\bibinfo  {journal} {Ann.
  Phys.},\ }\textbf {\bibinfo {volume} {4}},\ \bibinfo {pages} {553} (\bibinfo
  {year} {1951})}\BibitemShut {NoStop}%
\bibitem [{\citenamefont {Casimir}(1948)}]{Casimir48}%
  \BibitemOpen
  \bibfield  {author} {\bibinfo {author} {\bibfnamefont {H.~B.~G.}\
  \bibnamefont {Casimir}},\ }\href@noop {} {\bibfield  {journal} {\bibinfo
  {journal} {Proc. K. Ned. Akad. Wet.},\ }\textbf {\bibinfo {volume} {51}},\
  \bibinfo {pages} {793} (\bibinfo {year} {1948})}\BibitemShut {NoStop}%
\bibitem [{\citenamefont {Lifshitz}(1956)}]{Lifshitz56}%
  \BibitemOpen
  \bibfield  {author} {\bibinfo {author} {\bibfnamefont {E.~M.}\ \bibnamefont
  {Lifshitz}},\ }\href@noop {} {\bibfield  {journal} {\bibinfo  {journal} {Sov.
  Phys. JETP},\ }\textbf {\bibinfo {volume} {2}},\ \bibinfo {pages} {73}
  (\bibinfo {year} {1956})}\BibitemShut {NoStop}%
\bibitem [{\citenamefont {Milonni}(1994)}]{Milonni}%
  \BibitemOpen
  \bibfield  {author} {\bibinfo {author} {\bibfnamefont {P.~W.}\ \bibnamefont
  {Milonni}},\ }\href@noop {} {\emph {\bibinfo {title} {The Quantum Vacuum}}}\
  (\bibinfo  {publisher} {Academic Press},\ \bibinfo {address} {San Diego},\
  \bibinfo {year} {1994})\BibitemShut {NoStop}%
\bibitem [{\citenamefont {Emig}\ \emph {et~al.}(2007)\citenamefont {Emig},
  \citenamefont {Graham}, \citenamefont {Jaffe},\ and\ \citenamefont
  {Kardar}}]{Emig07}%
  \BibitemOpen
  \bibfield  {author} {\bibinfo {author} {\bibfnamefont {T.}~\bibnamefont
  {Emig}}, \bibinfo {author} {\bibfnamefont {N.}~\bibnamefont {Graham}},
  \bibinfo {author} {\bibfnamefont {R.~L.}\ \bibnamefont {Jaffe}}, \ and\
  \bibinfo {author} {\bibfnamefont {M.}~\bibnamefont {Kardar}},\ }\href@noop {}
  {\bibfield  {journal} {\bibinfo  {journal} {Phys. Rev. Lett.},\ }\textbf
  {\bibinfo {volume} {99}},\ \bibinfo {pages} {170403} (\bibinfo {year}
  {2007})}\BibitemShut {NoStop}%
\bibitem [{\citenamefont {Neto}\ \emph {et~al.}(2008)\citenamefont {Neto},
  \citenamefont {Lambrecht},\ and\ \citenamefont {Reynaud}}]{Neto08}%
  \BibitemOpen
  \bibfield  {author} {\bibinfo {author} {\bibfnamefont {P.~A.}\
  \bibnamefont {Maia Neto}}, \bibinfo {author} {\bibfnamefont {A.}~\bibnamefont
  {Lambrecht}}, \ and\ \bibinfo {author} {\bibfnamefont {S.}~\bibnamefont
  {Reynaud}},\ }\href@noop {} {\bibfield  {journal} {\bibinfo  {journal}
  {{Phys. Rev. A}},\ }\textbf {\bibinfo {volume} {{78}}},\ \bibinfo {pages}
  {{012115}} (\bibinfo {year} {{2008}})}\BibitemShut {NoStop}%
\bibitem [{\citenamefont {Rahi}\ \emph {et~al.}(2008)\citenamefont {Rahi},
  \citenamefont {Emig}, \citenamefont {Graham}, \citenamefont {Jaffe},\ and\
  \citenamefont {Kardar}}]{Rahi09}%
  \BibitemOpen
  \bibfield  {author} {\bibinfo {author} {\bibfnamefont {S.~J.}\ \bibnamefont
  {Rahi}}, \bibinfo {author} {\bibfnamefont {T.}~\bibnamefont {Emig}}, \bibinfo
  {author} {\bibfnamefont {N.}~\bibnamefont {Graham}}, \bibinfo {author}
  {\bibfnamefont {R.~L.}\ \bibnamefont {Jaffe}}, \ and\ \bibinfo {author}
  {\bibfnamefont {M.}~\bibnamefont {Kardar}},\ }\href@noop {} {\bibfield
  {journal} {\bibinfo  {journal} {Phys. Rev. D},\ }\textbf {\bibinfo {volume}
  {80}},\ \bibinfo {pages} {085021} (\bibinfo {year} {2009})}\BibitemShut
  {NoStop}%
\bibitem [{\citenamefont {Rahi}\ \emph {et~al.}(2010)\citenamefont {Rahi},
  \citenamefont {Kardar},\ and\ \citenamefont {Emig}}]{Rahi10}%
  \BibitemOpen
  \bibfield  {author} {\bibinfo {author} {\bibfnamefont {S.~J.}\ \bibnamefont
  {Rahi}}, \bibinfo {author} {\bibfnamefont {M.}~\bibnamefont {Kardar}}, \ and\
  \bibinfo {author} {\bibfnamefont {T.}~\bibnamefont {Emig}},\ }\href@noop {}
  {\bibfield  {journal} {\bibinfo  {journal} {Phys Rev. Lett.},\ }\textbf
  {\bibinfo {volume} {105}},\ \bibinfo {pages} {070404} (\bibinfo {year}
  {2010})}\BibitemShut {NoStop}%
\bibitem [{\citenamefont {Levin}\ \emph {et~al.}(2010)\citenamefont {Levin},
  \citenamefont {McCauley}, \citenamefont {Rodriguez}, \citenamefont {Reid},\
  and\ \citenamefont {Johnson}}]{LevinMc10}%
  \BibitemOpen
  \bibfield  {author} {\bibinfo {author} {\bibfnamefont {M.}~\bibnamefont
  {Levin}}, \bibinfo {author} {\bibfnamefont {A.~P.}\ \bibnamefont {McCauley}},
  \bibinfo {author} {\bibfnamefont {A.~W.}\ \bibnamefont {Rodriguez}}, \bibinfo
  {author} {\bibfnamefont {M.~T.~H.}\ \bibnamefont {Reid}}, \ and\ \bibinfo
  {author} {\bibfnamefont {S.~G.}\ \bibnamefont {Johnson}},\ }\href@noop {}
  {\bibfield  {journal} {\bibinfo  {journal} {Phys. Rev. Lett.},\ }\textbf
  {\bibinfo {volume} {105}},\ \bibinfo {pages} {090403} (\bibinfo {year}
  {2010})}\BibitemShut {NoStop}%
\bibitem [{\citenamefont {Bordag}\ \emph {et~al.}(2009)\citenamefont {Bordag},
  \citenamefont {Klimchitskaya}, \citenamefont {Mohideen},\ and\ \citenamefont
  {Mostepanenko}}]{Bordag}%
  \BibitemOpen
  \bibfield  {author} {\bibinfo {author} {\bibfnamefont {M.}~\bibnamefont
  {Bordag}}, \bibinfo {author} {\bibfnamefont {G.~L.}\ \bibnamefont
  {Klimchitskaya}}, \bibinfo {author} {\bibfnamefont {U.}~\bibnamefont
  {Mohideen}}, \ and\ \bibinfo {author} {\bibfnamefont {V.~M.}\ \bibnamefont
  {Mostepanenko}},\ }\href@noop {} {\emph {\bibinfo {title} {Advances in the
  Casimir effect}}}\ (\bibinfo  {publisher} {Oxford University Press},\
  \bibinfo {address} {Oxford},\ \bibinfo {year} {2009})\BibitemShut {NoStop}%
\bibitem [{\citenamefont {Antezza}\ \emph {et~al.}(2008)\citenamefont
  {Antezza}, \citenamefont {Pitaevskii}, \citenamefont {Stringari},\ and\
  \citenamefont {Svetovoy}}]{Antezza08}%
  \BibitemOpen
  \bibfield  {author} {\bibinfo {author} {\bibfnamefont {M.}~\bibnamefont
  {Antezza}}, \bibinfo {author} {\bibfnamefont {L.~P.}\ \bibnamefont
  {Pitaevskii}}, \bibinfo {author} {\bibfnamefont {S.}~\bibnamefont
  {Stringari}}, \ and\ \bibinfo {author} {\bibfnamefont {V.~B.}\ \bibnamefont
  {Svetovoy}},\ }\Doi {10.1103/PhysRevA.77.022901} {\bibfield  {journal}
  {\bibinfo  {journal} {Phys. Rev. A},\ }\textbf {\bibinfo {volume} {77}},\
  \bibinfo {pages} {022901} (\bibinfo {year} {2008})}\BibitemShut {NoStop}%
\bibitem [{\citenamefont {Dalvit}\ \emph {et~al.}(2010)\citenamefont {Dalvit},
  \citenamefont {Neto},\ and\ \citenamefont {Mazzitelli}}]{dalvit10}%
  \BibitemOpen
  \bibfield  {author} {\bibinfo {author} {\bibfnamefont {D.~A.~R.}\
  \bibnamefont {Dalvit}}, \bibinfo {author} {\bibfnamefont {P.~A.~M.}\
  \bibnamefont {Neto}}, \ and\ \bibinfo {author} {\bibfnamefont {F.~D.}\
  \bibnamefont {Mazzitelli}},\ }\href@noop {} {} (\bibinfo {year} {2010}),\
  \bibinfo {note} {arXiv:1006.4790}\BibitemShut {NoStop}%
\bibitem [{\citenamefont {Rytov}\ \emph {et~al.}(1989)\citenamefont {Rytov},
  \citenamefont {Kravtsov},\ and\ \citenamefont {Tatarskii}}]{Rytov3}%
  \BibitemOpen
  \bibfield  {author} {\bibinfo {author} {\bibfnamefont {S.~M.}\ \bibnamefont
  {Rytov}}, \bibinfo {author} {\bibfnamefont {Y.~A.}\ \bibnamefont {Kravtsov}},
  \ and\ \bibinfo {author} {\bibfnamefont {V.~I.}\ \bibnamefont {Tatarskii}},\
  }\href@noop {} {\emph {\bibinfo {title} {Principles of statistical
  radiophysics 3}}}\ (\bibinfo  {publisher} {Springer},\ \bibinfo {address}
  {Berlin},\ \bibinfo {year} {1989})\BibitemShut {NoStop}%
\bibitem [{\citenamefont {Bimonte}(2009)}]{Bimonte09}%
  \BibitemOpen
  \bibfield  {author} {\bibinfo {author} {\bibfnamefont {G.}~\bibnamefont
  {Bimonte}},\ }\href@noop {} {\bibfield  {journal} {\bibinfo  {journal} {Phys.
  Rev. A},\ }\textbf {\bibinfo {volume} {80}},\ \bibinfo {pages} {042102}
  (\bibinfo {year} {2009})}\BibitemShut {NoStop}%
\bibitem [{\citenamefont {Henkel}\ \emph {et~al.}(2002)\citenamefont {Henkel},
  \citenamefont {Joulain}, \citenamefont {Mulet},\ and\ \citenamefont
  {Greffet}}]{Henkel02}%
  \BibitemOpen
  \bibfield  {author} {\bibinfo {author} {\bibfnamefont {C.}~\bibnamefont
  {Henkel}}, \bibinfo {author} {\bibfnamefont {K.}~\bibnamefont {Joulain}},
  \bibinfo {author} {\bibfnamefont {J.-P.}\ \bibnamefont {Mulet}}, \ and\
  \bibinfo {author} {\bibfnamefont {J.-J.}\ \bibnamefont {Greffet}},\
  }\href@noop {} {\bibfield  {journal} {\bibinfo  {journal} {J. Opt. A Pure
  Appl. Opt.},\ }\textbf {\bibinfo {volume} {4}},\ \bibinfo {pages} {S109}
  (\bibinfo {year} {2002})}\BibitemShut {NoStop}%
\bibitem [{\citenamefont {Antezza}\ \emph {et~al.}(2005)\citenamefont
  {Antezza}, \citenamefont {Pitaevskii},\ and\ \citenamefont
  {Stringari}}]{Antezza05}%
  \BibitemOpen
  \bibfield  {author} {\bibinfo {author} {\bibfnamefont {M.}~\bibnamefont
  {Antezza}}, \bibinfo {author} {\bibfnamefont {L.~P.}\ \bibnamefont
  {Pitaevskii}}, \ and\ \bibinfo {author} {\bibfnamefont {S.}~\bibnamefont
  {Stringari}},\ }\Doi {10.1103/PhysRevLett.95.113202} {\bibfield  {journal}
  {\bibinfo  {journal} {Phys. Rev. Lett.},\ }\textbf {\bibinfo {volume} {95}},\
  \bibinfo {pages} {113202} (\bibinfo {year} {2005})}\BibitemShut {NoStop}%
\bibitem [{\citenamefont {Ellingsen}\ \emph {et~al.}(2010)\citenamefont
  {Ellingsen}, \citenamefont {Sherkunov}, \citenamefont {Buhmann},\ and\
  \citenamefont {Scheel}}]{Ellingsen10}%
  \BibitemOpen
  \bibfield  {author} {\bibinfo {author} {\bibfnamefont {S.~{\AA}.}\
  \bibnamefont {Ellingsen}}, \bibinfo {author} {\bibfnamefont {Y.}~\bibnamefont
  {Sherkunov}}, \bibinfo {author} {\bibfnamefont {S.~Y.}\ \bibnamefont
  {Buhmann}}, \ and\ \bibinfo {author} {\bibfnamefont {S.}~\bibnamefont
  {Scheel}},\ }\href@noop {} {\bibfield  {journal} {\bibinfo  {journal}
  {Proceed. of QFEXT09 (World Scientific)},\ \bibinfo {pages} {168}} (\bibinfo
  {year} {2010})}\BibitemShut {NoStop}%
\bibitem [{\citenamefont {Kweon}\ and\ \citenamefont
  {Lawandy}(1993)}]{Kweon93}%
  \BibitemOpen
  \bibfield  {author} {\bibinfo {author} {\bibfnamefont {G.-i.}\ \bibnamefont
  {Kweon}}\ and\ \bibinfo {author} {\bibfnamefont {N.~M.}\ \bibnamefont
  {Lawandy}},\ }\href@noop {} {\bibfield  {journal} {\bibinfo  {journal} {Phys.
  Rev. A},\ }\textbf {\bibinfo {volume} {47}},\ \bibinfo {pages} {4513}
  (\bibinfo {year} {1993})}\BibitemShut {NoStop}%
\bibitem [{\citenamefont {Power}\ and\ \citenamefont
  {Thirunamachandran}(1995)}]{Power94}%
  \BibitemOpen
  \bibfield  {author} {\bibinfo {author} {\bibfnamefont {E.~A.}\ \bibnamefont
  {Power}}\ and\ \bibinfo {author} {\bibfnamefont {T.}~\bibnamefont
  {Thirunamachandran}},\ }\href@noop {} {\bibfield  {journal} {\bibinfo
  {journal} {Phys. Rev. A},\ }\textbf {\bibinfo {volume} {51}},\ \bibinfo
  {pages} {3660} (\bibinfo {year} {1995})}\BibitemShut {NoStop}%
\bibitem [{\citenamefont {Cohen}\ and\ \citenamefont
  {Mukamel}(2003)}]{Cohen03}%
  \BibitemOpen
  \bibfield  {author} {\bibinfo {author} {\bibfnamefont {A.~E.}\ \bibnamefont
  {Cohen}}\ and\ \bibinfo {author} {\bibfnamefont {S.}~\bibnamefont
  {Mukamel}},\ }\href@noop {} {\bibfield  {journal} {\bibinfo  {journal} {Phys.
  Rev. Lett.},\ }\textbf {\bibinfo {volume} {91}},\ \bibinfo {pages} {233202}
  (\bibinfo {year} {2003})}\BibitemShut {NoStop}%
\bibitem [{\citenamefont {Sherkunov}(2009)}]{Sherkunov09}%
  \BibitemOpen
  \bibfield  {author} {\bibinfo {author} {\bibfnamefont {Y.}~\bibnamefont
  {Sherkunov}},\ }\href@noop {} {\bibfield  {journal} {\bibinfo  {journal}
  {Phys. Rev. A},\ }\textbf {\bibinfo {volume} {79}},\ \bibinfo {pages}
  {032101} (\bibinfo {year} {2009})}\BibitemShut {NoStop}%
\bibitem [{\citenamefont {Behunin}\ and\ \citenamefont {Hu}(2010)}]{Behunin10}%
  \BibitemOpen
  \bibfield  {author} {\bibinfo {author} {\bibfnamefont {R.~O.}\ \bibnamefont
  {Behunin}}\ and\ \bibinfo {author} {\bibfnamefont {B.-L.}\ \bibnamefont
  {Hu}},\ }\href@noop {} {\bibfield  {journal} {\bibinfo  {journal} {Phys. Rev.
  A},\ }\textbf {\bibinfo {volume} {82}},\ \bibinfo {pages} {022507} (\bibinfo
  {year} {2010})}\BibitemShut {NoStop}%
\bibitem [{\citenamefont {Haakh}\ \emph {et~al.}()\citenamefont {Haakh},
  \citenamefont {Schiefele},\ and\ \citenamefont {Henkel}}]{Haakh11}%
  \BibitemOpen
  \bibfield  {author} {\bibinfo {author} {\bibfnamefont {H.~R.}\ \bibnamefont
  {Haakh}}, \bibinfo {author} {\bibfnamefont {J.}~\bibnamefont {Schiefele}}, \
  and\ \bibinfo {author} {\bibfnamefont {C.}~\bibnamefont {Henkel}},\
  }\href@noop {} {}\bibinfo {note} {ArXiv:1111.3748}\BibitemShut {NoStop}%
\bibitem [{\citenamefont {Zurita-S{\'a}nchez}\ and\ \citenamefont
  {Henkel}(2012)}]{Sanchez12}%
  \BibitemOpen
  \bibfield  {author} {\bibinfo {author} {\bibfnamefont {J.~R.}\ \bibnamefont
  {Zurita-S{\'a}nchez}}\ and\ \bibinfo {author} {\bibfnamefont
  {C.}~\bibnamefont {Henkel}},\ }\href@noop {} {\bibfield  {journal} {\bibinfo
  {journal} {Europhys. Lett.},\ }\textbf {\bibinfo {volume} {97}},\ \bibinfo
  {pages} {43002} (\bibinfo {year} {2012})}\BibitemShut {NoStop}%
\bibitem [{\citenamefont {Messina}\ and\ \citenamefont
  {Antezza}(2011){\natexlab{a}}}]{Messina}%
  \BibitemOpen
  \bibfield  {author} {\bibinfo {author} {\bibfnamefont {R.}~\bibnamefont
  {Messina}}\ and\ \bibinfo {author} {\bibfnamefont {M.}~\bibnamefont
  {Antezza}},\ }\href@noop {} {\bibfield  {journal} {\bibinfo  {journal}
  {Europhys. Lett.},\ }\textbf {\bibinfo {volume} {95}},\ \bibinfo {pages}
  {61002} (\bibinfo {year} {2011}{\natexlab{a}})}\BibitemShut {NoStop}%
\bibitem [{\citenamefont {Kr\"uger}\ \emph
  {et~al.}(2011){\natexlab{a}}\citenamefont {Kr\"uger}, \citenamefont {Emig},\
  and\ \citenamefont {Kardar}}]{Kruger11}%
  \BibitemOpen
  \bibfield  {author} {\bibinfo {author} {\bibfnamefont {M.}~\bibnamefont
  {Kr\"uger}}, \bibinfo {author} {\bibfnamefont {T.}~\bibnamefont {Emig}}, \
  and\ \bibinfo {author} {\bibfnamefont {M.}~\bibnamefont {Kardar}},\
  }\href@noop {} {\bibfield  {journal} {\bibinfo  {journal} {Phys. Rev.
  Lett.},\ }\textbf {\bibinfo {volume} {106}},\ \bibinfo {pages} {210404}
  (\bibinfo {year} {2011}{\natexlab{a}})}\BibitemShut {NoStop}%
\bibitem [{\citenamefont {Kr\"uger}\ \emph
  {et~al.}(2011){\natexlab{b}}\citenamefont {Kr\"uger}, \citenamefont {Emig},
  \citenamefont {Bimonte},\ and\ \citenamefont {Kardar}}]{Kruger11b}%
  \BibitemOpen
  \bibfield  {author} {\bibinfo {author} {\bibfnamefont {M.}~\bibnamefont
  {Kr\"uger}}, \bibinfo {author} {\bibfnamefont {T.}~\bibnamefont {Emig}},
  \bibinfo {author} {\bibfnamefont {G.}~\bibnamefont {Bimonte}}, \ and\
  \bibinfo {author} {\bibfnamefont {M.}~\bibnamefont {Kardar}},\ }\href@noop {}
  {\bibfield  {journal} {\bibinfo  {journal} {Europhys. Lett.},\ }\textbf
  {\bibinfo {volume} {95}},\ \bibinfo {pages} {21002} (\bibinfo {year}
  {2011}{\natexlab{b}})}\BibitemShut {NoStop}%
\bibitem [{\citenamefont {Messina}\ and\ \citenamefont
  {Antezza}(2011){\natexlab{b}}}]{Messina11b}%
  \BibitemOpen
  \bibfield  {author} {\bibinfo {author} {\bibfnamefont {R.}~\bibnamefont
  {Messina}}\ and\ \bibinfo {author} {\bibfnamefont {M.}~\bibnamefont
  {Antezza}},\ }\Doi {10.1103/PhysRevA.84.042102} {\bibfield  {journal}
  {\bibinfo  {journal} {Phys. Rev. A},\ }\textbf {\bibinfo {volume} {84}},\
  \bibinfo {pages} {042102} (\bibinfo {year} {2011}{\natexlab{b}})}\BibitemShut
  {NoStop}%
\bibitem [{\citenamefont {Bimonte}\ \emph {et~al.}(2011)\citenamefont
  {Bimonte}, \citenamefont {Emig}, \citenamefont {Kr\"uger},\ and\
  \citenamefont {Kardar}}]{Bimonte11}%
  \BibitemOpen
  \bibfield  {author} {\bibinfo {author} {\bibfnamefont {G.}~\bibnamefont
  {Bimonte}}, \bibinfo {author} {\bibfnamefont {T.}~\bibnamefont {Emig}},
  \bibinfo {author} {\bibfnamefont {M.}~\bibnamefont {Kr\"uger}}, \ and\
  \bibinfo {author} {\bibfnamefont {M.}~\bibnamefont {Kardar}},\ }\Doi
  {10.1103/PhysRevA.84.042503} {\bibfield  {journal} {\bibinfo  {journal}
  {Phys. Rev. A},\ }\textbf {\bibinfo {volume} {84}},\ \bibinfo {pages}
  {042503} (\bibinfo {year} {2011})}\BibitemShut {NoStop}%
\bibitem [{\citenamefont {Golyk}\ \emph
  {et~al.}(2012){\natexlab{a}}\citenamefont {Golyk}, \citenamefont {Kr\"uger},
  \citenamefont {Reid},\ and\ \citenamefont {Kardar}}]{Golyk12}%
  \BibitemOpen
  \bibfield  {author} {\bibinfo {author} {\bibfnamefont {V.~A.}\ \bibnamefont
  {Golyk}}, \bibinfo {author} {\bibfnamefont {M.}~\bibnamefont {Kr\"uger}},
  \bibinfo {author} {\bibfnamefont {M.~T.~H.}\ \bibnamefont {Reid}}, \ and\
  \bibinfo {author} {\bibfnamefont {M.}~\bibnamefont {Kardar}},\ }\href@noop {}
  {\bibfield  {journal} {\bibinfo  {journal} {Phys. Rev. D},\ }\textbf
  {\bibinfo {volume} {85}},\ \bibinfo {pages} {065011} (\bibinfo {year}
  {2012}{\natexlab{a}})}\BibitemShut {NoStop}%
\bibitem [{\citenamefont {Dean}\ \emph {et~al.}(2012)\citenamefont {Dean},
  \citenamefont {D\'emery}, \citenamefont {Parsegian},\ and\ \citenamefont
  {Podgornik}}]{Dean12}%
  \BibitemOpen
  \bibfield  {author} {\bibinfo {author} {\bibfnamefont {D.~S.}\ \bibnamefont
  {Dean}}, \bibinfo {author} {\bibfnamefont {V.}~\bibnamefont {D\'emery}},
  \bibinfo {author} {\bibfnamefont {V.~A.}\ \bibnamefont {Parsegian}}, \ and\
  \bibinfo {author} {\bibfnamefont {R.}~\bibnamefont {Podgornik}},\ }\href@noop
  {} {\bibfield  {journal} {\bibinfo  {journal} {Phys. Rev. E},\ }\textbf
  {\bibinfo {volume} {85}},\ \bibinfo {pages} {031108} (\bibinfo {year}
  {2012})}\BibitemShut {NoStop}%
\bibitem [{\citenamefont {Sheng}\ \emph {et~al.}(2009)\citenamefont {Sheng},
  \citenamefont {Narayanaswamy},\ and\ \citenamefont {Chen}}]{Sheng09}%
  \BibitemOpen
  \bibfield  {author} {\bibinfo {author} {\bibfnamefont {S.}~\bibnamefont
  {Sheng}}, \bibinfo {author} {\bibfnamefont {A.}~\bibnamefont
  {Narayanaswamy}}, \ and\ \bibinfo {author} {\bibfnamefont {G.}~\bibnamefont
  {Chen}},\ }\href@noop {} {\bibfield  {journal} {\bibinfo  {journal} {Nano
  Lett.},\ }\textbf {\bibinfo {volume} {9}},\ \bibinfo {pages} {2909} (\bibinfo
  {year} {2009})}\BibitemShut {NoStop}%
\bibitem [{\citenamefont {Rousseau}\ \emph {et~al.}(2009)\citenamefont
  {Rousseau}, \citenamefont {Siria}, \citenamefont {Jourdan}, \citenamefont
  {Volz}, \citenamefont {Comin}, \citenamefont {Chevrier},\ and\ \citenamefont
  {Greffet}}]{Rousseau09}%
  \BibitemOpen
  \bibfield  {author} {\bibinfo {author} {\bibfnamefont {E.}~\bibnamefont
  {Rousseau}}, \bibinfo {author} {\bibfnamefont {A.}~\bibnamefont {Siria}},
  \bibinfo {author} {\bibfnamefont {G.}~\bibnamefont {Jourdan}}, \bibinfo
  {author} {\bibfnamefont {S.}~\bibnamefont {Volz}}, \bibinfo {author}
  {\bibfnamefont {F.}~\bibnamefont {Comin}}, \bibinfo {author} {\bibfnamefont
  {J.}~\bibnamefont {Chevrier}}, \ and\ \bibinfo {author} {\bibfnamefont
  {J.-J.}\ \bibnamefont {Greffet}},\ }\href@noop {} {\bibfield  {journal}
  {\bibinfo  {journal} {Nature Photon.},\ }\textbf {\bibinfo {volume} {3}},\
  \bibinfo {pages} {514} (\bibinfo {year} {2009})}\BibitemShut {NoStop}%
\bibitem [{\citenamefont {Ottens}\ \emph {et~al.}(2011)\citenamefont {Ottens},
  \citenamefont {Quetschke}, \citenamefont {Wise}, \citenamefont {Alemi},
  \citenamefont {Lundock}, \citenamefont {Mueller}, \citenamefont {Reitze},
  \citenamefont {Tanner},\ and\ \citenamefont {Whiting}}]{Ottens11}%
  \BibitemOpen
  \bibfield  {author} {\bibinfo {author} {\bibfnamefont {R.~S.}\ \bibnamefont
  {Ottens}}, \bibinfo {author} {\bibfnamefont {V.}~\bibnamefont {Quetschke}},
  \bibinfo {author} {\bibfnamefont {S.}~\bibnamefont {Wise}}, \bibinfo {author}
  {\bibfnamefont {A.~A.}\ \bibnamefont {Alemi}}, \bibinfo {author}
  {\bibfnamefont {R.}~\bibnamefont {Lundock}}, \bibinfo {author} {\bibfnamefont
  {G.}~\bibnamefont {Mueller}}, \bibinfo {author} {\bibfnamefont {D.~H.}\
  \bibnamefont {Reitze}}, \bibinfo {author} {\bibfnamefont {D.~B.}\
  \bibnamefont {Tanner}}, \ and\ \bibinfo {author} {\bibfnamefont {B.~F.}\
  \bibnamefont {Whiting}},\ }\href@noop {} {\bibfield  {journal} {\bibinfo
  {journal} {Phys. Rev. Lett.},\ }\textbf {\bibinfo {volume} {107}},\ \bibinfo
  {pages} {014301} (\bibinfo {year} {2011})}\BibitemShut {NoStop}%
\bibitem [{\citenamefont {Polder}\ and\ \citenamefont
  {Van~Hove}(1971)}]{Polder71}%
  \BibitemOpen
  \bibfield  {author} {\bibinfo {author} {\bibfnamefont {D.}~\bibnamefont
  {Polder}}\ and\ \bibinfo {author} {\bibfnamefont {M.}~\bibnamefont
  {Van~Hove}},\ }\href@noop {} {\bibfield  {journal} {\bibinfo  {journal}
  {Phys. Rev. B},\ }\textbf {\bibinfo {volume} {4}},\ \bibinfo {pages} {3303}
  (\bibinfo {year} {1971})}\BibitemShut {NoStop}%
\bibitem [{\citenamefont {Volokitin}\ and\ \citenamefont
  {Persson}(2001)}]{Volokitin01}%
  \BibitemOpen
  \bibfield  {author} {\bibinfo {author} {\bibfnamefont {A.~I.}\ \bibnamefont
  {Volokitin}}\ and\ \bibinfo {author} {\bibfnamefont {B.~N.~J.}\ \bibnamefont
  {Persson}},\ }\href@noop {} {\bibfield  {journal} {\bibinfo  {journal} {Phys.
  Rev. B},\ }\textbf {\bibinfo {volume} {63}},\ \bibinfo {pages} {205404}
  (\bibinfo {year} {2001})}\BibitemShut {NoStop}%
\bibitem [{\citenamefont {Narayanaswamy}\ and\ \citenamefont
  {Chen}(2008)}]{Narayanaswamy08}%
  \BibitemOpen
  \bibfield  {author} {\bibinfo {author} {\bibfnamefont {A.}~\bibnamefont
  {Narayanaswamy}}\ and\ \bibinfo {author} {\bibfnamefont {G.}~\bibnamefont
  {Chen}},\ }\Doi {10.1103/PhysRevB.77.075125} {\bibfield  {journal} {\bibinfo
  {journal} {Phys. Rev. B},\ }\textbf {\bibinfo {volume} {77}},\ \bibinfo
  {pages} {075125} (\bibinfo {year} {2008})}\BibitemShut {NoStop}%
\bibitem [{\citenamefont {Sasihithlu}\ and\ \citenamefont
  {Narayanaswamy}(2011){\natexlab{a}}}]{Sasihithlu11}%
  \BibitemOpen
  \bibfield  {author} {\bibinfo {author} {\bibfnamefont {K.}~\bibnamefont
  {Sasihithlu}}\ and\ \bibinfo {author} {\bibfnamefont {A.}~\bibnamefont
  {Narayanaswamy}},\ }\Doi {10.1103/PhysRevB.83.161406} {\bibfield  {journal}
  {\bibinfo  {journal} {Phys. Rev. B},\ }\textbf {\bibinfo {volume} {83}},\
  \bibinfo {pages} {161406} (\bibinfo {year} {2011}{\natexlab{a}})}\BibitemShut
  {NoStop}%
\bibitem [{\citenamefont {Otey}\ and\ \citenamefont {Fan}(2011)}]{Otey}%
  \BibitemOpen
  \bibfield  {author} {\bibinfo {author} {\bibfnamefont {C.}~\bibnamefont
  {Otey}}\ and\ \bibinfo {author} {\bibfnamefont {S.}~\bibnamefont {Fan}},\
  }\href@noop {} {\bibfield  {journal} {\bibinfo  {journal} {Phys. Rev. B},\
  }\textbf {\bibinfo {volume} {84}},\ \bibinfo {pages} {245431} (\bibinfo
  {year} {2011})},\ \bibinfo {note} {arXiv:1103.2668}\BibitemShut {NoStop}%
\bibitem [{\citenamefont {McCauley}\ \emph {et~al.}(2012)\citenamefont
  {McCauley}, \citenamefont {Reid}, \citenamefont {Kr\"uger},\ and\
  \citenamefont {Johnson}}]{McCauley}%
  \BibitemOpen
  \bibfield  {author} {\bibinfo {author} {\bibfnamefont {A.~P.}\ \bibnamefont
  {McCauley}}, \bibinfo {author} {\bibfnamefont {M.~T.~H.}\ \bibnamefont
  {Reid}}, \bibinfo {author} {\bibfnamefont {M.}~\bibnamefont {Kr\"uger}}, \
  and\ \bibinfo {author} {\bibfnamefont {S.~G.}\ \bibnamefont {Johnson}},\
  }\href@noop {} {\bibfield  {journal} {\bibinfo  {journal} {Phys. Rev. B},\
  }\textbf {\bibinfo {volume} {85}},\ \bibinfo {pages} {165104} (\bibinfo
  {year} {2012})}\BibitemShut {NoStop}%
\bibitem [{\citenamefont {Rodriguez}\ \emph {et~al.}(2011)\citenamefont
  {Rodriguez}, \citenamefont {Ilic}, \citenamefont {Bermel}, \citenamefont
  {Celanovic}, \citenamefont {Joannopoulos}, \citenamefont {Solja\ifmmode
  \check{c}\else \v{c}\fi{}i\ifmmode~\acute{c}\else \'{c}\fi{}},\ and\
  \citenamefont {Johnson}}]{Rodriguez11}%
  \BibitemOpen
  \bibfield  {author} {\bibinfo {author} {\bibfnamefont {A.~W.}\ \bibnamefont
  {Rodriguez}}, \bibinfo {author} {\bibfnamefont {O.}~\bibnamefont {Ilic}},
  \bibinfo {author} {\bibfnamefont {P.}~\bibnamefont {Bermel}}, \bibinfo
  {author} {\bibfnamefont {I.}~\bibnamefont {Celanovic}}, \bibinfo {author}
  {\bibfnamefont {J.~D.}\ \bibnamefont {Joannopoulos}}, \bibinfo {author}
  {\bibfnamefont {M.}~\bibnamefont {Solja\ifmmode \check{c}\else
  \v{c}\fi{}i\ifmmode~\acute{c}\else \'{c}\fi{}}}, \ and\ \bibinfo {author}
  {\bibfnamefont {S.~G.}\ \bibnamefont {Johnson}},\ }\Doi
  {10.1103/PhysRevLett.107.114302} {\bibfield  {journal} {\bibinfo  {journal}
  {Phys. Rev. Lett.},\ }\textbf {\bibinfo {volume} {107}},\ \bibinfo {pages}
  {114302} (\bibinfo {year} {2011})}\BibitemShut {NoStop}%
\bibitem [{\citenamefont {Rodriguez}\ \emph {et~al.}()\citenamefont
  {Rodriguez}, \citenamefont {Reid},\ and\ \citenamefont
  {Johnson}}]{Rodriguez}%
  \BibitemOpen
  \bibfield  {author} {\bibinfo {author} {\bibfnamefont {A.~W.}\ \bibnamefont
  {Rodriguez}}, \bibinfo {author} {\bibfnamefont {M.~T.~H.}\ \bibnamefont
  {Reid}}, \ and\ \bibinfo {author} {\bibfnamefont {S.~G.}\ \bibnamefont
  {Johnson}},\ }\href@noop {} {}\bibinfo {note} {ArXiv:1206.1772}\BibitemShut
  {NoStop}%
\bibitem [{\citenamefont {Chapuis}\ \emph {et~al.}(2008)\citenamefont
  {Chapuis}, \citenamefont {Volz}, \citenamefont {Henkel}, \citenamefont
  {Joulain},\ and\ \citenamefont {Greffet}}]{Chapuis08}%
  \BibitemOpen
  \bibfield  {author} {\bibinfo {author} {\bibfnamefont {P.-O.}\ \bibnamefont
  {Chapuis}}, \bibinfo {author} {\bibfnamefont {S.}~\bibnamefont {Volz}},
  \bibinfo {author} {\bibfnamefont {C.}~\bibnamefont {Henkel}}, \bibinfo
  {author} {\bibfnamefont {K.}~\bibnamefont {Joulain}}, \ and\ \bibinfo
  {author} {\bibfnamefont {J.-J.}\ \bibnamefont {Greffet}},\ }\href@noop {}
  {\bibfield  {journal} {\bibinfo  {journal} {Phys. Rev. B},\ }\textbf
  {\bibinfo {volume} {77}},\ \bibinfo {pages} {035431} (\bibinfo {year}
  {2008})}\BibitemShut {NoStop}%
\bibitem [{\citenamefont {Biehs}\ \emph {et~al.}(2011)\citenamefont {Biehs},
  \citenamefont {Rosa},\ and\ \citenamefont {Ben-Abdallah}}]{Biehs11}%
  \BibitemOpen
  \bibfield  {author} {\bibinfo {author} {\bibfnamefont {S.-A.}\ \bibnamefont
  {Biehs}}, \bibinfo {author} {\bibfnamefont {F.~S.~S.}\ \bibnamefont {Rosa}},
  \ and\ \bibinfo {author} {\bibfnamefont {P.}~\bibnamefont {Ben-Abdallah}},\
  }\href@noop {} {\bibfield  {journal} {\bibinfo  {journal} {Appl. Phys.
  Lett.},\ }\textbf {\bibinfo {volume} {98}},\ \bibinfo {pages} {243102}
  (\bibinfo {year} {2011})}\BibitemShut {NoStop}%
\bibitem [{\citenamefont {Ben-Abdallah}\ \emph {et~al.}(2011)\citenamefont
  {Ben-Abdallah}, \citenamefont {Biehs},\ and\ \citenamefont
  {Joulain}}]{Ben-Abdallah11}%
  \BibitemOpen
  \bibfield  {author} {\bibinfo {author} {\bibfnamefont {P.}~\bibnamefont
  {Ben-Abdallah}}, \bibinfo {author} {\bibfnamefont {S.-A.}\ \bibnamefont
  {Biehs}}, \ and\ \bibinfo {author} {\bibfnamefont {K.}~\bibnamefont
  {Joulain}},\ }\Doi {10.1103/PhysRevLett.107.114301} {\bibfield  {journal}
  {\bibinfo  {journal} {Phys. Rev. Lett.},\ }\textbf {\bibinfo {volume}
  {107}},\ \bibinfo {pages} {114301} (\bibinfo {year} {2011})}\BibitemShut
  {NoStop}%
\bibitem [{\citenamefont {Gamble}\ \emph {et~al.}(2011)\citenamefont {Gamble},
  \citenamefont {Friesen}, \citenamefont {Joynt},\ and\ \citenamefont
  {Coppersmith}}]{Gamble11}%
  \BibitemOpen
  \bibfield  {author} {\bibinfo {author} {\bibfnamefont {J.~K.}\ \bibnamefont
  {Gamble}}, \bibinfo {author} {\bibfnamefont {M.}~\bibnamefont {Friesen}},
  \bibinfo {author} {\bibfnamefont {R.}~\bibnamefont {Joynt}}, \ and\ \bibinfo
  {author} {\bibfnamefont {S.~N.}\ \bibnamefont {Coppersmith}},\ }\href@noop {}
  {\bibfield  {journal} {\bibinfo  {journal} {Phys. Rev. B},\ }\textbf
  {\bibinfo {volume} {84}},\ \bibinfo {pages} {125321} (\bibinfo {year}
  {2011})}\BibitemShut {NoStop}%
\bibitem [{\citenamefont {Biehs}\ \emph {et~al.}()\citenamefont {Biehs},
  \citenamefont {Tschikin},\ and\ \citenamefont {Ben-Abdallah}}]{Biehs}%
  \BibitemOpen
  \bibfield  {author} {\bibinfo {author} {\bibfnamefont {S.-A.}\ \bibnamefont
  {Biehs}}, \bibinfo {author} {\bibfnamefont {M.}~\bibnamefont {Tschikin}}, \
  and\ \bibinfo {author} {\bibfnamefont {P.}~\bibnamefont {Ben-Abdallah}},\
  }\href@noop {} {}\bibinfo {note} {ArXiv:1112.4966}\BibitemShut {NoStop}%
\bibitem [{\citenamefont {Ben-Abdallah}\ \emph {et~al.}()\citenamefont
  {Ben-Abdallah}, \citenamefont {Rosa}, \citenamefont {Tschikin},\ and\
  \citenamefont {Biehs}}]{Abdallah}%
  \BibitemOpen
  \bibfield  {author} {\bibinfo {author} {\bibfnamefont {P.}~\bibnamefont
  {Ben-Abdallah}}, \bibinfo {author} {\bibfnamefont {F.~S.~S.}\ \bibnamefont
  {Rosa}}, \bibinfo {author} {\bibfnamefont {M.}~\bibnamefont {Tschikin}}, \
  and\ \bibinfo {author} {\bibfnamefont {S.-A.}\ \bibnamefont {Biehs}},\
  }\href@noop {} {}\bibinfo {note} {ArXiv:1112.3470}\BibitemShut {NoStop}%
\bibitem [{\citenamefont {Gu\'erout}\ \emph {et~al.}()\citenamefont
  {Gu\'erout}, \citenamefont {Lussange}, \citenamefont {Rosa}, \citenamefont
  {Hugonin}, \citenamefont {Dalvit}, \citenamefont {Greffet}, \citenamefont
  {Lambrecht},\ and\ \citenamefont {Reynaud}}]{Guerout12}%
  \BibitemOpen
  \bibfield  {author} {\bibinfo {author} {\bibfnamefont {R.}~\bibnamefont
  {Gu\'erout}}, \bibinfo {author} {\bibfnamefont {J.}~\bibnamefont {Lussange}},
  \bibinfo {author} {\bibfnamefont {F.~S.~S.}\ \bibnamefont {Rosa}}, \bibinfo
  {author} {\bibfnamefont {J.-P.}\ \bibnamefont {Hugonin}}, \bibinfo {author}
  {\bibfnamefont {D.~A.~R.}\ \bibnamefont {Dalvit}}, \bibinfo {author}
  {\bibfnamefont {J.~J.}\ \bibnamefont {Greffet}}, \bibinfo {author}
  {\bibfnamefont {A.}~\bibnamefont {Lambrecht}}, \ and\ \bibinfo {author}
  {\bibfnamefont {S.}~\bibnamefont {Reynaud}},\ }\href@noop {} {}\bibinfo
  {note} {ArXiv:1203.1496}\BibitemShut {NoStop}%
\bibitem [{\citenamefont {Rytov}(1959)}]{Rytovc}%
  \BibitemOpen
  \bibfield  {author} {\bibinfo {author} {\bibfnamefont {S.~M.}\ \bibnamefont
  {Rytov}},\ }\href@noop {} {\emph {\bibinfo {title} {Theory of electric
  fluctuations and thermal radiation}}}\ (\bibinfo  {publisher} {Electronics
  Research Directorate, Air Force Cambridge Research Center, Air Research and
  Development Command, U.S. Air Force},\ \bibinfo {address} {Bedford, Mass.},\
  \bibinfo {year} {1959})\BibitemShut {NoStop}%
\bibitem [{\citenamefont {Kattawar}\ and\ \citenamefont
  {Eisner}(1970)}]{Kattawar70}%
  \BibitemOpen
  \bibfield  {author} {\bibinfo {author} {\bibfnamefont {G.~W.}\ \bibnamefont
  {Kattawar}}\ and\ \bibinfo {author} {\bibfnamefont {M.}~\bibnamefont
  {Eisner}},\ }\href@noop {} {\bibfield  {journal} {\bibinfo  {journal} {Appl.
  Opt.},\ }\textbf {\bibinfo {volume} {9}},\ \bibinfo {pages} {2685} (\bibinfo
  {year} {1970})}\BibitemShut {NoStop}%
\bibitem [{\citenamefont {Bohren}\ and\ \citenamefont
  {Huffmann}(2004)}]{Bohren}%
  \BibitemOpen
  \bibfield  {author} {\bibinfo {author} {\bibfnamefont {C.~F.}\ \bibnamefont
  {Bohren}}\ and\ \bibinfo {author} {\bibfnamefont {D.~R.}\ \bibnamefont
  {Huffmann}},\ }\href@noop {} {\emph {\bibinfo {title} {Absorption and
  scattering of light by small particles}}}\ (\bibinfo  {publisher} {Wiley},\
  \bibinfo {address} {Weinheim},\ \bibinfo {year} {2004})\BibitemShut {NoStop}%
\bibitem [{\citenamefont {Hansen}\ and\ \citenamefont
  {Campbell}(1998)}]{Hansen98}%
  \BibitemOpen
  \bibfield  {author} {\bibinfo {author} {\bibfnamefont {K.}~\bibnamefont
  {Hansen}}\ and\ \bibinfo {author} {\bibfnamefont {E.~E.~B.}\ \bibnamefont
  {Campbell}},\ }\href@noop {} {\bibfield  {journal} {\bibinfo  {journal}
  {Phys. Rev. E},\ }\textbf {\bibinfo {volume} {58}},\ \bibinfo {pages} {5477}
  (\bibinfo {year} {1998})}\BibitemShut {NoStop}%
\bibitem [{\citenamefont {Bimonte}\ \emph {et~al.}(2009)\citenamefont
  {Bimonte}, \citenamefont {Cappellin}, \citenamefont {Carugno}, \citenamefont
  {Ruoso},\ and\ \citenamefont {Saadeh}}]{Bimonte09b}%
  \BibitemOpen
  \bibfield  {author} {\bibinfo {author} {\bibfnamefont {G.}~\bibnamefont
  {Bimonte}}, \bibinfo {author} {\bibfnamefont {L.}~\bibnamefont {Cappellin}},
  \bibinfo {author} {\bibfnamefont {G.}~\bibnamefont {Carugno}}, \bibinfo
  {author} {\bibfnamefont {G.}~\bibnamefont {Ruoso}}, \ and\ \bibinfo {author}
  {\bibfnamefont {D.}~\bibnamefont {Saadeh}},\ }\href@noop {} {\bibfield
  {journal} {\bibinfo  {journal} {New J. Phys.},\ }\textbf {\bibinfo {volume}
  {11}},\ \bibinfo {pages} {033014} (\bibinfo {year} {2009})}\BibitemShut
  {NoStop}%
\bibitem [{\citenamefont {Golyk}\ \emph
  {et~al.}(2012){\natexlab{b}}\citenamefont {Golyk}, \citenamefont {Kr\"uger},\
  and\ \citenamefont {Kardar}}]{Golyk11}%
  \BibitemOpen
  \bibfield  {author} {\bibinfo {author} {\bibfnamefont {V.~A.}\ \bibnamefont
  {Golyk}}, \bibinfo {author} {\bibfnamefont {M.}~\bibnamefont {Kr\"uger}}, \
  and\ \bibinfo {author} {\bibfnamefont {M.}~\bibnamefont {Kardar}},\
  }\href@noop {} {\bibfield  {journal} {\bibinfo  {journal} {Phys. Rev. E},\
  }\textbf {\bibinfo {volume} {85}},\ \bibinfo {pages} {046603} (\bibinfo
  {year} {2012}{\natexlab{b}})}\BibitemShut {NoStop}%
\bibitem [{\citenamefont {{\"O}hman}(1961)}]{Ohman61}%
  \BibitemOpen
  \bibfield  {author} {\bibinfo {author} {\bibfnamefont {Y.}~\bibnamefont
  {{\"O}hman}},\ }\href@noop {} {\bibfield  {journal} {\bibinfo  {journal}
  {Nature},\ }\textbf {\bibinfo {volume} {192}},\ \bibinfo {pages} {254}
  (\bibinfo {year} {1961})}\BibitemShut {NoStop}%
\bibitem [{\citenamefont {Li}\ \emph {et~al.}(2003)\citenamefont {Li},
  \citenamefont {Jiang}, \citenamefont {Liu}, \citenamefont {Li}, \citenamefont
  {Fan},\ and\ \citenamefont {Sun}}]{li03}%
  \BibitemOpen
  \bibfield  {author} {\bibinfo {author} {\bibfnamefont {P.}~\bibnamefont
  {Li}}, \bibinfo {author} {\bibfnamefont {K.}~\bibnamefont {Jiang}}, \bibinfo
  {author} {\bibfnamefont {M.}~\bibnamefont {Liu}}, \bibinfo {author}
  {\bibfnamefont {Q.}~\bibnamefont {Li}}, \bibinfo {author} {\bibfnamefont
  {S.}~\bibnamefont {Fan}}, \ and\ \bibinfo {author} {\bibfnamefont
  {J.}~\bibnamefont {Sun}},\ }\href@noop {} {\bibfield  {journal} {\bibinfo
  {journal} {Appl. Phys. Lett.},\ }\textbf {\bibinfo {volume} {82}},\ \bibinfo
  {pages} {1763} (\bibinfo {year} {2003})}\BibitemShut {NoStop}%
\bibitem [{\citenamefont {Fan}\ \emph {et~al.}(2009)\citenamefont {Fan},
  \citenamefont {Singer}, \citenamefont {Bergstrom},\ and\ \citenamefont
  {Regan}}]{Fan09}%
  \BibitemOpen
  \bibfield  {author} {\bibinfo {author} {\bibfnamefont {Y.}~\bibnamefont
  {Fan}}, \bibinfo {author} {\bibfnamefont {S.~B.}\ \bibnamefont {Singer}},
  \bibinfo {author} {\bibfnamefont {R.}~\bibnamefont {Bergstrom}}, \ and\
  \bibinfo {author} {\bibfnamefont {B.~C.}\ \bibnamefont {Regan}},\ }\Doi
  {10.1103/PhysRevLett.102.187402} {\bibfield  {journal} {\bibinfo  {journal}
  {Phys. Rev. Lett.},\ }\textbf {\bibinfo {volume} {102}},\ \bibinfo {pages}
  {187402} (\bibinfo {year} {2009})}\BibitemShut {NoStop}%
\bibitem [{\citenamefont {Singer}\ \emph
  {et~al.}(2011){\natexlab{a}}\citenamefont {Singer}, \citenamefont
  {Mecklenburg}, \citenamefont {White},\ and\ \citenamefont
  {Regan}}]{Singer11}%
  \BibitemOpen
  \bibfield  {author} {\bibinfo {author} {\bibfnamefont {S.~B.}\ \bibnamefont
  {Singer}}, \bibinfo {author} {\bibfnamefont {M.}~\bibnamefont {Mecklenburg}},
  \bibinfo {author} {\bibfnamefont {E.~R.}\ \bibnamefont {White}}, \ and\
  \bibinfo {author} {\bibfnamefont {B.~C.}\ \bibnamefont {Regan}},\ }\href@noop
  {} {\bibfield  {journal} {\bibinfo  {journal} {Phys. Rev. B},\ }\textbf
  {\bibinfo {volume} {83}},\ \bibinfo {pages} {233404} (\bibinfo {year}
  {2011}{\natexlab{a}})}\BibitemShut {NoStop}%
\bibitem [{\citenamefont {Singer}\ \emph
  {et~al.}(2011){\natexlab{b}}\citenamefont {Singer}, \citenamefont
  {Mecklenburg}, \citenamefont {White},\ and\ \citenamefont
  {Regan}}]{Singer11b}%
  \BibitemOpen
  \bibfield  {author} {\bibinfo {author} {\bibfnamefont {S.~B.}\ \bibnamefont
  {Singer}}, \bibinfo {author} {\bibfnamefont {M.}~\bibnamefont {Mecklenburg}},
  \bibinfo {author} {\bibfnamefont {E.~R.}\ \bibnamefont {White}}, \ and\
  \bibinfo {author} {\bibfnamefont {B.~C.}\ \bibnamefont {Regan}},\ }\href@noop
  {} {\bibfield  {journal} {\bibinfo  {journal} {Phys. Rev. B},\ }\textbf
  {\bibinfo {volume} {84}},\ \bibinfo {pages} {195468} (\bibinfo {year}
  {2011}{\natexlab{b}})}\BibitemShut {NoStop}%
\bibitem [{\citenamefont {Manjavacas}\ and\ \citenamefont {Garc\'ia~de
  Abajo}(2010)}]{Manjavacas10}%
  \BibitemOpen
  \bibfield  {author} {\bibinfo {author} {\bibfnamefont {A.}~\bibnamefont
  {Manjavacas}}\ and\ \bibinfo {author} {\bibfnamefont {F.~J.}\ \bibnamefont
  {Garc\'ia~de Abajo}},\ }\Doi {10.1103/PhysRevA.82.063827} {\bibfield
  {journal} {\bibinfo  {journal} {Phys. Rev. A},\ }\textbf {\bibinfo {volume}
  {82}},\ \bibinfo {pages} {063827} (\bibinfo {year} {2010})}\BibitemShut
  {NoStop}%
\bibitem [{\citenamefont {Maghrebi}\ \emph {et~al.}(2012)\citenamefont
  {Maghrebi}, \citenamefont {Jaffe},\ and\ \citenamefont
  {Kardar}}]{Maghrebi11b}%
  \BibitemOpen
  \bibfield  {author} {\bibinfo {author} {\bibfnamefont {M.~F.}\ \bibnamefont
  {Maghrebi}}, \bibinfo {author} {\bibfnamefont {R.~L.}\ \bibnamefont {Jaffe}},
  \ and\ \bibinfo {author} {\bibfnamefont {M.}~\bibnamefont {Kardar}},\
  }\href@noop {} {\bibfield  {journal} {\bibinfo  {journal} {Phys. Rev.
  Lett.},\ }\textbf {\bibinfo {volume} {108}},\ \bibinfo {pages} {230403}
  (\bibinfo {year} {2012})}\BibitemShut {NoStop}%
\bibitem [{\citenamefont {Beenakker}(1999)}]{Beenakker99}%
  \BibitemOpen
  \bibfield  {author} {\bibinfo {author} {\bibfnamefont {C.~W.~J.}\
  \bibnamefont {Beenakker}},\ }\href@noop {} {} (\bibinfo {year} {1999}),\
  \bibinfo {note} {in: Diffuse Waves in Complex Media, edited by J.-P. Fouque,
  NATO Science Series C531 (Kluwer, Dordrecht, 1999): pp.
  137–164.}\BibitemShut {Stop}%
\bibitem [{\citenamefont {Eckhardt}(1984)}]{Eckhardt83}%
  \BibitemOpen
  \bibfield  {author} {\bibinfo {author} {\bibfnamefont {W.}~\bibnamefont
  {Eckhardt}},\ }\href@noop {} {\bibfield  {journal} {\bibinfo  {journal}
  {Phys. Rev. A},\ }\textbf {\bibinfo {volume} {29}},\ \bibinfo {pages} {1991}
  (\bibinfo {year} {1984})}\BibitemShut {NoStop}%
\bibitem [{\citenamefont {Lippmann}\ and\ \citenamefont
  {Schwinger}(1950)}]{Lippmann50}%
  \BibitemOpen
  \bibfield  {author} {\bibinfo {author} {\bibfnamefont {B.~A.}\ \bibnamefont
  {Lippmann}}\ and\ \bibinfo {author} {\bibfnamefont {J.}~\bibnamefont
  {Schwinger}},\ }\href@noop {} {\bibfield  {journal} {\bibinfo  {journal}
  {Phys. Rev.},\ }\textbf {\bibinfo {volume} {79}},\ \bibinfo {pages} {469}
  (\bibinfo {year} {1950})}\BibitemShut {NoStop}%
\bibitem [{\citenamefont {Jackson}(1998)}]{Jackson}%
  \BibitemOpen
  \bibfield  {author} {\bibinfo {author} {\bibfnamefont {J.~D.}\ \bibnamefont
  {Jackson}},\ }\href@noop {} {\emph {\bibinfo {title} {Classical
  Electrodynamics}}}\ (\bibinfo  {publisher} {Wiley},\ \bibinfo {address} {New
  York},\ \bibinfo {year} {1998})\BibitemShut {NoStop}%
\bibitem [{\citenamefont {Maghrebi}(2011)}]{Maghrebi11}%
  \BibitemOpen
  \bibfield  {author} {\bibinfo {author} {\bibfnamefont {M.~F.}\ \bibnamefont
  {Maghrebi}},\ }\href@noop {} {\bibfield  {journal} {\bibinfo  {journal}
  {Phys. Rev. D},\ }\textbf {\bibinfo {volume} {83}},\ \bibinfo {pages}
  {045004} (\bibinfo {year} {2011})}\BibitemShut {NoStop}%
\bibitem [{\citenamefont {Reid}\ \emph {et~al.}(2011)\citenamefont {Reid},
  \citenamefont {White},\ and\ \citenamefont {Johnson}}]{Reid}%
  \BibitemOpen
  \bibfield  {author} {\bibinfo {author} {\bibfnamefont {M.~T.~H.}\
  \bibnamefont {Reid}}, \bibinfo {author} {\bibfnamefont {J.}~\bibnamefont
  {White}}, \ and\ \bibinfo {author} {\bibfnamefont {S.~G.}\ \bibnamefont
  {Johnson}},\ }\href@noop {} {\bibfield  {journal} {\bibinfo  {journal} {Phys.
  Rev. A},\ }\textbf {\bibinfo {volume} {84}},\ \bibinfo {pages} {010503(R)}
  (\bibinfo {year} {2011})}\BibitemShut {NoStop}%
\bibitem [{\citenamefont {Emig}\ \emph {et~al.}(2008)\citenamefont {Emig},
  \citenamefont {Graham}, \citenamefont {Jaffe},\ and\ \citenamefont
  {Kardar}}]{Emig08}%
  \BibitemOpen
  \bibfield  {author} {\bibinfo {author} {\bibfnamefont {T.}~\bibnamefont
  {Emig}}, \bibinfo {author} {\bibfnamefont {N.}~\bibnamefont {Graham}},
  \bibinfo {author} {\bibfnamefont {R.~L.}\ \bibnamefont {Jaffe}}, \ and\
  \bibinfo {author} {\bibfnamefont {M.}~\bibnamefont {Kardar}},\ }\href@noop {}
  {\bibfield  {journal} {\bibinfo  {journal} {{Phys. Rev. D}},\ }\textbf
  {\bibinfo {volume} {{77}}},\ \bibinfo {pages} {025005}
  (\bibinfo {year} {{2008}})}\BibitemShut {NoStop}%
\bibitem [{\citenamefont {Tsang}\ \emph {et~al.}(2000)\citenamefont {Tsang},
  \citenamefont {Kong},\ and\ \citenamefont {Ding}}]{Tsang}%
  \BibitemOpen
  \bibfield  {author} {\bibinfo {author} {\bibfnamefont {L.}~\bibnamefont
  {Tsang}}, \bibinfo {author} {\bibfnamefont {J.~A.}\ \bibnamefont {Kong}}, \
  and\ \bibinfo {author} {\bibfnamefont {K.-H.}\ \bibnamefont {Ding}},\
  }\href@noop {} {\emph {\bibinfo {title} {Scattering of Electromagnetic
  Waves}}}\ (\bibinfo  {publisher} {Wiley},\ \bibinfo {address} {New York},\
  \bibinfo {year} {2000})\BibitemShut {NoStop}%
\bibitem [{\citenamefont {Wittmann}(1988)}]{Wittmann88}%
  \BibitemOpen
  \bibfield  {author} {\bibinfo {author} {\bibfnamefont {R.~C.}\ \bibnamefont
  {Wittmann}},\ }\href@noop {} {\bibfield  {journal} {\bibinfo  {journal} {IEEE
  Transactions on antennas and propagation},\ }\textbf {\bibinfo {volume}
  {36}},\ \bibinfo {pages} {1078} (\bibinfo {year} {1988})}\BibitemShut
  {NoStop}%
\bibitem [{\citenamefont {Zandi}\ \emph {et~al.}(2010)\citenamefont {Zandi},
  \citenamefont {Emig},\ and\ \citenamefont {Mohideen}}]{Zandi10}%
  \BibitemOpen
  \bibfield  {author} {\bibinfo {author} {\bibfnamefont {R.}~\bibnamefont
  {Zandi}}, \bibinfo {author} {\bibfnamefont {T.}~\bibnamefont {Emig}}, \ and\
  \bibinfo {author} {\bibfnamefont {U.}~\bibnamefont {Mohideen}},\ }\href@noop
  {} {\bibfield  {journal} {\bibinfo  {journal} {Phys. Rev. B},\ }\textbf
  {\bibinfo {volume} {81}},\ \bibinfo {pages} {195423} (\bibinfo {year}
  {2010})}\BibitemShut {NoStop}%
\bibitem [{\citenamefont {Sasihithlu}\ and\ \citenamefont
  {Narayanaswamy}(2011){\natexlab{b}}}]{Sasihithlu11a}%
  \BibitemOpen
  \bibfield  {author} {\bibinfo {author} {\bibfnamefont {K.}~\bibnamefont
  {Sasihithlu}}\ and\ \bibinfo {author} {\bibfnamefont {A.}~\bibnamefont
  {Narayanaswamy}},\ }\Doi {10.1364/OE.19.00A772} {\bibfield  {journal}
  {\bibinfo  {journal} {Opt. Express},\ }\textbf {\bibinfo {volume} {19}},\
  \bibinfo {pages} {A772} (\bibinfo {year} {2011}{\natexlab{b}})}\BibitemShut
  {NoStop}%
\bibitem [{\citenamefont {Landau}\ and\ \citenamefont
  {Lifshitz}(1984)}]{Landauel}%
  \BibitemOpen
  \bibfield  {author} {\bibinfo {author} {\bibfnamefont {L.~D.}\ \bibnamefont
  {Landau}}\ and\ \bibinfo {author} {\bibfnamefont {E.~M.}\ \bibnamefont
  {Lifshitz}},\ }\href@noop {} {\emph {\bibinfo {title} {Electrodynamics of
  continuous media}}}\ (\bibinfo  {publisher} {Pergamon},\ \bibinfo {address}
  {Oxford},\ \bibinfo {year} {1984})\BibitemShut {NoStop}%
\bibitem [{\citenamefont {Noruzifar}\ \emph {et~al.}(2011)\citenamefont
  {Noruzifar}, \citenamefont {Emig},\ and\ \citenamefont
  {Zandi}}]{Noruzifar11}%
  \BibitemOpen
  \bibfield  {author} {\bibinfo {author} {\bibfnamefont {E.}~\bibnamefont
  {Noruzifar}}, \bibinfo {author} {\bibfnamefont {T.}~\bibnamefont {Emig}}, \
  and\ \bibinfo {author} {\bibfnamefont {R.}~\bibnamefont {Zandi}},\
  }\href@noop {} {\bibfield  {journal} {\bibinfo  {journal} {{Phys. Rev. A}},\
  }\textbf {\bibinfo {volume} {{84}}}, \bibinfo {pages} {042501} (\bibinfo {year} {{2011}})}\BibitemShut
  {NoStop}%
\bibitem [{\citenamefont {Crichton}\ and\ \citenamefont
  {Marston}(2000)}]{Crichton00}%
  \BibitemOpen
  \bibfield  {author} {\bibinfo {author} {\bibfnamefont {J.~H.}\ \bibnamefont
  {Crichton}}\ and\ \bibinfo {author} {\bibfnamefont {P.~L.}\ \bibnamefont
  {Marston}},\ }\href@noop {} {\bibfield  {journal} {\bibinfo  {journal}
  {Electronic Journal of Differential Equations},\ }\textbf {\bibinfo {volume}
  {04}},\ \bibinfo {pages} {37} (\bibinfo {year} {2000})}\BibitemShut {NoStop}%
\bibitem [{\citenamefont {Antezza}\ \emph {et~al.}(2004)\citenamefont
  {Antezza}, \citenamefont {Pitaevskii},\ and\ \citenamefont
  {Stringari}}]{Antezza04}%
  \BibitemOpen
  \bibfield  {author} {\bibinfo {author} {\bibfnamefont {M.}~\bibnamefont
  {Antezza}}, \bibinfo {author} {\bibfnamefont {L.~P.}\ \bibnamefont
  {Pitaevskii}}, \ and\ \bibinfo {author} {\bibfnamefont {S.}~\bibnamefont
  {Stringari}},\ }\href@noop {} {\bibfield  {journal} {\bibinfo  {journal}
  {Phys. Rev. A},\ }\textbf {\bibinfo {volume} {70}},\ \bibinfo {pages}
  {053619} (\bibinfo {year} {2004})}\BibitemShut {NoStop}%
\bibitem [{\citenamefont {Zeman}\ and\ \citenamefont {Schatz}(1987)}]{Zeman87}%
  \BibitemOpen
  \bibfield  {author} {\bibinfo {author} {\bibfnamefont {E.~J.}\ \bibnamefont
  {Zeman}}\ and\ \bibinfo {author} {\bibfnamefont {G.~C.}\ \bibnamefont
  {Schatz}},\ }\href@noop {} {\bibfield  {journal} {\bibinfo  {journal} {J.
  Phys. Chem.},\ }\textbf {\bibinfo {volume} {91}},\ \bibinfo {pages} {634}
  (\bibinfo {year} {1987})}\BibitemShut {NoStop}%
\bibitem [{\citenamefont {Spitzer}\ \emph {et~al.}(1959)\citenamefont
  {Spitzer}, \citenamefont {Kleinmann},\ and\ \citenamefont
  {Walsh}}]{Spitzer59}%
  \BibitemOpen
  \bibfield  {author} {\bibinfo {author} {\bibfnamefont {W.~G.}\ \bibnamefont
  {Spitzer}}, \bibinfo {author} {\bibfnamefont {D.}~\bibnamefont {Kleinmann}},
  \ and\ \bibinfo {author} {\bibfnamefont {D.}~\bibnamefont {Walsh}},\
  }\href@noop {} {\bibfield  {journal} {\bibinfo  {journal} {Phys. Rev.},\
  }\textbf {\bibinfo {volume} {113}},\ \bibinfo {pages} {127} (\bibinfo {year}
  {1959})}\BibitemShut {NoStop}%
\bibitem [{\citenamefont {Kittel}(2005)}]{Kittel}%
  \BibitemOpen
  \bibfield  {author} {\bibinfo {author} {\bibfnamefont {C.}~\bibnamefont
  {Kittel}},\ }\href@noop {} {\emph {\bibinfo {title} {Introduction to Solid
  State Physics}}}\ (\bibinfo  {publisher} {Wiley},\ \bibinfo {year}
  {2005})\BibitemShut {NoStop}%
\end{thebibliography}
\end{document}